\documentclass[twocolumn]{aastex631}

\usepackage{graphicx}	% Including figure files
\usepackage{amsmath}	% Advanced maths commands
\usepackage{amssymb}	% Extra maths symbols

\usepackage{amsmath}
\usepackage{bm}
\usepackage{newtxtext,newtxmath}
\usepackage{graphicx}	% Including figure files
\usepackage{grffile}
\usepackage{amsmath}	% Advanced maths commands
\usepackage{amssymb}	% Extra maths symbols

\usepackage{epsfig,amsmath,natbib}
\usepackage{color,subfigure}
\usepackage{xcolor}

\usepackage{mathrsfs,amssymb,amstext}
\usepackage{ulem}
\usepackage{url}
\usepackage{bm}
\usepackage[T1]{fontenc}

\DeclareRobustCommand{\VAN}[3]{#2}
\let\VANthebibliography\thebibliography
\def\thebibliography{\DeclareRobustCommand{\VAN}[3]{##3}\VANthebibliography}

\def\mean#1{\left< #1 \right>}

\newcommand{\blue}[1]{\textcolor{blue}{#1}}

\begin{document}

\title{Velocity acoustic oscillations on Cosmic Dawn 21 cm power spectrum as a probe of small-scale density fluctuations}

\correspondingauthor{Bin Yue; Yan Gong; Xuelei Chen}
\email{yuebin@nao.cas.cn; gongyan@nao.cas.cn; xuelei@cosmology.bao.ac.cn}

\author{Xin Zhang}
\affiliation{National Astronomical Observatories, Chinese Academy of Sciences, 20A Datun Road, Chaoyang District, Beijing 100101, China}

\affiliation{School of Astronomy and Space Science, University of Chinese Academy of Sciences, Beijing 100049, China}

\author{Hengjie Lin}
\affiliation{National Astronomical Observatories, Chinese Academy of Sciences, 20A Datun Road, Chaoyang District, Beijing 100101, China}

\affiliation{School of Astronomy and Space Science, University of Chinese Academy of Sciences, Beijing 100049, China}

\author{Meng Zhang}
\affiliation{National Astronomical Observatories, Chinese Academy of Sciences, 20A Datun Road, Chaoyang District, Beijing 100101, China}

\affiliation{School of Astronomy and Space Science, University of Chinese Academy of Sciences, Beijing 100049, China}

\author{Bin Yue}
\affiliation{National Astronomical Observatories, Chinese Academy of Sciences, 20A Datun Road, Chaoyang District, Beijing 100101, China}

\affiliation{Key Laboratory of Radio Astronomy and Technology, Chinese Academy of Sciences, 20A Datun
Road, Chaoyang District, Beijing 100101, China}

\author{Yan Gong}
\affiliation{National Astronomical Observatories, Chinese Academy of Sciences, 20A Datun Road, Chaoyang District, Beijing 100101, China}

\affiliation{Key Laboratory of Radio Astronomy and Technology, Chinese Academy of Sciences, 20A Datun
Road, Chaoyang District, Beijing 100101, China} 

\author{Yidong Xu}
\affiliation{National Astronomical Observatories, Chinese Academy of Sciences, 20A Datun Road, Chaoyang District, Beijing 100101, China}

\affiliation{Key Laboratory of Radio Astronomy and Technology, Chinese Academy of Sciences, 20A Datun
Road, Chaoyang District, Beijing 100101, China}

\author{Xuelei Chen}
\affiliation{National Astronomical Observatories, Chinese Academy of Sciences, 20A Datun Road, Chaoyang District, Beijing 100101, China}

\affiliation{School of Astronomy and Space Science, University of Chinese Academy of Sciences, Beijing 100049, China}

\affiliation{Key Laboratory of Radio Astronomy and Technology, Chinese Academy of Sciences, 20A Datun
Road, Chaoyang District, Beijing 100101, China}

\affiliation{Key Laboratory of Cosmology and Astrophysics (Liaoning) \& College of Sciences, Northeastern
University, Shenyang 110819, China}

\affiliation{ Center of High Energy Physics, Peking University, Beijing 100871, China}

\shorttitle{VAO \& small-scale fluctuations}
\shortauthors{Zhang et al.}



\begin{abstract}
We investigate the feasibility of using the velocity acoustic oscillations (VAO) features on the Cosmic Dawn 21 cm power spectrum to probe small-scale density fluctuations. In the standard cold dark matter (CDM) model, Pop III stars form in minihalos and affect the 21 cm signal through Ly$\alpha$ and X-ray radiation. Such a process is modulated by the relative motion between dark matter and baryons, generating the VAO wiggles on the 21 cm power spectrum. In the fuzzy or warm dark matter models for which the number of minihalos is reduced, the VAO wiggles are weaker or even fully invisible. We investigate the wiggle features in the CDM with different astrophysical models and in different dark matter models. We find: 1) In the CDM model the relative streaming velocities can generate the VAO wiggles for broad ranges of parameters $f_*$, $\zeta_X$ and $f_{\rm esc,LW}\zeta_{\rm LW}$, though for different parameters the wiggles would appear at different redshifts and have different amplitudes. 2) For the axion model with $m_{\rm a} \lesssim10^{-19}$ eV, the VAO wiggles are negligible. In the mixed model, the VAO signal is sensitive to the axion fraction. For example, the wiggles almost disappear when $f_{\rm a} \gtrsim 10\%$ for $m_{\rm a}=10^{-21}$ eV. Therefore, the VAO signal can be an effective  
indicator for small-scale density fluctuations and a useful probe of the nature of dark matter. 
The SKA-low with $\sim$2000 hour observation time has the ability to detect the VAO signal and constraint dark matter models.
\end{abstract}



\section{Introduction}

The hyperfine transition line of the neutral Hydrogen, named 21 cm line, is one of the most powerful tools that can efficiently map the 3D large-scale structures of the Universe from the recombination era to present day. This signal depends on the density of neutral Hydrogen, the spin temperature and the CMB temperature. The brightness temperature of the signal writes (e.g. \citealt{Mesinger2011MNRAS,Barkana2001PhR,Furlanetto2006PhR,Pritchard2012RPPh})
\begin{equation}
\delta T_{\rm b}=27x_{\rm HI}(1+\delta) \left( \frac{T_{\rm s}-T_{\rm CMB}}{T_{\rm s}}  \right) \left( \frac{1+z}{10}\right)^{1/2}\left( \frac{0.15}{\Omega_{\rm m} h^2} \right)^{1/2} \left( \frac{\Omega_{\rm b}h^2}{0.023}\right)~~~{\rm mK},
\label{eq:delta_T_b}
\end{equation}
where $x_{\rm HI}$ is the neutral fraction of the hydrogen; $\delta$ is the over-density of the gas; $T_{\rm CMB}$ is the CMB temperature at redshift $z$;  $\Omega_{\rm m}$, $\Omega_{\rm b}$ and $h$ are the matter relative density and Hubble constant respectively. The spin temperature of Hydrogen atoms is determined jointly by the scattering with CMB photons and Ly$\alpha$ photons, and the collisions between baryonic particles,
\begin{equation}
T_{\rm s}^{-1}=\frac{T_{\rm CMB}^{-1}+x_\alpha T_{\rm k}^{-1}+x_{\rm c} T_{\rm k}^{-1}}{1+x_\alpha+x_{\rm c}},
\end{equation}
where $x_\alpha$ and $x_{\rm c}$ are Ly$\alpha$ coupling (the Wouthuysen-Field effect, \citealt{Wouthuysen1952AJ,Field1958PIRE}) and collision coupling coefficients respectively; $T_{\rm k}$ is the kinetic temperature of the inter-galactic medium (IGM).
The 21 cm signal depends on the Ly$\alpha$, the X-ray, and the ionizing radiation fields. 
At the Cosmic Dawn, these fields are built up mainly by the formation of Pop III stars in minihalos. Therefore the 21 cm signal reflects not only the large-scale distribution of the neutral hydrogen, but also the mechanisms that suppress or promote the formation of minihalos and the star formation therein.

After the decoupling of dark matter, and before the recombination era, the baryons are coupled tightly with the photons, while the dark matter evolves almost independently. As a result, there are large-scale relative motions between the dark matter and baryonic matter.
The root mean square (rms) of the relative streaming velocities $\approx 30$ km s$^{-1}$  at the recombination era \citep{Tseliakhovich2010PhRvD}. After recombination, the rms of the relative streaming velocities decays as $\sigma_{\rm rms}=\mean{v^2_{\rm db}}^{1/2}\approx 30(1+z)/(1+z_{\rm rec})$ km s$^{-1}$ as the Universe expands,  where $z_{\rm rec}$ is the redshift of recombination.
Because of the baryon acoustic oscillations before recombination, the relative streaming velocity was also coherently oscillating, and it imprints oscillatory wiggles on the power spectrum of the large-scale streaming velocity field after recombination (e.g. \citealt{Tseliakhovich2011MNRAS}).

The relative streaming velocity varies in space, and it can prevent the gas from collapsing into dark matter halos with circular velocities smaller than or comparable to the local streaming velocity. Many simulations show that Pop III stars form first in minihalos with mass $\sim10^5~M_\odot$ through H$_2$ cooling (e.g. \citealt{Haiman1996ApJa,Tegmark1997ApJ,Yoshida2008Sci}). Such H$_2$-cooling minihalos have typical circular velocities of $\sim 3$ km s$^{-1}$, comparable with the relative velocity which has $\sigma_{\rm rms} \approx 0.6$ km s$^{-1}$ at $z=20$. 
The formation of Pop III stars in those minihalos 
located in regions with streaming velocities $\gtrsim 1\sigma_{\rm rms}$ 
would be significantly suppressed \citep{Greif2011ApJ,Fialkov2012MNRAS}. The spatial distribution of the star-forming minihalos will reflect the large-scale features of the relative streaming velocity field \citep{Tseliakhovich2010PhRvD,Tseliakhovich2011MNRAS}.

The radiation fields of the Pop III stars (i.e. Ly$\alpha$, X-ray, and ionizing photons) should show large-scale features corresponding to the structures of the streaming velocity field, and finally this would in turn affect the large-scale distribution of the 21 cm signal \citep{Dalal2010JCAP,Visbal2012Natur,Fialkov2013MNRAS,Ali-Haimoud2014PhRvD,Munoz2019PhRvD,Munoz2020PhRvD,Munoz2022MNRAS,Schauer2023ApJ,Conaboy2022arXiv,Hegde2023arXiv}.
Furthermore, the dark matter/baryon relative motion also suppresses the growth of small-scale perturbations, and influences the 21 cm forest \citep{Shimabukuro2023PhRvD}.
Basically, the overall shape and amplitude of the 21 cm power spectrum will be changed by the relative streaming velocities. 
More importantly, the large-scale coherently oscillating 
structures would leave imprints on the 21 cm signal \--- generate some wiggles on the 21 cm power spectrum \citep{McQuinn2012ApJ,Ali-Haimoud2014PhRvD}. Such velocity acoustic oscillations (VAO) features are useful tools for constraining the cosmology (e.g. \citealt{Munoz2019PhRvL,Kovez2023PhRvD}) and astrophysical characteristics of Pop III stars. 

Although the cold dark matter (CDM) model has been taken as the standard model of cosmology, the dark matter could alternatively be warm or fuzzy (e.g. \citealt{Newton2021JCAP,Dekker2022PhRvD,Dayal2023arXiv,Vid2017PhRvL,2021PhRvL.126i1101N,Marsh2019PhRvL}). A distinguishing character of these models is their 
small-scale density fluctuations.
In the warm dark matter (WDM) or fuzzy dark matter (FDM) models, small-scale fluctuations are suppressed and the abundance of minihalos are significantly reduced. As a result, the build-up of Ly$\alpha$, X-ray, and ionizing radiation fields is delayed, and the global evolution and spatial fluctuations of the 21 cm signal change. Therefore the 21 cm signal can be a useful tool to distinguish the different dark matter models (e.g. \citealt{Shao2023NatAs,2022ApJ...929..151H}). 
In the WDM or FDM model, the number of small halos is reduced, their contribution to the ionizing photons budget is also reduced. To be consistent with the constraints on reionization history, massive halos may hold a larger amount of stellar mass. This can explain the high-$z$ stellar mass excess \citep{Gong2023ApJ,Lin2023arXiv}.   
Note however, since the clumpiness of the IGM is also reduced in the WDM model, the reionization process is not necessarily delayed \citep{Yue2012ApJ}.

In addition to the relative streaming velocities, the Lyman-Werner (LW) radiation (e.g. \citealt{Omukai2001ApJ,Machacek2001ApJ,Wise2007ApJ,OShea2008ApJ}) and X-ray radiation (e.g. \citealt{Ricotti2016MNRAS,Park2021aMNRAS,Park2021bMNRAS,Park2023MNRAS}) from previously-formed Pop III stars can also suppress the Pop III stars formation in minihalos, boosting the critical mass of minihalos that can host Pop III stars. Since the influence of relative streaming velocities on the 21 cm signal is more sensitive to the smaller minihalos, the LW and X-ray feedback effects may significantly reduce the influence of relative motion.

\citet{Sarkar2022PhRvD} investigated the 21 cm global spectrum and power spectrum in the FDM model, in the presence of dark matter/baryon relative motion and LW feedback, and the CMB and Ly$\alpha$ heating. They predicted that HERA should be able to constrain the FDM masses up to $\sim10^{-19}-10^{-18}$ eV. 
\citet{Flitter2022PhRvD} further proposed that HERA can detect the FDM  down to a fraction of $\sim1\%$ in the mixed dark matter model. 
In \citet{Vanzan2023arXiv} they investigated the feasibility of using the 21 cm angular power spectrum, influenced by both small-scale suppression and the relative motion, to constrain the ultra-light axions for various array configurations.

The large-scale VAO wiggles on the 21 cm power spectrum are modulated by the formation of Pop III stars in minihalos, whose abundance is rather sensitive to the small-scale fluctuations of the density field. Such wiggles, regardless of the overall shape and amplitude of the 21 cm power spectrum, can 
be a good indicator for small-scale density fluctuations. \citet{Hotinli2022PhRvD} proposed that the large-scale VAO wiggles on the 21 cm power spectrum can be used to measure the mass of ultra-light axions, and predicted that the HERA experiment can detect such signal when the axion mass $m_{\rm a}\gtrsim 10^{-18}$ eV, much more sensitive than current constraints.

The LW feedback is the main negative mechanism that reduces the amplitude of the VAO wiggles. Sometimes it even makes the VAO signal less efficient in distinguishing the CDM model with strong feedback, and the WDM/FDM model. 
The effect depends on the properties of Pop III stars. Currently, there are only indirect constraints on the properties of Pop III stars (e.g. \citealt{Xing2023Natur,Maiolino2023arXiv,Mebane2020MNRAS,Chatterjee2020MNRAS,Bevins2022NatAs}). 
It is therefore worthwhile to consider the evolution and amplitude of the VAO signal for wide ranges of Pop III star parameters, and investigate the feasibility of distinguishing CDM and non-CDM dark matter models (e.g. the FDM or WDM), or constraining the fraction of the non-CDM component in mixed dark matter models, by the upcoming large radio interferometer arrays such as the SKA-low telescope. This is the basic motivation of this work.

The outline of this paper is: in Sec. \ref{sec:method} we introduce the methods to include  
the streaming velocivites and LW feedback effects in calculating the 21 cm signal from the Cosmic Dawn. In Sec. \ref{sec:results} we present our results of the VAO signal on the 21 cm power spectrum for various astrophysical parameters and axion masses. In Sec. \ref{sec:conclusion} we give the summary and conclusion.

\section{Methods}\label{sec:method}

\subsection{The collapse fraction field modulated by relative streaming velocities} 

The power spectrum of dark matter (d) or baryon (b)
\begin{equation}
P_{\rm d/b}(k,z)=A_s k^{n_s} T_{\rm d/b}^2(k,z),
\end{equation}
where $P_{\rm pri}=A_s k^{n_s}$ is the power spectrum of the primordial overdensity $\delta_{\rm pri}$; $T_{\rm d/b}(k,z)$ is the transfer function of dark matter or baryon at redshift $z$. Throughout this paper, we compute the transfer functions using the {\tt AxionCAMB} program \citep{axionCAMB}, which is a modified version of the cosmology code {\tt CAMB} \citep{CAMB}.

According to the linearized continuity equation
\begin{equation}
\dot{\delta}+a^{-1}\boldsymbol{\nabla}\cdot{\boldsymbol{v}}=0,
\end{equation}
the velocity of dark matter ($\boldsymbol{v}_{\rm d}$) or baryon ($\boldsymbol{v}_{\rm b}$) in Fourier space is  
\begin{equation}
\boldsymbol{v}_{\rm d/b}({\boldsymbol{k},a})=-i\frac{a\boldsymbol{k}}{k^2}\dot{\delta}_{\rm d/b}({\boldsymbol{k},a})=-i\frac{a\boldsymbol{k}}{k^2}\dot{T}_{\rm d/b}({\boldsymbol{k},a})
\delta_{\rm pri}({\boldsymbol{k}}),
\label{vk}
\end{equation}
where $a=1/(1+z)$ is the scale factor.
To make predictions of the 21cm signal before reionization, we modify the original version of 
{\tt 21cmFAST} \citep{Mesinger2011MNRAS} to include the effect of relative motion between dark matter and baryon.
The velocity fields of both the dark matter and the baryon is obtained according to Eq. (\ref{vk}), then the speed of the relative motion $\boldsymbol{v}_{\rm db}(\boldsymbol{x},z) =\boldsymbol{v}_{\rm d}(\boldsymbol{x},z) -\boldsymbol{v}_{\rm b}(\boldsymbol{x},z)$.
We generate random density and velocity fields on $120^3$ cells in a box with comoving side length 2000~Mpc. By comparing the results of simulations with the same box size but different resolutions, we find that such a choice is enough for capturing the converged large-scale VAO wiggles, see the Sec. \ref{sec:convergence}.

The streaming velocities between dark matter and baryon could prevent the IGM gas from being
accreted into smaller minihalos, so that Pop III stars can never form in
such halos (e.g. \citealt{Fialkov2012MNRAS}). Moreover, the LW radiation also suppresses the star formation in smaller minihalos and reduces the VAO wiggles on the 21 cm signal (e.g. \citealt{Fialkov2013MNRAS}).
\citet{Schauer2021MNRAS} and \citet{Kulkarni2021ApJ} first performed high-resolution numerical simulations to find the critical minihalo mass for Pop III stars formation in the presence of LW feedback and relative streaming velocities simultaneously. Since \citet{Kulkarni2021ApJ} explored the LW specific intensity range wider than \citet{Schauer2021MNRAS},
we adopt their critical halo mass as the fiducial model. We will discuss the results using \citet{Schauer2021MNRAS} critical mass and other critical masses in Sec. \ref{sec:critical-mass}.

The \citet{Kulkarni2021ApJ}  critical mass writes
\begin{equation}
M_{\rm crit}(J_{\rm LW,21}, v^{\rm rec}_{\rm db}, z)=M_{z=20}(J_{\rm LW,21},v_{\rm db}^{\rm rec}) \left( \frac{1+z}{21}\right)^{-\alpha(J_{\rm LW,21},v_{\rm bc}^{\rm rec})},
\label{eq:M_crit_K21}
\end{equation}
where $J_{\rm LW,21}$ is the specific intensity of LW radiation in units of $10^{-21}$ erg s$^{-1}$ cm$^{-2}$Hz$^{-1}$sr$^{-1}$, $v_{\rm db}^{\rm rec}$ is the dark matter-baryon streaming velocity at recombination era.  
$M_{z=20}$ and $\alpha$ are fitted well by
\begin{equation}
M_{z=20}=(M_{z=20})_0(1+J_{\rm LW,21})^{\beta_1} (1+v_{\rm db}^{\rm rec}/30)^{\beta_2}(1+J_{\rm LW,21} v_{\rm db}^{\rm rec}/3)^{\beta_3},
\end{equation}
and
\begin{equation}
\alpha=\alpha_0(1+J_{\rm LW,21})^{\gamma_1} (1+v_{\rm db}^{\rm rec}/30)^{\gamma_2}(1+J_{\rm LW,21} v_{\rm db}^{\rm rec}/3)^{\gamma_3},
\end{equation}
with parameters: $(M_{z=20})_0=1.96\times10^5~M_\odot$, $\beta_1=0.80$, $\beta_2=1.83$, $\beta_3=-0.06$, and $\alpha_0=1.64$, $\gamma_1=0.36$, $\gamma_2=-0.62$, $\gamma_3=0.13$.

The halo collapse fraction for a cell with overdensity $\delta_{\rm cell}$, mass $M_{\rm cell}$ and 
streaming velocity $v_{\rm db}$ is 
\begin{equation}
f_{\rm coll}(z|\delta_{\rm cell},M_{\rm cell},v_{\rm db})=\frac{1}{\bar{\rho}_{\rm m}}\int_{M_{\rm crit}}^\infty dM M\frac{dn}{dM}(M,z|\delta_{\rm cell},M_{\rm cell}),
\label{fcoll}
\end{equation}
where $\bar{\rho}_{\rm m}$ is the mean density of the total matter in the Universe.
The conditional halo mass function $dn/dM$ in the cell is calculated from the Extended Press-Schechter formalism \citep{Lacey1993MNRAS,Mo1996MNRAS,Sheth2002MNRAS}, 
\begin{align}
\frac{dn}{dM}(M,z|\delta_{\rm cell},M_{\rm cell})&=\frac{1}{\sqrt{2\pi}} \frac{\bar{\rho}_{\rm m}}{M} \frac{ \delta_c(z) -\delta_{\rm cell}/D_{\rm lin}(z)  }{ [ S(M)-S(M_{\rm cell}) ]^{3/2} } \nonumber \\
&\exp\left(-\frac{ [\delta_c(z) -\delta_{\rm cell}/D_{\rm lin}(z)]^2  }{ 2[ S(M)-S(M_{\rm cell}) ] }\right)\frac{dS}{dM},
\end{align}
where $D_{\rm lin}(z)$ is the linear growth factor and $S$ is the variance of density fluctuations at mass scale $M$. In the standard CDM model the collapse barrier is simply mass-independent, $\delta_c(z)=1.686/D_{\rm lin}(z)$. In the non-CDM model, however, the collapse barrier depends on mass. We follow the \citet{Marsh2014MNRAS} to define a mass-dependent critical overdensity for the non-CDM (axion or axion-CDM mixed) model, 
\begin{equation}
\delta^{\rm non-CDM}_c(M,z)=\frac{ 1.686}{D_{\rm non-CDM}(M,z)},
\end{equation}
with the growth factor 
\begin{equation}
D_{\rm non-CDM}(M,z)=\frac{T_{\rm non-CDM}(k_M,z) T_{\rm non-CDM}(k_0,z_h)  }{T_{\rm non-CDM}(k_M,z_h) T_{\rm non-CDM}(k_0,z)},
\end{equation}
where $k_M=\pi/[3M/(4\pi)]^{1/3}$, $k_0=0.002h$ Mpc$^{-1}$ and $z_h=300$. The transfer function $T_{\rm non-CDM}$ is calculated by {\tt axionCAMB}.

Different from the original steps of {\tt 21cmFAST} where the density field is first smoothed and then the collapse fraction is calculated by the smoothed density, here we first calculate the collapse fraction for each cell of our random field by the EPS formalism 
\citep{Lacey1993MNRAS,Mo1996MNRAS,Sheth2002MNRAS}, then we smooth the collapse fraction field  
on different scales directly by Fast Fourier Transform (FFT), to calculate the flux of X-ray and Ly$\alpha$ photons from Pop III stars.

The specific intensity of the LW radiation at redshift $z$
\begin{equation}
J_{\rm LW}(z)=\frac{(1+z)^3}{4\pi} f_{\rm esc,LW}\int_z^{z_{\rm up}} \epsilon_{\rm LW}(z')\frac{c dz'}{H(z')(1+z')},
\end{equation}
where $f_{\rm esc,LW}$ is the escape fraction of the LW radiation, $z_{\rm up}=13.6/11.2(1+z)-1$, $H(z)$ is the Hubble parameter and $c$ is the light speed. The LW emissivity is given by
\begin{align}
\epsilon_{\rm LW}(z')=\zeta_{\rm LW} n_b f_* \frac{\Omega_{\rm b}}{\Omega_{\rm m}} \frac{ d f_{\rm coll}}{dt}(z'), 
\end{align}
where $n_b$ is the number density of baryon atoms in the present Universe, $\zeta_{\rm LW}$ is the total released LW energy per stellar atom throughout the stellar lifetime, divided by the LW bandwidth. {\tt 21cmFAST} adopted the Pop III star spectrum in \citet{Barkana2005ApJ}, the corresponding $\zeta^{\rm BL05}_{\rm LW}\sim 1.1\times 10^{-21}$ erg/Hz. But we will explore the VAO signal for a wide range of $\zeta_{\rm LW}$ values. Finally, we calculate the LW intensity of each cell using 
\begin{equation}
J_{\rm LW}(z)=\frac{(1+z)^3}{4\pi} f_{\rm esc,LW}\zeta_{\rm LW} n_b f_* \frac{\Omega_{\rm b}}{\Omega_{\rm m}} c \int_z^{z_{\rm up}}\frac{d f_{\rm coll}}{dz'}dz'.
\end{equation}
Regarding the $f_{\rm esc,LW}$, it depends on both stellar mass and host halo mass \citep{Schauer2015MNRAS,Schauer2017MNRAS}. The result of the moderate model in \citet{Schauer2015MNRAS}  is $f_{\rm esc,LW}\sim 0.5$. In this paper, we treat $f_{\rm esc,LW}\zeta_{\rm LW}$ as a joint parameter.

In addition to the LW radiation, the X-ray radiation also influences the formation of Pop III stars. Different from the LW radiation that almost purely plays a negative role, the effect of X-ray radiation is more complicated. Strong X-ray radiation will, of course, suppress the formation of Pop III stars, since it heats and ionizes the IGM. However, moderate X-ray radiation may promote the formation because it generates lots of free electrons that help H$_2$ to form (e.g. \citealt{Haiman2000ApJ,Ricotti2016MNRAS,Park2021aMNRAS,Park2021bMNRAS,Park2023MNRAS} and references therein). For this reason, we will not model X-ray feedback on Pop III star formation in our paper. Instead, we limit our investigation in the redshift range where the X-ray feedback is not the dominant effect, say $T_{\rm k}\lesssim 10^3$ K. In all our cases, when the VAO wiggles reach maximum amplitude, $T_{\rm k}$ is still smaller than or comparable to the CMB temperature. We therefore expect the X-ray feedback will not change our conclusion on the VAO signal.

\begin{figure*}
\centering{
\subfigure{\includegraphics[width=0.45\textwidth]{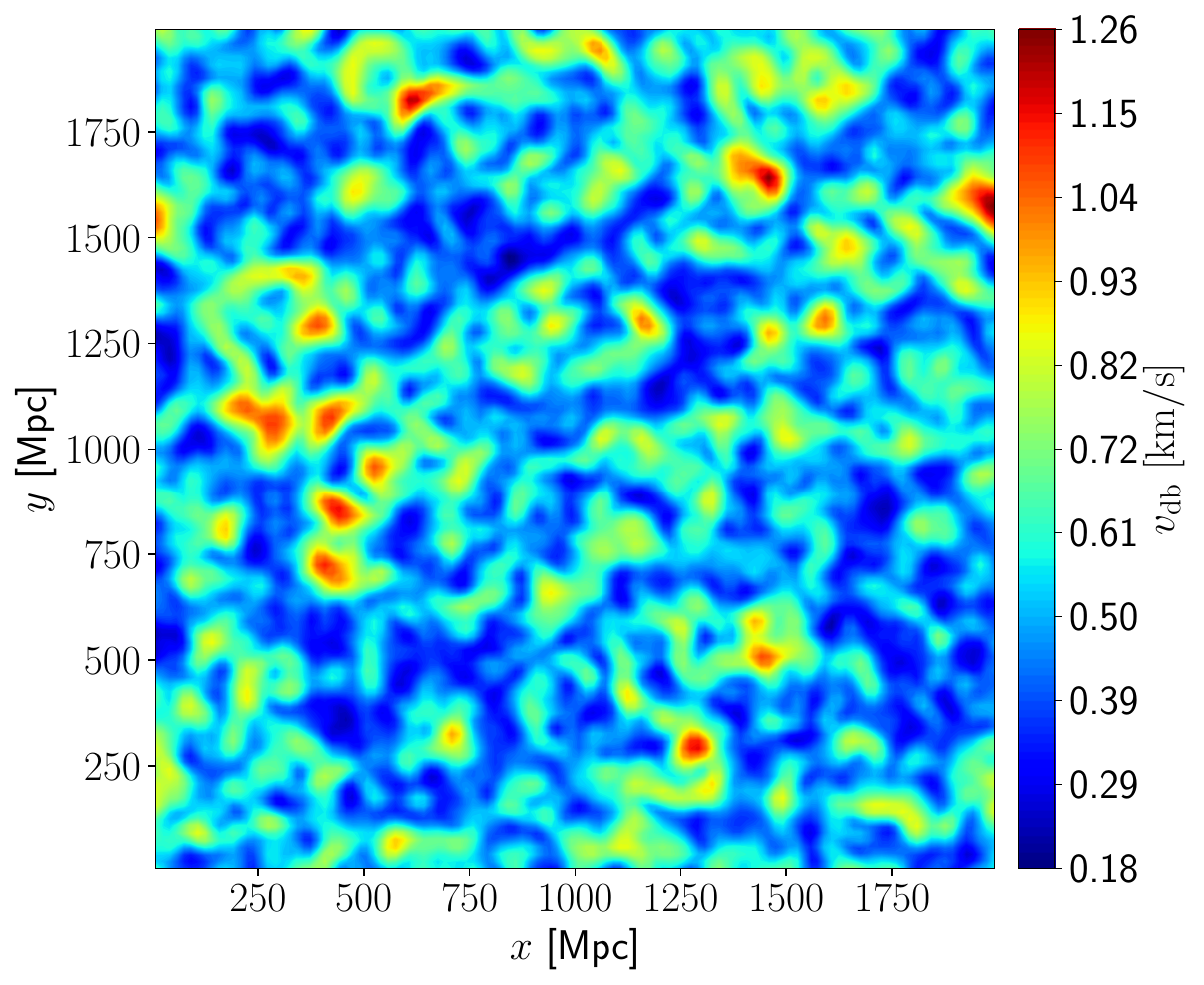}}
\subfigure{\includegraphics[width=0.45\textwidth]{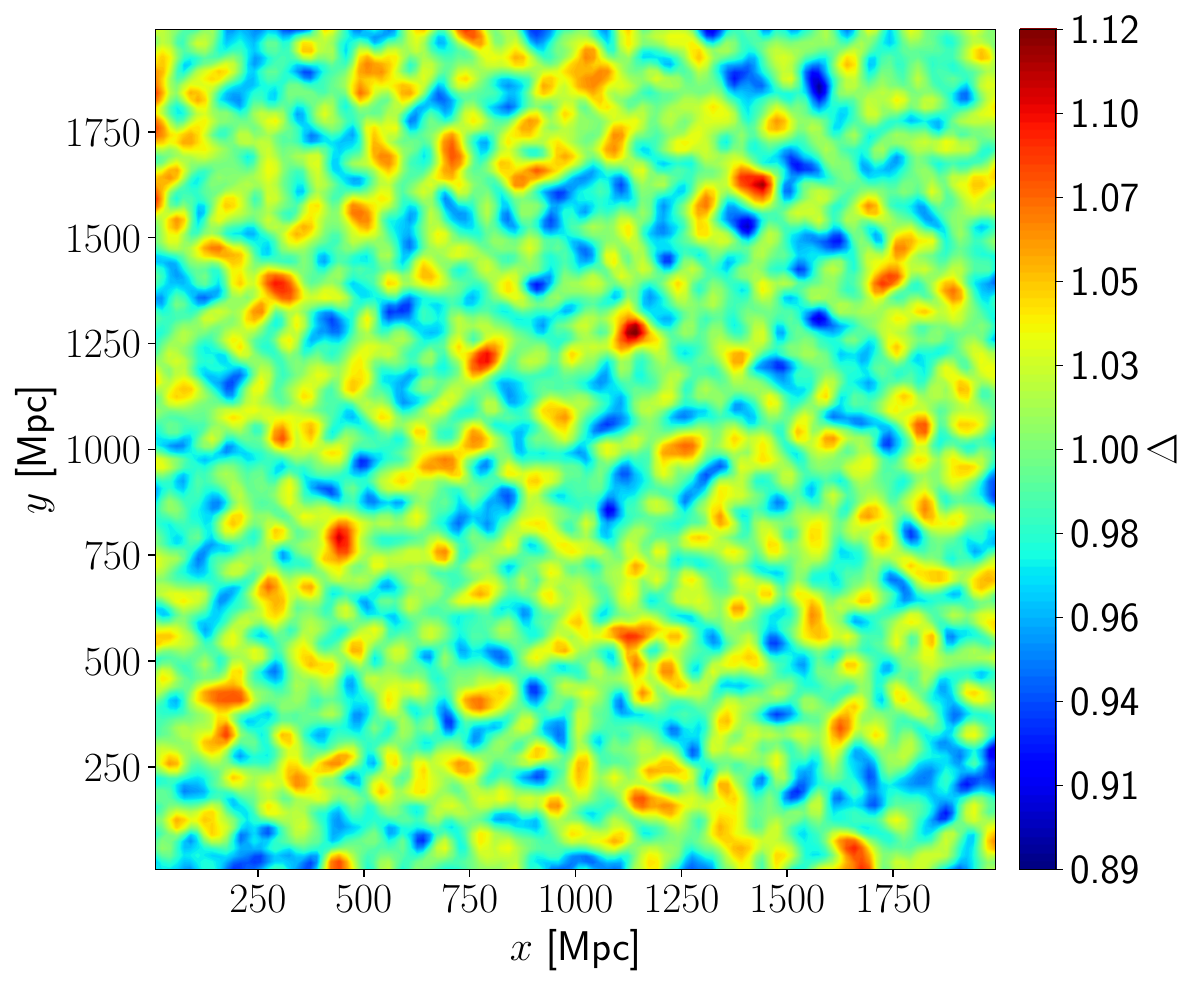}}
\subfigure{\includegraphics[width=0.45\textwidth]{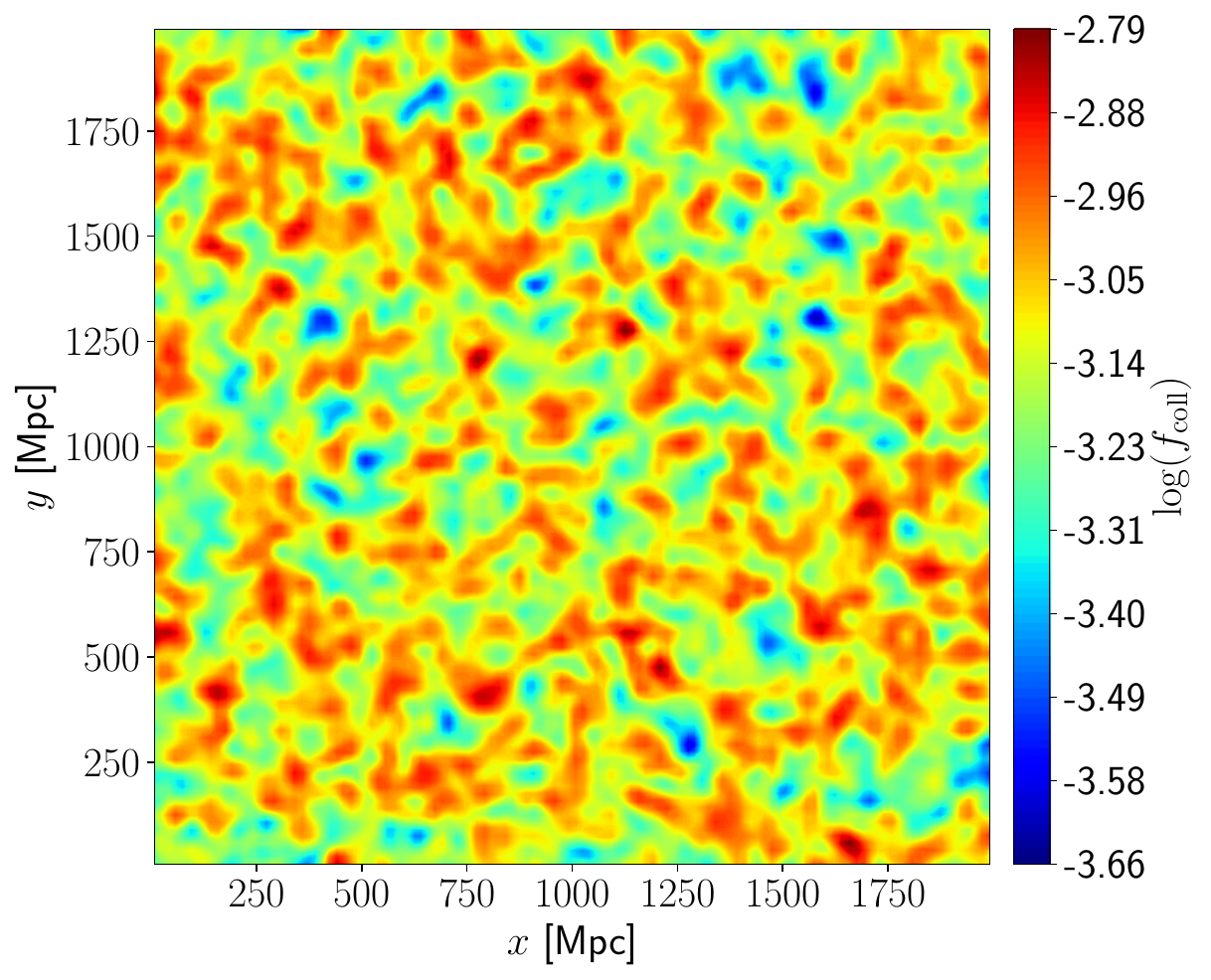}}
\subfigure{\includegraphics[width=0.45\textwidth]{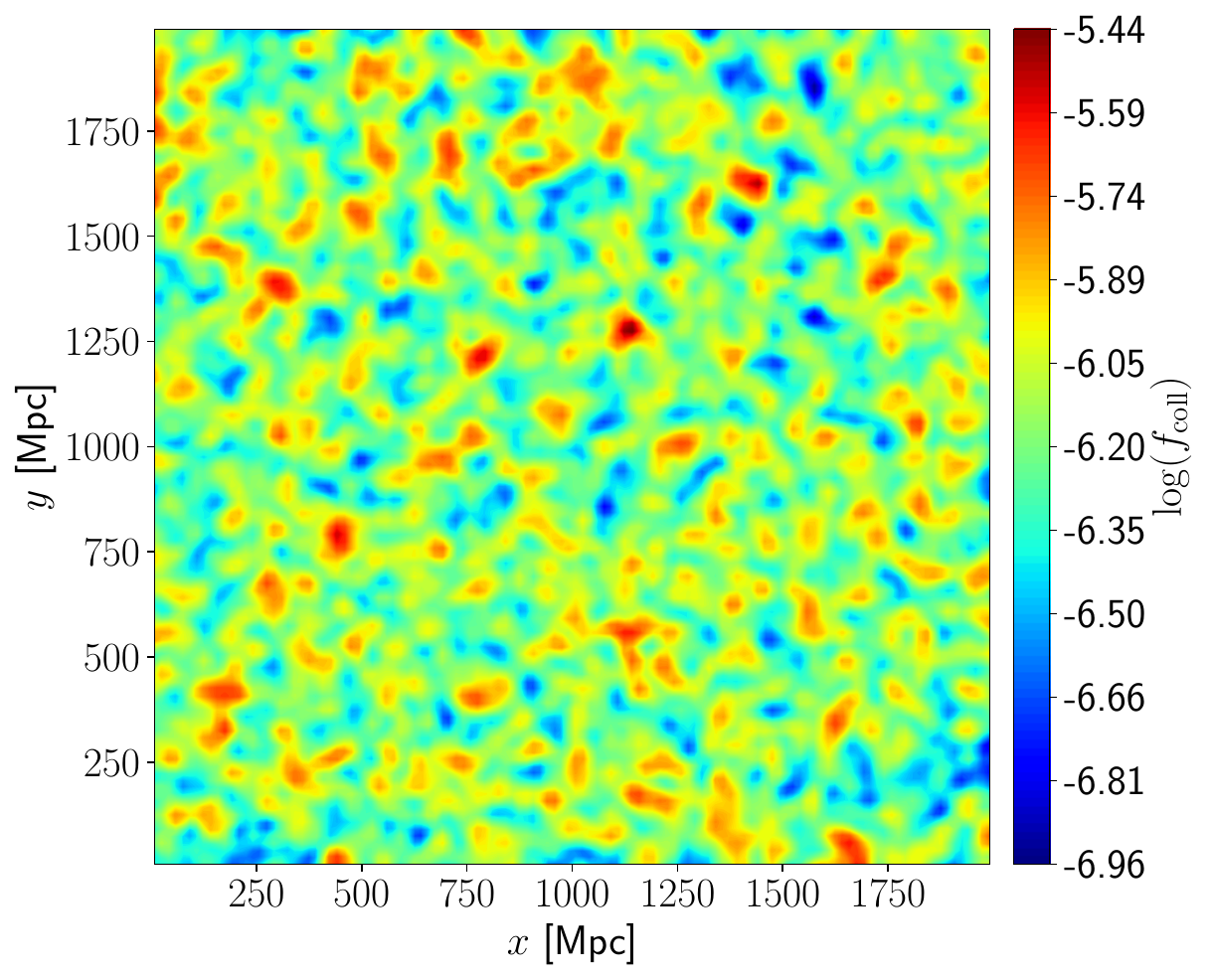}}
\caption{
The streaming velocity field (top left), density field (top right), and collapse fraction field  in the CDM model (bottom left), and in the axion model with $m_{\rm a}=10^{-20}$ eV (bottom right), at $z=20$.
}
\label{fig:fields}
}
\end{figure*}

In Fig. \ref{fig:fields} we plot slices of the velocity field $|\boldsymbol{v}_{\rm db}(\boldsymbol{x},z)|$ and density field $\Delta(\boldsymbol{x},z)$ at $z=20$, for the CDM model. The results of axion models are quite similar, as they have quite similar power spectra on large scales. We also plot the collapse fraction $f_{\rm coll}(>M_{\rm crit},\boldsymbol{x},z)$ fields of the same slice for the CDM model and axion model with $m_{\rm a}=10^{-20}$ eV, ignoring the LW feedback. The collapse fraction field is modulated by both the density field and the relative velocity field, so in principle, it should be tightly correlated with both of them. However, the latter mechanism relies on the contribution of halos with mass $\sim M_{\rm crit}$ in $f_{\rm coll}$. For the axion model, small halos are suppressed, therefore the VAO feature in the collapse fraction field is much less obvious than in the CDM model.
In the CDM model, the correlation coefficient between  $|\boldsymbol{v}_{\rm db}(\boldsymbol{x},z)|$ field and  $f_{\rm coll}(\boldsymbol{x},z)$ field is $-0.44$, and that
between $\Delta (\boldsymbol{x},z)$ field and $f_{\rm coll}(\boldsymbol{x},z)$ field is 0.84. In the axion model with $m_{\rm a}=10^{-20}$ eV, the correlation coefficients are -0.03 and 0.82 respectively. Indeed, for such an axion model, the collapse fractions are almost independent of the relative streaming velocities.

\subsection{The radio telescope sensitivity}

The rms of noise flux per beam for an interferometer array with $N$ antennas is
\begin{align}
\sigma_{\rm noise}&=\frac{2k_{\rm B} T_{\rm sys}}{A_{\rm e,dish} \sqrt{N(N-1)\delta \nu t_{\rm s} }} \nonumber \\
&=\frac{2k_{\rm B} T_{\rm sys}}{A_{\rm e,array} \sqrt{\delta \nu t_{\rm s} }}
\end{align}
where $k_{\rm B}$ is the Boltzmann constant, $T_{\rm sys}$ is the system temperature, $A_{\rm e}$ is the effective area of either a single dish or the full array, $\delta \nu$ is the frequency channel width and $t_{\rm s}$ is the on-source integration time. Using the total observation time $t_{\rm obs}\approx t_s\frac{\Omega_{\rm survey}}{\Omega_{\rm FoV}}$, where $\Omega_{\rm survey}$ is the survey area and $\Omega_{\rm FoV}$ is the field of view of the array, and the survey speed figure of merit (SSFOM) \citep{Braun2019}, the noise is 
\begin{equation}
\sigma_{\rm noise}= 2k_{\rm B}\sqrt{ \frac{ \Omega_{\rm survey} }{ {\rm SSFOM\times} \delta \nu t_{\rm obs}}}.
\end{equation}

The  noise power spectrum 
\begin{equation}
P_{\rm noise}=\left( \frac{ \sigma_{\rm noise}}{\theta^2_{\rm beam}} \right)^2V_{\rm vox},
\end{equation}
where $\theta_{\rm beam}=\lambda/d_{\rm max}$ is the beam width, $d_{\rm max}$ is the maximum baseline length, and 
\begin{equation}
V_{\rm vox}\approx D^2(z)\theta^2_{\rm beam}\left[\frac{c(1+z)}{H(z)}\frac{\delta \nu}{\nu}\right]
\end{equation}
is the comoving volume of the real space voxel. $D(z)$ is the comoving distance up to $z$. Throughout this paper, we adopt $d_{\rm max}=2000$ m, as most of the SKA-low stations are located within such distance\footnote{\url{https://indico.skatelescope.org/event/384/attachments/3008/3961/SKA1_Low_Configuration_V4a.pdf},
\url{https://www.skao.int/sites/default/files/documents/d17-SKA-TEL-SKO-0000557_01_-DesignConstraints-1.pdf}
}.

For SKA1-low and SKA2-low, $A_{\rm e,dish}/T_{\rm sys}$ and $A_{\rm e,array}/T_{\rm sys}$ are given in \cite{Braun2019}. For example, at 100 MHz for SKA1-low $A_{\rm e,array}/T_{\rm sys}=441.3~{\rm m^2/K}$, and for SKA2-low $A_{\rm e,array}/T_{\rm sys}\approx 4400~{\rm m^2/K}$. If we take $ \theta_{\rm beam}=300''$ (corresponding to baseline length 2000 m at 100 MHz) and $\delta \nu=0.2$ MHz, then $V_{\rm vox}=920$ Mpc$^3$. 
For $t_{\rm obs}=2000$ hr and a survey area of 20 deg$^2$, using the SSFOM in \citet{Braun2019}, the noise power spectrum is $P_{\rm noise}=1.1\times10^5$ (mK)$^2$Mpc$^3$ for SKA1-low, and $P_{\rm noise}=1.2\times10^3$ (mK)$^2$Mpc$^3$ for SKA2-low, respectively. 

The variance of the measured 21 cm power spectrum is 
\begin{equation}
\sigma^2_P=\frac{1}{N_m(k)}[P_{21}(k)+P_{\rm noise}W^{-2}(k)]^2,
\label{eq:sigma_P}
\end{equation}
where $N_m(k)$ is the number of $k$-modes in the $k$-bin; the window function \citep{Battye2013MNRAS}
\begin{equation}
W(k)=\exp\left[-\frac{1}{2}k^2 D^2(z) \frac{\theta^2_{\rm beam}}{4}   \right]
\end{equation}
denotes the rapid decline of the measured power spectrum below the resolution.

Suppose a survey has area $\Omega_{\rm survey}$ and bandwidth $\Delta \nu$, then the survey volume
\begin{equation}
    V_{\rm survey} \approx D^{2}(z) \Omega_{\rm survey} \left[\frac{c(1+z)}{H(z)} \frac{\Delta \nu}{\nu} \right],
\end{equation}
and
\begin{equation}
\label{eq:k mode number}
    N_{\rm m} (k) \approx 2 \pi k^{3} d \ln{k} \frac{V_{\rm survey}}{(2\pi)^{3}},
\end{equation}
where $d\ln k$ is the relative width of the selected $k$-bin and we take $d\ln k=0.12$.

\section{results}\label{sec:results}

\subsection{The modulations of spin temperature by Ly$\alpha$ scattering and X-ray heating}

Since the VAO wiggles reflect the inhomogeneity of the Ly$\alpha$ radiation and X-ray heating, according to Eq. (\ref{eq:delta_T_b}) and the formation history of Pop III stars, in principle, we expect to see the VAO signal in the following scenarios: 

\begin{enumerate}

    \item If the Ly$\alpha$ coupling effect starts to work while the IGM is still neutral and unheated, say $x_\alpha\sim 1$, $T_{\rm k}\sim T_{\rm k}^{\rm ad}$ and $x_{\rm HI}\sim1$, where $T_{\rm k}^{\rm ad}$ is the kinetic temperature of the IGM following adiabatic evolution. In this case, the 21 cm power spectrum reflects the structures of the Ly$\alpha$ radiation field. This is the pure Ly$\alpha$-modulated mode;
    
    \item If the Ly$\alpha$ coupling has already been saturated, the IGM is obviously (but not heavily) heated but still not yet reionized, say $x_\alpha \gg 1$, $T_{\rm k}^{\rm ad} < T_{\rm k} \sim \mathcal{O}(T_{\rm CMB})$ and $x_{\rm HI}\sim1$. In this case, the 21 cm power spectrum reflects the structures of the X-ray radiation field. This is the pure X-ray-modulated mode;
    
    \item  If both the Ly$\alpha$ coupling and IGM heating have already been saturated, and the IGM has already been obviously (but not fully) reionized, say $x_\alpha\gg 1$, $T_{\rm k}\gg T_{\rm CMB}$ and $x_{\rm HI} < 1$. In this case, the 21 cm power spectrum reflects the structures of the ionizing radiation field. This is the pure reionization\blue{-}modulated mode.
    
\end{enumerate}
   
If these modulations work at distinct and well separated epochs, we would observe the following sequence: at the beginning when both the Ly$\alpha$ and X-ray radiation are weak, there is no VAO signal. Then the VAO signal appears when Ly$\alpha$ modulation starts to work. This signal disappears later on when Ly$\alpha$ coupling saturates, i.e. $x_\alpha \gg 1$. The VAO signal will appear again when the X-ray modulation works. And disappears again when $T_{\rm k} \gg T_{\rm CMB}$. Next, the VAO signal may appear again when the reionization modulation works, and finally disappears at the end of the reionization era. 
In practice, however, these three modulation mechanisms may work together, showing joint effects, depending on the choice of parameters of star formation efficiency $f_*$, X-ray production efficiency $\zeta_X$ that refers to the cumulative number of X-ray photons produced per Solar mass, and the ionization efficiency $\zeta_{\rm ion}$.

\begin{figure}
\centering{
\subfigure{\includegraphics[width=0.45\textwidth]{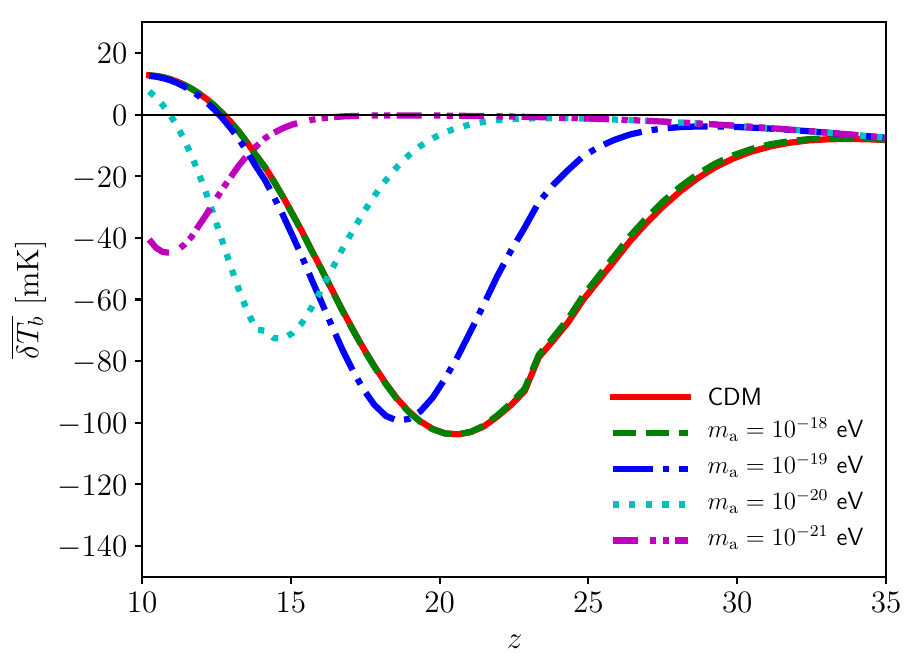}}
\caption{
The global spectrum of the 21 cm signal in different dark matter models. 
}
\label{fig:21cm_global}
}
\end{figure}

The 21 cm signal is modulated by star formation in minihalos. When the IGM is heated to $T_{\rm k} \gtrsim 10^3$ K, the Jeans mass \citep{Ripamonti2007MNRAS} exceeds the atomic cooling mass, and star formation in minihalos is fully suppressed and the VAO signal disappears soon. In this paper, we limit our investigation in the stage with $T_{\rm k} \lesssim 10^3$ K. 
In Fig. \ref{fig:21cm_global} we plot the 21 cm global spectrum in different dark matter models. 
Here we set $f_*=0.005$ since it is generally believed that the star formation is less efficient in minihalos (e.g. \citealt{Trenti2009ApJ}). 
We adopt $\zeta_{\rm X}=5\times 10^{55}~M_\odot^{-1}$ , roughly corresponding to $f_{\rm X}=0.05$ in \cite{Furlanetto2006PhR}. It is seen that, for such star formation efficiency and X-ray efficiency, the 21 cm global signal achieves the maximum absorption $\sim -100$ mK at $z\sim 20$ in the CDM model. In the next paragraph, we will see that the VAO wiggles reach the maximum amplitude at $z\sim 17$. 
We note that, however, the value of $\zeta_X$ for Pop III stars is still quite uncertain. For example, the SARAS-2 result \citep{Singh2017ApJ} disfavors models with very low X-ray efficiency.  
In practice, generally, theoretical models with X-ray efficiency spanning a wide range are allowed  
\citep{Cohen2017MNRAS}. Moreover, we set $f_{\rm esc,LW}\zeta_{\rm LW}=\zeta_{\rm LW}^{\rm BL05}$.
For the axion model with $m_{\rm a}=10^{-18}$ eV, the 21 cm global spectrum is almost identical to the CDM. However, for $m_{\rm a}\lesssim 10^{-19}$ eV, the global evolution of the  21 cm signal is delayed.

\begin{figure}
\centering{
\subfigure{\includegraphics[width=0.45\textwidth]{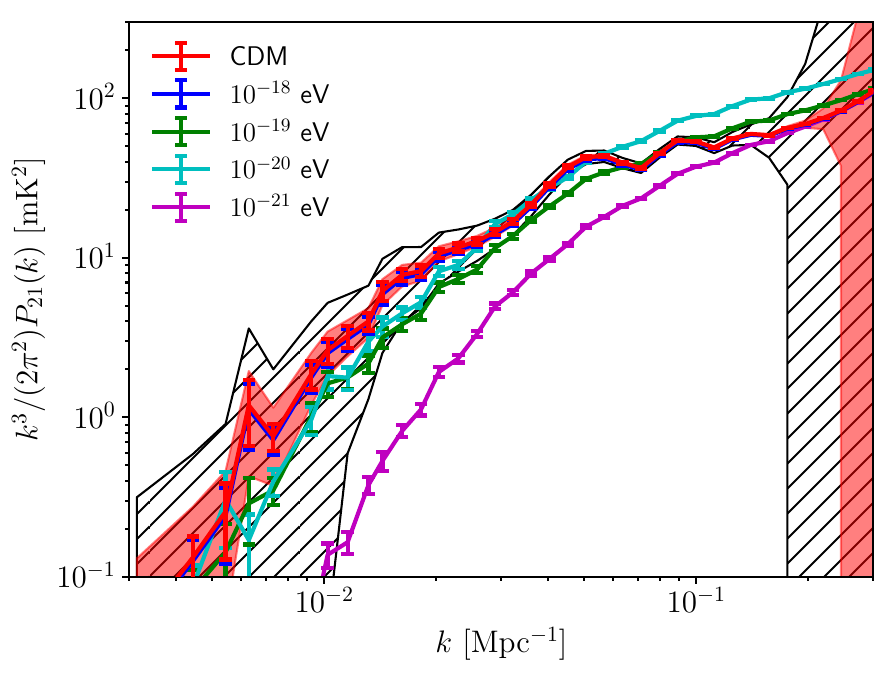}}

\subfigure{\includegraphics[width=0.45\textwidth]{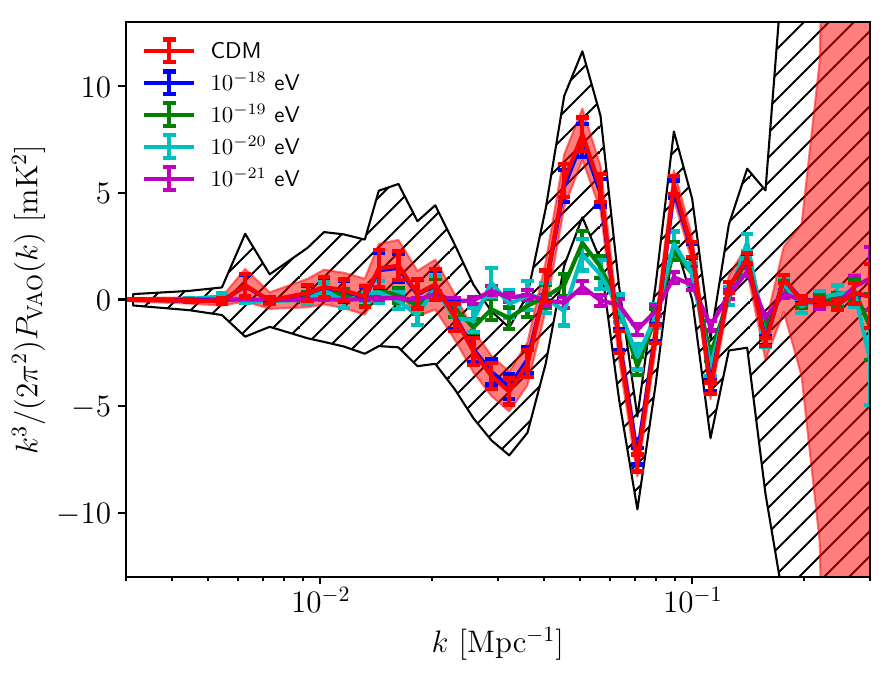}}
\caption{
{\it Top:} The 21 cm power spectra for different dark matter models. 
Errorbars refer to the sample variance of the simulated signal.
The hatched regions around the CDM model are the SKA1-low uncertainties for $\Omega_{\rm survey}=20$ deg$^2$, while filled regions are  SKA2-low uncertainties for $\Omega_{\rm survey}=200$ deg$^2$.
The observation time $t_{\rm obs}=2000$ hour. 
{\it Bottom:} The pure wiggles signal extracted from the 21 cm power spectrum by removing a polynomial component.
}
\label{fig:21cm_ps}
}
\end{figure}

In the top panel of Fig. \ref{fig:21cm_ps} we plot the 21 cm power spectrum at redshifts when $x_\alpha\sim1.0$, for the CDM model and axion models with different masses. It is clearly seen that in the CDM model and models with massive axion particles (i.e. $m_{\rm a}\gtrsim10^{-18}$ eV), the streaming velocities produce wiggles on the 21 cm power spectrum, through both Ly$\alpha$ scattering and X-ray heating. To clearly show the wiggles signal, we fit a smooth component of the  21 cm power spectrum by a polynomial \citep{Munoz2019PhRvD,Hotinli2022PhRvD}\footnote{One can also fit the wiggles by the analytical formula proposed by \citet{Munoz2019PhRvD}.}, 
\begin{equation}
\ln\left(\frac{k^3}{2\pi^2}P^{\rm smooth}_{21}(k)\right)=\sum_{i=0}^5 c_i(z) (\ln k)^i.
\end{equation}
Then we subtract this smooth component from the power spectrum, to get the pure VAO wiggles, $P_{\rm VAO}(k)=P_{21}(k)-P^{\rm smooth}_{21}(k)$. The results are shown in the bottom panel of Fig. \ref{fig:21cm_ps}. The first and strongest wiggle is at $k\approx 0.05$ Mpc$^{-1}$. The amplitude is $\sim 8$ mK in the CDM model.
For models with $m_{\rm a} \lesssim 10^{-19}$ eV, the wiggles are substantially reduced.
The location of the wiggles does not evolve with redshift.

We also check the detectability by using the uncertainties estimated from Eq. (\ref{eq:sigma_P}). We ignore the foreground here but present discussions of its influence in Sec. \ref{sec:foreground}. The VAO signal in the CDM model is marginally detectable for SKA1-low with  $\Omega_{\rm survey}=20~{\rm deg}^2$ and
$t_{\rm obs}=2000$ hour. The total signal-to-noise ratio is $\approx 6$. See the hatched regions in the bottom panel of Fig. \ref{fig:21cm_ps}.
Since the VAO signal appears at large scale, the cosmic variance dominates the uncertainties when the survey is as small as 20 deg$^2$. However, for SKA1-low this cannot be reduced by simply increasing the survey area, because given the total observation time $t_{\rm obs}$, if the survey area is larger, then the on-source integration time is smaller. So larger area survey may have smaller uncertainties at large scale where the uncertainties are dominated by sample variance, but will have larger uncertainties at small scale where the uncertainties are dominated by the instrumental noise.
For this reason, we consider the SKA2-low that has higher sensitivity, and $\Omega_{\rm survey}=200$
 deg$^2$. See the filled regions in the bottom panel of Fig. \ref{fig:21cm_ps}. For such a survey, the total signal-to-noise increases to $\approx22$.

 In both CDM and axion models, there are relative motions between the dark matter and baryon. However, in the axion model, or any other dark matter lacking small-scale density fluctuations, there is a lack of minihalos and Pop III stars, therefore the relative motions have no chance to modulate the Ly$\alpha$ radiation field and X-ray field. Therefore the VAO features are not encoded in the 21 cm signal. Indeed, the VAO signal is a good indicator of the small-scale density fluctuations.

\subsection{The choice of $f_*$, $\zeta_X$ and $f_{\rm esc,LW}\zeta_{\rm LW}$}

\begin{figure*}
\centering{
\subfigure{\includegraphics[width=0.45\textwidth]{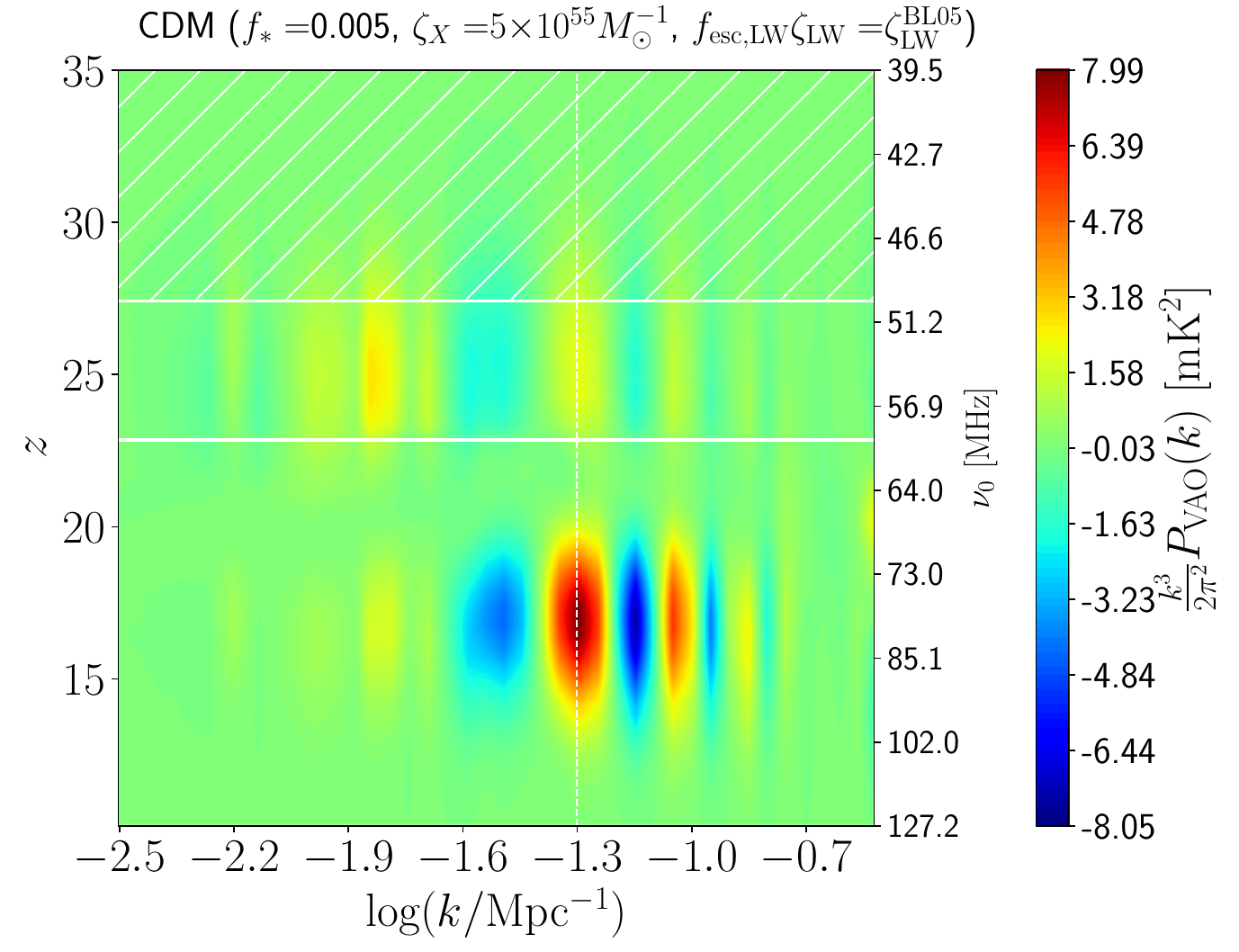}}
\subfigure{\includegraphics[width=0.45\textwidth]{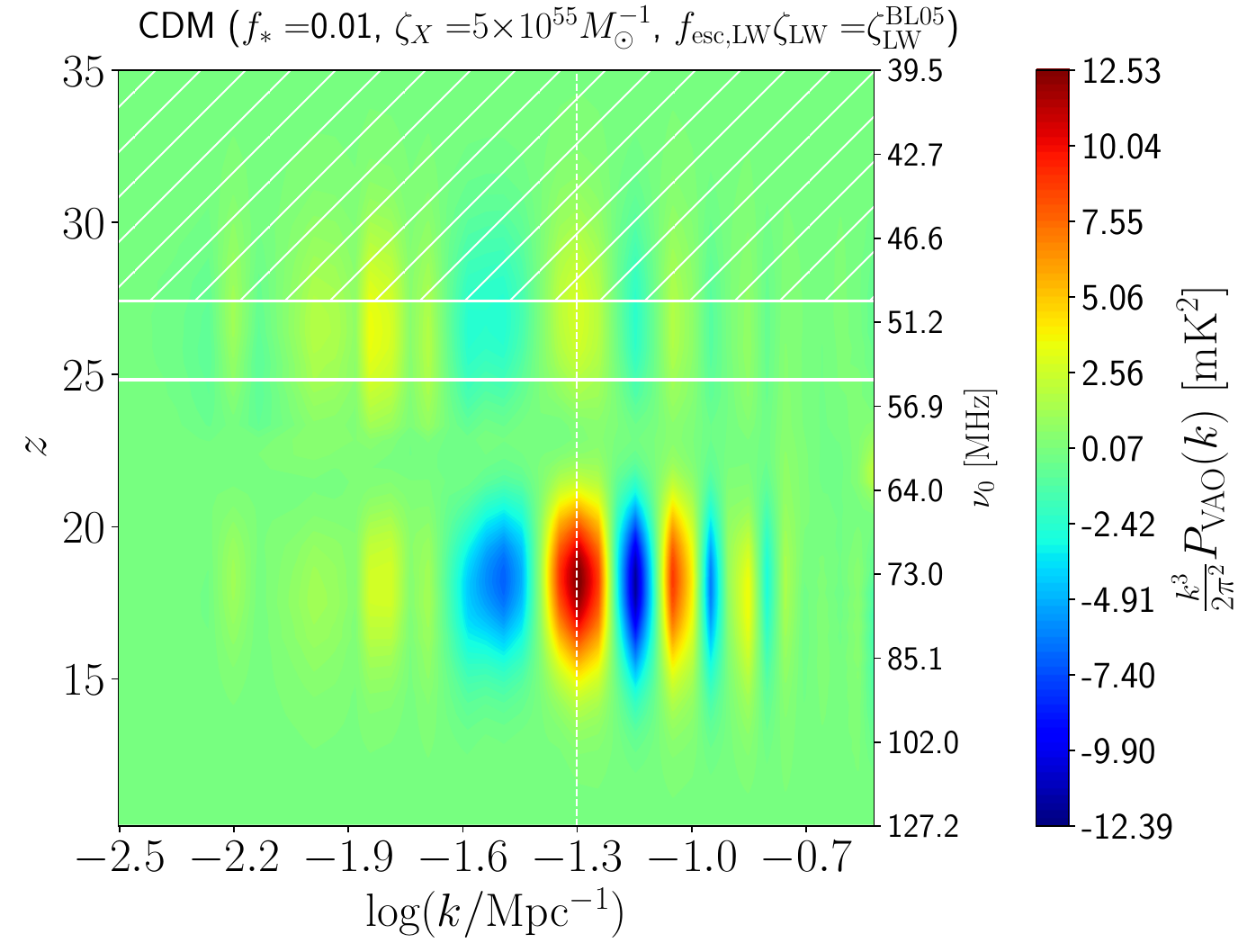}}
\subfigure{\includegraphics[width=0.45\textwidth]{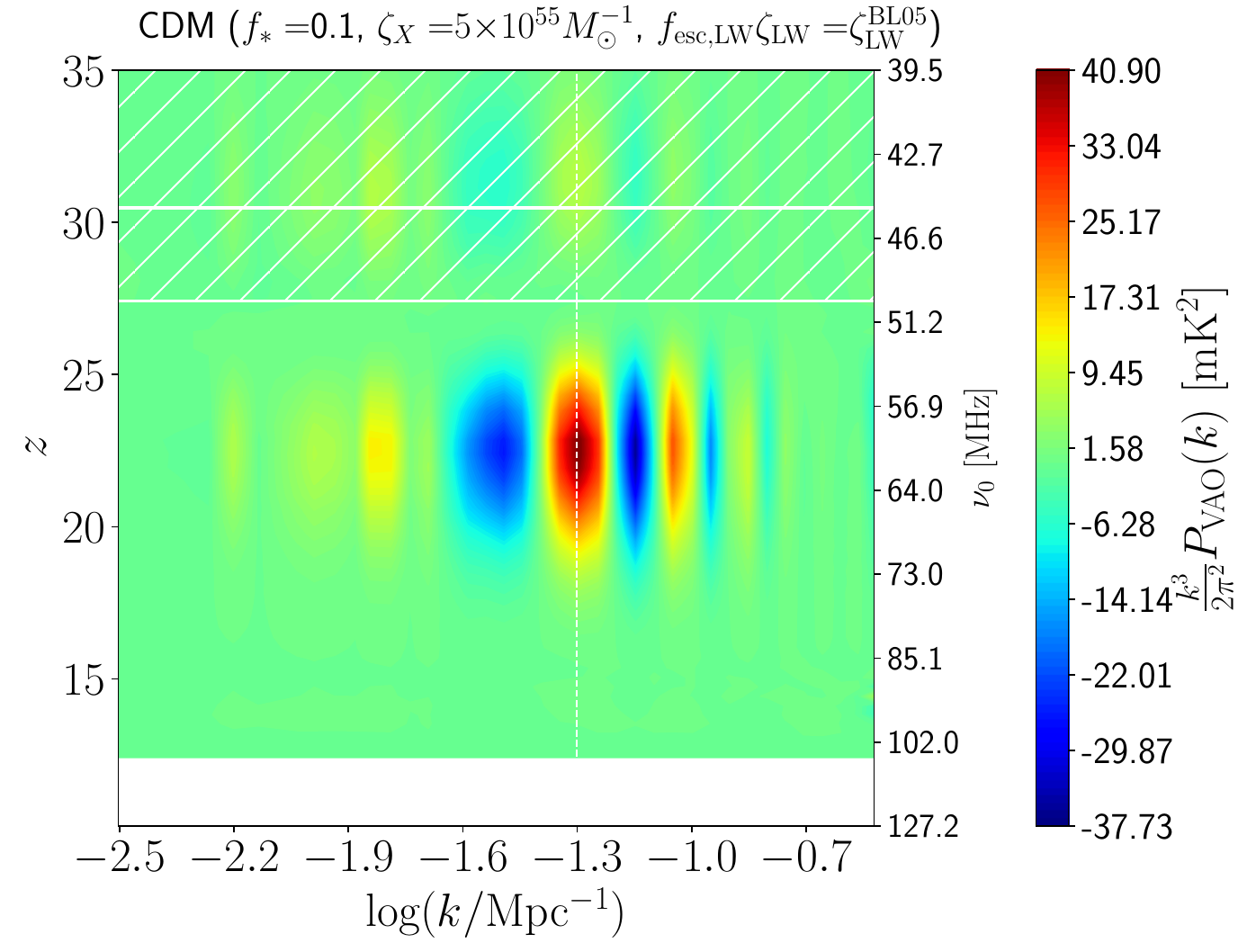}}
\caption{ 
The VAO wiggles extracted from 21 cm power spectra at different redshifts for CDM models with various $f_*$. We mark the 21 cm frequencies on the right $y$-axis. 
In each panel, the solid white line marks the redshift when $\bar{x}_\alpha\approx1$. The empty regions have $T_{\rm k}\ge 10^3$ K, where our models are not valid. The hatched regions are beyond the frequency range of the SKA-low band. The vertical dashed line marks $k=0.05$ Mpc$^{-1}$, the location of the strongest VAO peak.
}
\label{fig:ps_contour_CDM_F_STAR}
}
\end{figure*}

\begin{figure*}
\centering{
\subfigure{\includegraphics[width=0.45\textwidth]{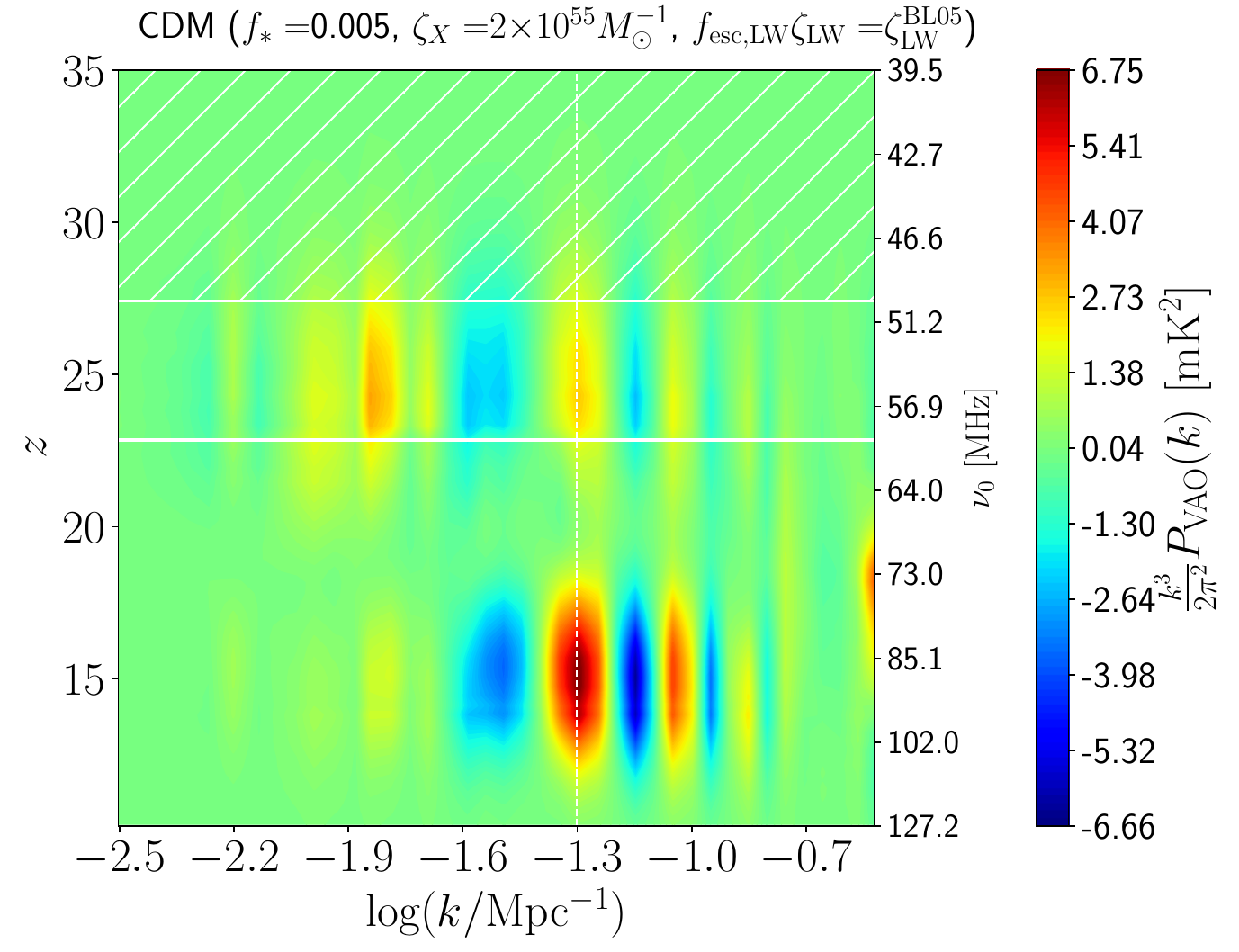}}
\subfigure{\includegraphics[width=0.45\textwidth]{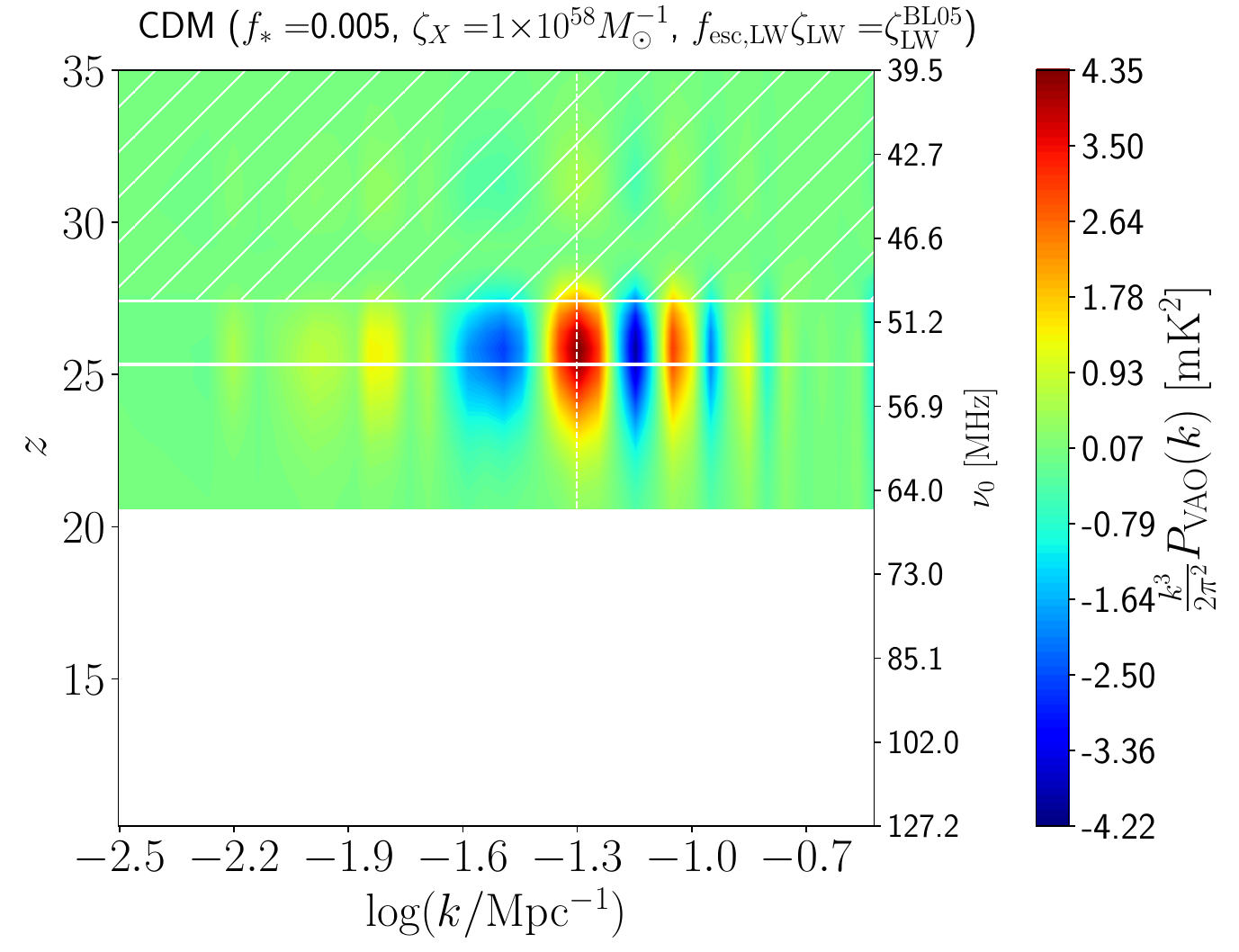}}
\subfigure{\includegraphics[width=0.45\textwidth]{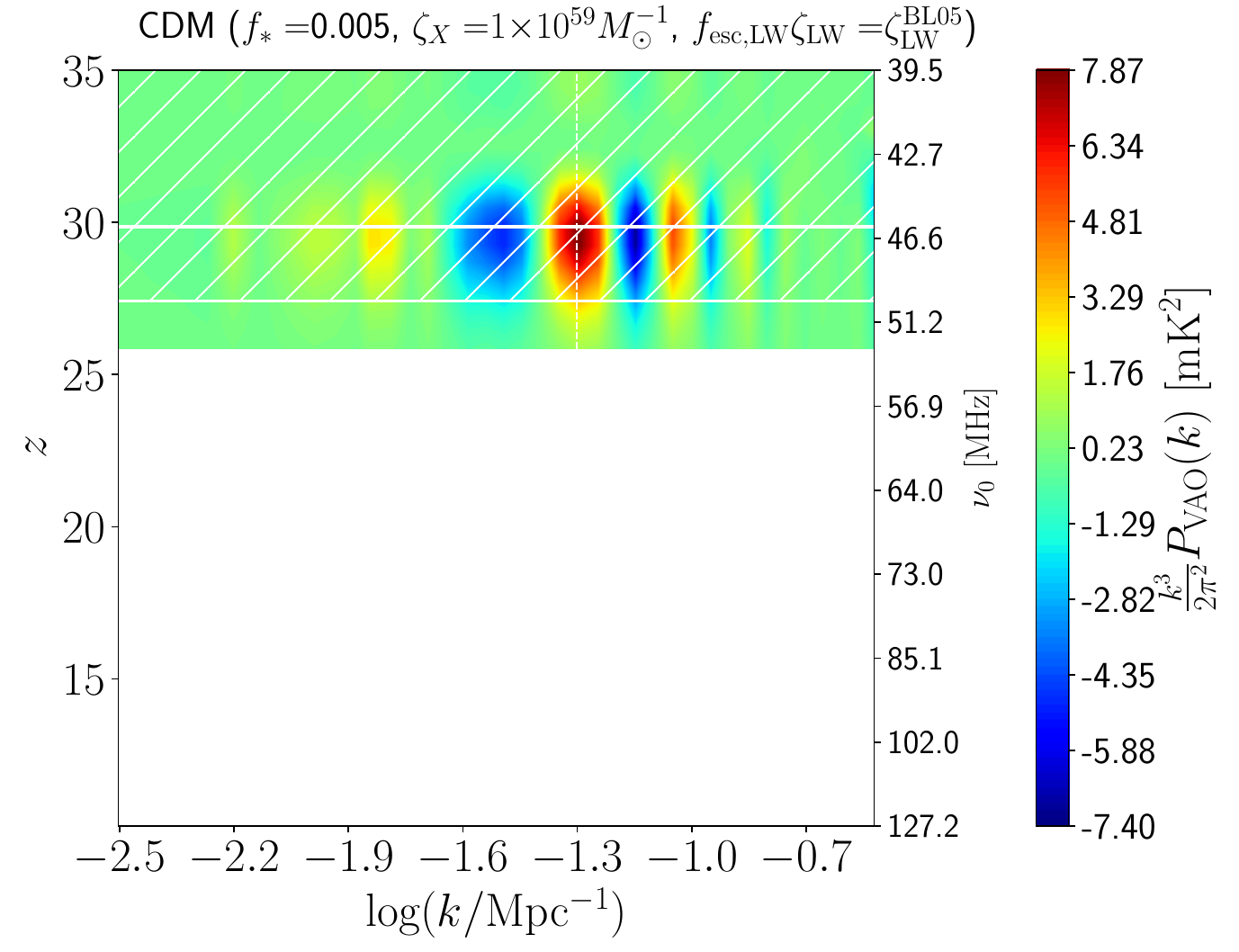}}
\caption{ 
Same to Fig. \ref{fig:ps_contour_CDM_F_STAR}, however here we vary the $\zeta_X$. 
}
\label{fig:ps_contour_CDM_ZETA_X}
}
\end{figure*}

\begin{figure*}
\centering{
\subfigure{\includegraphics[width=0.45\textwidth]{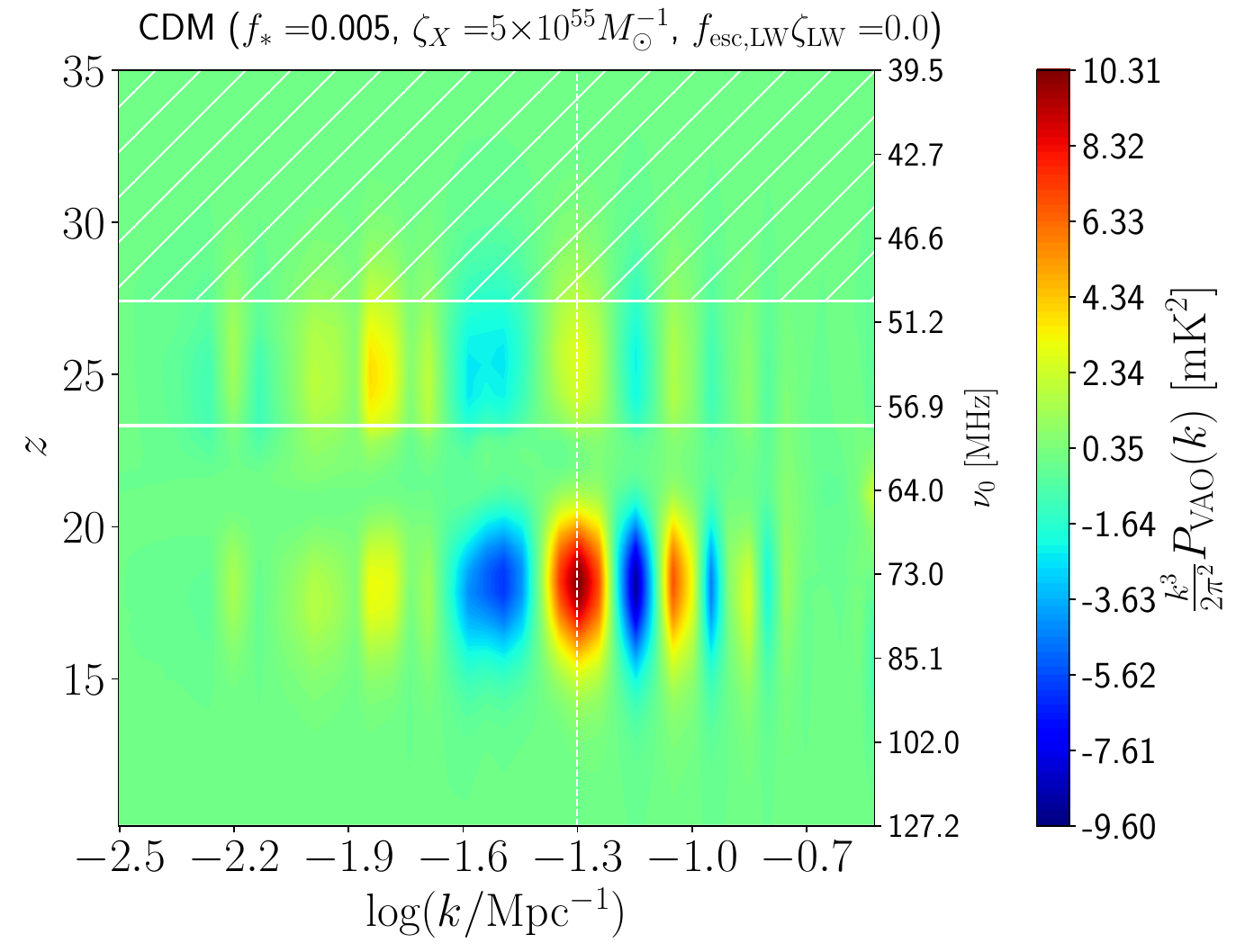}}
\subfigure{\includegraphics[width=0.45\textwidth]{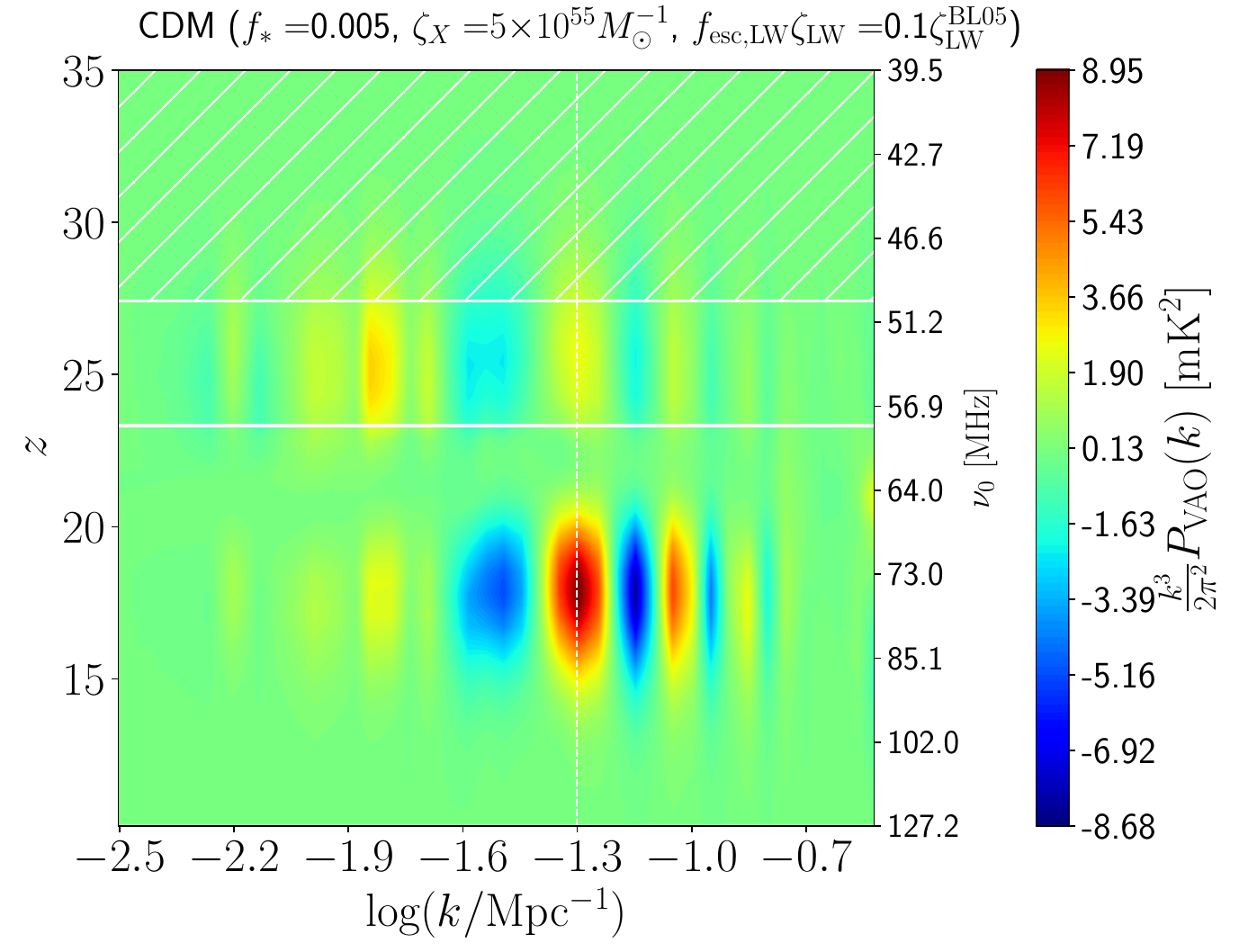}}
\subfigure{\includegraphics[width=0.45\textwidth]{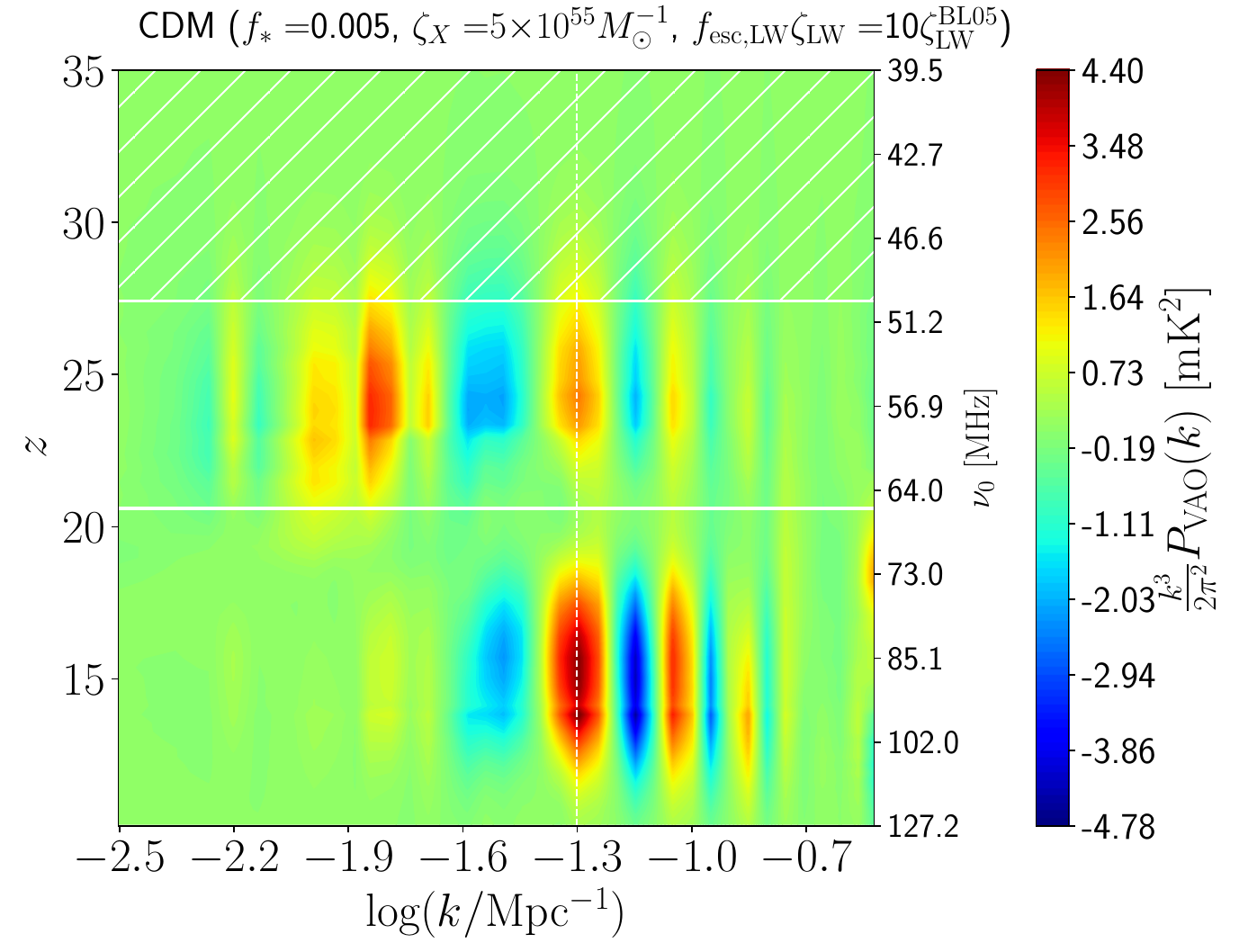}}
\caption{ 
Same to Fig. \ref{fig:ps_contour_CDM_F_STAR}, however here we vary the $f_{\rm esc,LW}\zeta_{\rm LW}$.
}
\label{fig:ps_contour_CDM_ZETA_LW}
}
\end{figure*}

For the CDM model, in Fig. \ref{fig:ps_contour_CDM_F_STAR}, \ref{fig:ps_contour_CDM_ZETA_X}, and \ref{fig:ps_contour_CDM_ZETA_LW} we show the extracted VAO wiggles vs. $k$ and $z$ for varying  $f_*$, $\zeta_X$ and $f_{\rm esc,LW}\zeta_{\rm LW}$ respectively. Parameters are given in each panel.
We see that the VAO signal increases with increasing $f_*$, and decreases with increasing LW feedback. 

Interestingly, for $\zeta_X$ the behaviors are more complicated. When $\zeta_X \lesssim 1\times10^{58}~M_\odot^{-1}$, the VAO signal decreases with increasing X-ray radiation. This is because as the IGM is heated, the spin temperature is more close to the CMB temperature, so the 21 cm signal becomes weaker. However, for higher $\zeta_X$ the VAO signal starts to increase again. This is because the X-ray is also a strong Ly$\alpha$ source. Roughly one-third of the X-ray energy is finally deposited into Ly$\alpha$ photons. Therefore X-ray can generate not only inhomogeneous heating but also an inhomogeneous Ly$\alpha$ background that promotes the VAO signal. Moreover, for higher $\zeta_X$ such a mechanism works earlier. This also promotes the VAO signal since at higher redshifts the relative streaming velocities are larger. 
Also for this reason, increasing the $\zeta_X$ will shift the VAO signal to higher redshifts, but cannot wipe out it. Note that here we show the results for $\zeta_X \gtrsim 1\times 10^{58}~M_\odot^{-1}$ just for the purpose of exploring the parameter space that allows the existence of the VAO wiggles. It is not clear if Pop III stars can really produce such amount of X-ray photons.

Even though a very strong  LW radiation field is artificially set, $f_{\rm esc,LW}\zeta_{\rm LW}$ is 10 times $\zeta_{\rm LW}^{\rm BL05}$, there are still VAO features on the 21 cm power spectrum (the bottom right panel in Fig. \ref{fig:ps_contour_CDM_ZETA_LW}). This is because: first, the LW radiation field is not just simply proportional to the $f_{\rm esc,LW}\zeta_{\rm LW}$, since stronger LW radiation will suppress the star formation in smaller halos, reducing the production amount of the LW radiation in return. Second, in the \citet{Kulkarni2021ApJ} model, the critical dark matter halo mass for H$_2$ cooling is less sensitive to the LW radiation at high redshifts. For example, at $z\sim25$, the critical mass is still just $\sim10^6~M_\odot$ even though $J_{\rm LW,21}$  is as high as $\sim 30$.

Nevertheless, for the CDM model, there is VAO signal for all $f_*$, $\zeta_{\rm X}$ and $f_{\rm esc,LW}\zeta_{\rm LW}$ parameters shown there. But if the signal appears too early, e.g. $f_*= 0.005$ and $\zeta_{\rm X}\gtrsim1\times10^{59}~M_\odot^{-1}$, all the frequencies are below the lower limit of the SKA-low band. Therefore, to investigate the detectability of the VAO signal, it is important to understand the full evolution of the VAO signal over different redshifts.

\begin{figure}
\centering{
\subfigure{\includegraphics[width=0.45\textwidth]{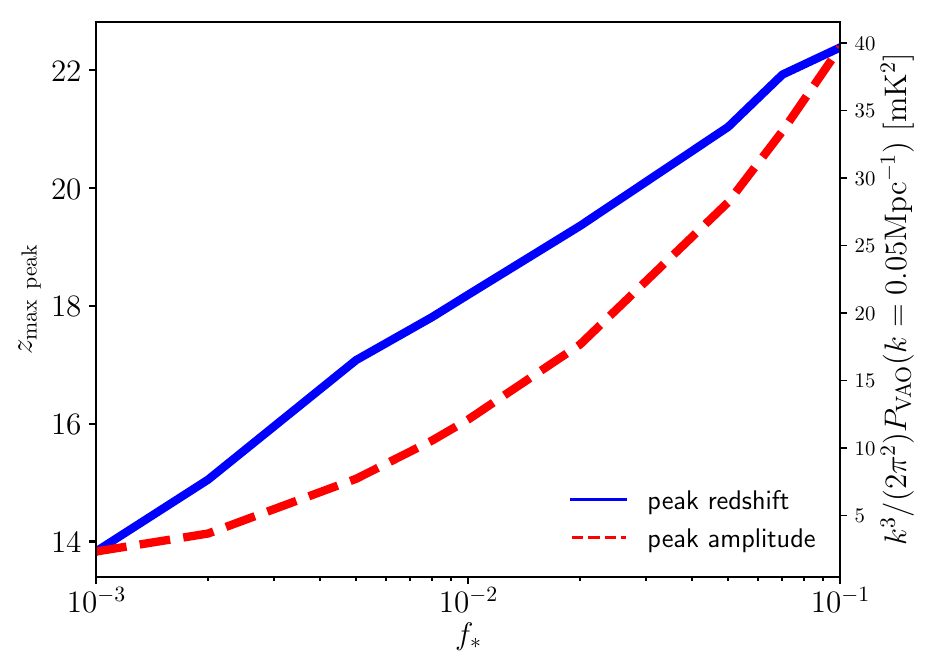}}
\caption{
The redshift at which the first peak at $k\approx 0.05$ Mpc$^{-1}$ reaches maximum (left $y$-axis), and the amplitude of the maximal peak (right $y$-axis), as a function of $f_*$.  
}
\label{fig:peak_redshift}
}
\end{figure}

For larger $f_*$, the Ly$\alpha$ coupling and X-ray heating occur earlier, therefore the VAO signal is more obvious, as the streaming velocities decline with decreasing redshift. The redshift at which the VAO signal first peaked can be a useful indicator of early star formation efficiency. In Fig. \ref{fig:peak_redshift} we plot the redshift at which the first peak at $k=0.05$ Mpc$^{-1}$ reaches maximum amplitude, as a function of $f_*$, see the left $y$-axis. We also plot the amplitude of the first peak, see the right $y$-axis.

On the other hand, reionization also sets limits on the $f_*$, since $\zeta_{\rm ion}=f_* f_{\rm esc,ion} N_{\rm ion} \frac{1}{1+\bar{n}_{\rm rec}}$. For Pop III stars the total released ionizing photons per stellar atom $N_{\rm ion}=4.4\times10^4$ \citep{Barkana2005ApJ}, the escape fraction of ionizing photons $f_{\rm esc,ion}\sim 20-40\%$ \citep{Kim2017MNRAS}, and the mean recombination times per ionized atom $\bar{n}_{\rm rec}\sim5-10$ \citep{Mao2020MNRAS}. Currently, the contribution to reionization from Pop III stars is not directly constrained, but it is generally believed that this is a small fraction, with the bulk part of the reionization produced by photons from the second generation stars (Pop II).   
In our model, for all $f_*$ and $\zeta_X$ values presented in this paper, the ionized fraction $\lesssim 10\%$ when the VAO signal reaches the maximum.   

\begin{figure*}
\centering{
\subfigure{\includegraphics[width=0.45\textwidth]{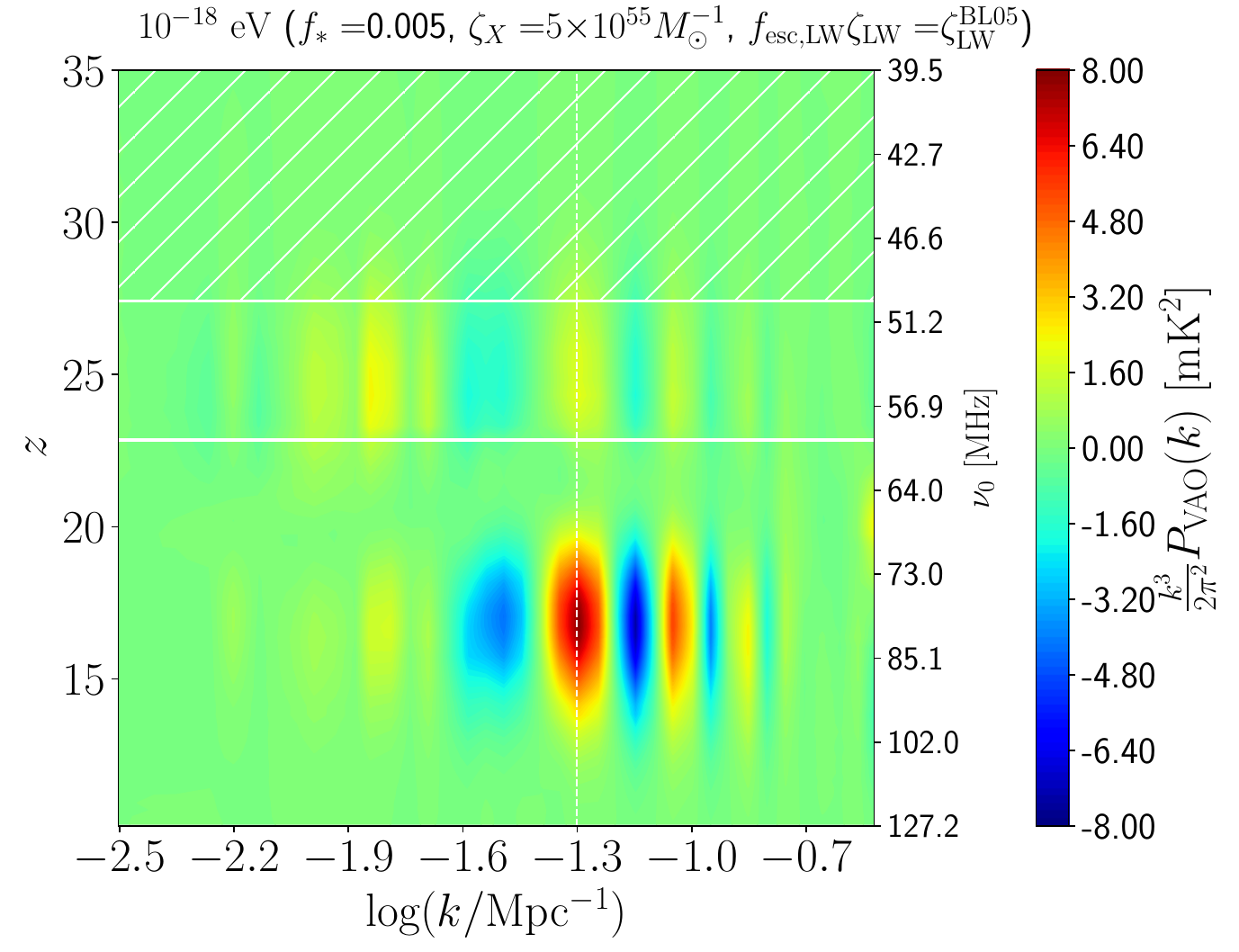}}
\subfigure{\includegraphics[width=0.45\textwidth]{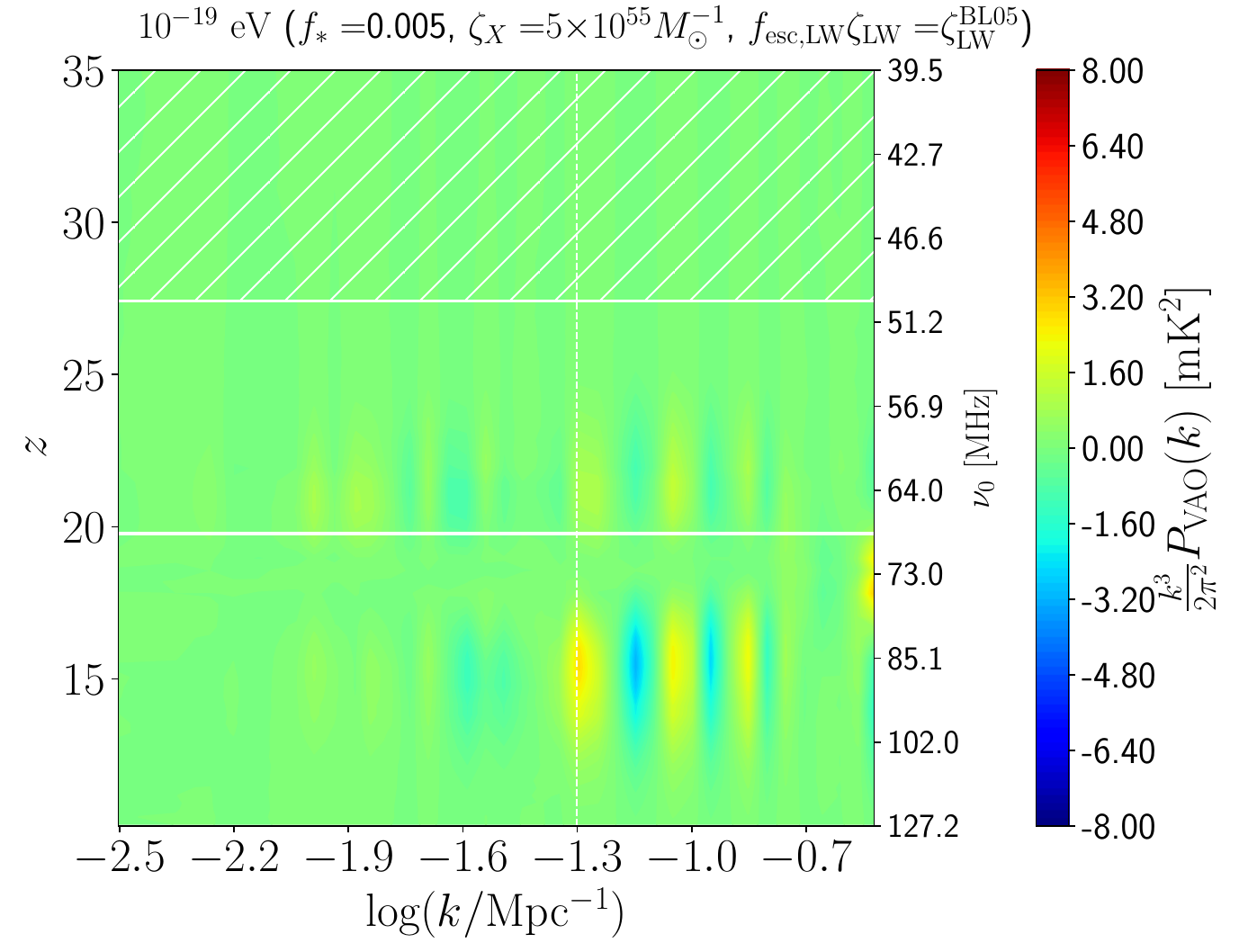}}
\subfigure{\includegraphics[width=0.45\textwidth]{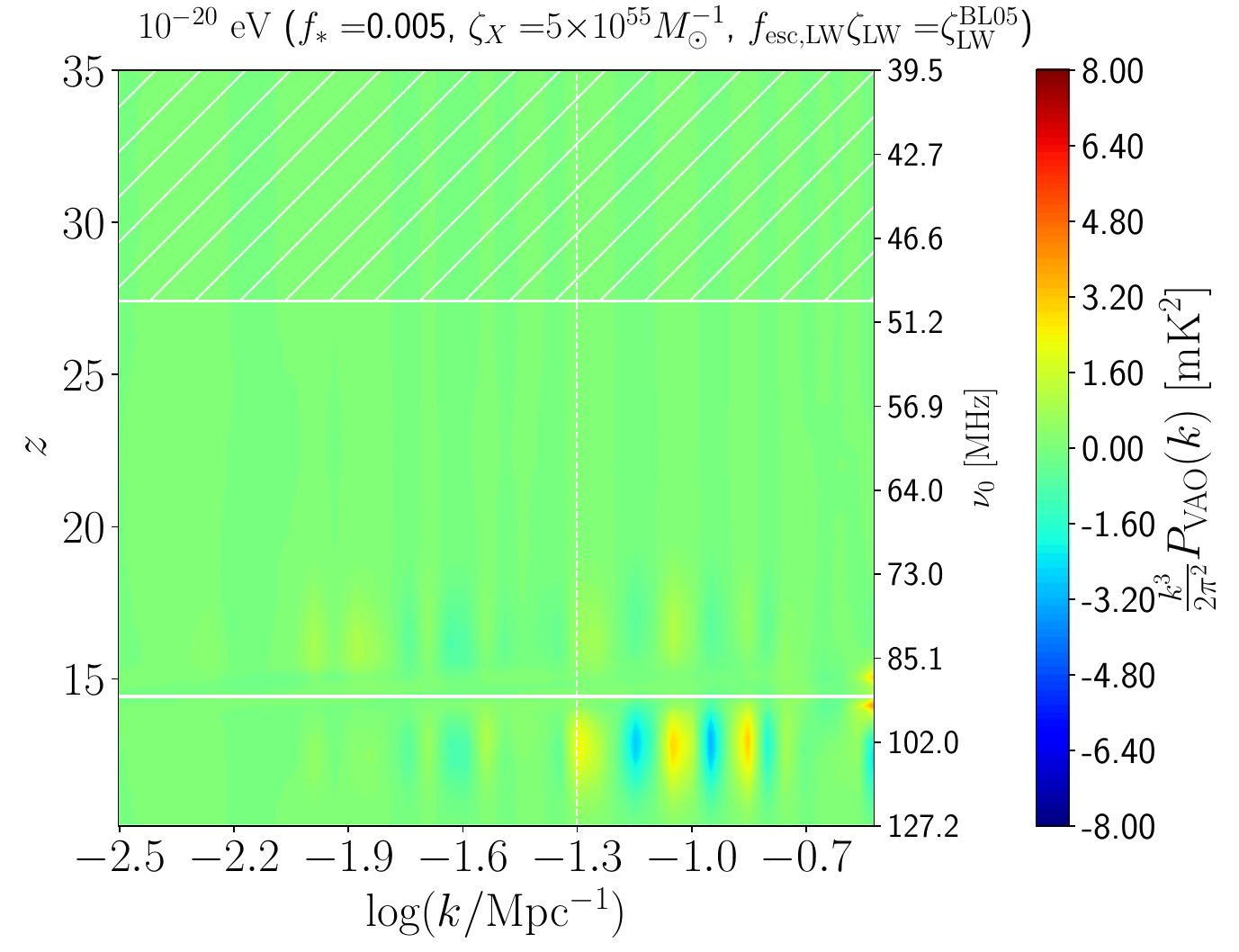}}
\subfigure{\includegraphics[width=0.45\textwidth]{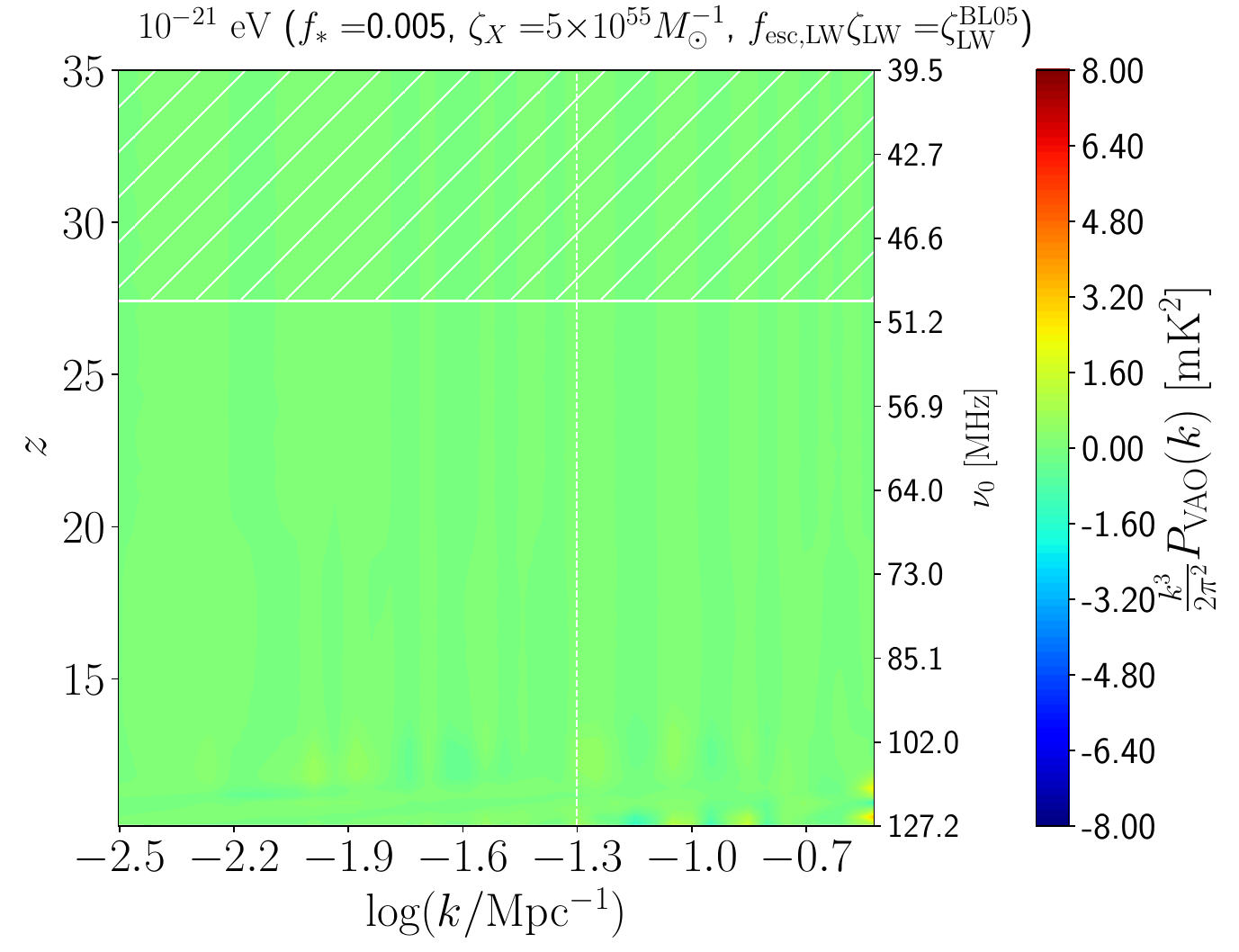}}
\caption{The extracted VAO wiggles in the axion model with various masses. For the purpose of doing comparison between different panels, we set the same colorbar ranges for all panels. 
}
\label{fig:ps_contour_axion}
}
\end{figure*}

In Fig. \ref{fig:ps_contour_axion} we plot the extracted VAO wiggles for various axion masses. We see that when $m_{\rm a}=10^{-18}$ eV, the VAO wiggles are still as obvious as that of the CDM model. However, for $m_{\rm a}\lesssim10^{-19}$ eV, the VAO signal is negligible at any redshifts, no matter what $f_*$ value we choose. The light ``wiggles'' in the redshift spectra plot for $m_{\rm a}\lesssim 10^{-19}$ eV are produced by pure density fields, i.e. the Baryon Acoustic Oscillations (BAO) effect. The $k$ locations of peaks and troughs of the BAO wiggles are similar (but not the same) to the VAO wiggles, however, the amplitudes are much smaller. The relative height of the BAO peak at $k=$0.05 Mpc$^{-1}$ is $\lesssim5\%$, while for the VAO signal the peak at $k=$0.05 Mpc$^{-1}$ can be as high as $\sim20\%$ in the CDM model.

In summary, in the CDM model, we always see the VAO features for wide ranges of astrophysical parameters. For some parameters, for example 
$f_*\sim0.005$ and $\zeta_X\sim1\times10^{59}~M_\odot^{-1}$, the frequencies of the VAO wiggles are below the lower bound of the SKA-low band. For stronger LW feedback, for example $f_{\rm esc,LW} \zeta_{\rm LW} \gg 10\zeta_{\rm LW}^{\rm BL05}$, Pop III star formation in minihalos with $\lesssim10^6~M_\odot$ could be fully suppressed and the VAO signal will be reduced significantly. In such cases, finding reasonable constraints on these parameters from other observations is important. For example, the 21 cm global spectrum can put constraints on the thermal history of the Universe that is tightly related to the Pop III stars \citep{Mirocha2018MNRAS}. Moreover, the unresolved Cosmic X-ray background and Cosmic Near-Infrared background can also put upper limits on  Pop III stars properties \citep{Xu2016ApJ}, although in practice it is challenging to distinguish the contribution of Pop III stars and the contributions of first Pop II galaxies and low-$z$ faint foreground galaxies \citep{Helgason2014ApJ}. 
Nevertheless, for most literatures, the adopted parameters $f_*\ll 0.1$, $\zeta_X\ll 1\times10^{59}~M_\odot^{-1}$ and  $f_{\rm esc,LW}\zeta_{\rm LW}\lesssim10\zeta_{\rm LW}^{\rm BL05}$  
(e.g. \citealt{Gilmore2012MNRAS,Gessey-Jones2022MNRAS,Chantavat2023,Bera2023JApA,Sun2021MNRAS,Qin2021MNRAS,Incatasciato2023MNRAS,Ventura2023MNRAS,Mittal2022MNRAS,Hegde2023arXiv}). That is to say, for parameters favored by most people, it is quite likely that the VAO signal does exist and its frequencies are in the SKA-low band.   

On the other hand, for the dark matter model that lacks small-scale density fluctuations and minihalos, e.g. axion models with $m_{\rm a}\lesssim10^{-19}$ eV, the VAO signal is negligible in all the frequency range. 
So indeed the VAO features are useful tools as a probe of small-scale density fluctuations. 

\subsection{The critical minihalo mass for Pop III stars formation}\label{sec:critical-mass}

So far we use the \citet{Kulkarni2021ApJ} critical mass (Eq. (\ref{eq:M_crit_K21}) ) as the threshold for Pop III star formation in minihalos in the presence of both LW radiation and relative streaming velocities. Here we check if the VAO signal still exists when we replace Eq. (\ref{eq:M_crit_K21}) with other definitions for the critical mass.

\citet{Schauer2021MNRAS} performed high-resolution numerical simulations to investigate the influence of LW feedback and relative streaming velocities on the Pop III stars formation in minihalos. Their result of the critical mass writes:
\begin{equation}
\log_{10} M_{\rm min}=\log_{10}M_0+s\times \frac{v_{\rm db}}{\sigma_{\rm rms}},
\label{eq:M_min_S21}
\end{equation}
where
\begin{equation}
\log_{10}M_0=5.562\times (1+0.279\times \sqrt{J_{\rm LW,21}}),
\end{equation}
and
\begin{equation}
s=0.614\times(1-0.560\times \sqrt{J_{\rm LW,21}}).
\end{equation}
They only performed simulations with $J_{\rm LW,21}\le 0.1$. Since our LW radiation field is inhomogeneous, we will use the above equations anyway. To avoid the risk, we only investigate the models with $f_{\rm esc,LW}\zeta_{\rm LW} \le \zeta_{\rm LW}^{\rm BL05}$ here.

\begin{figure*}
\centering{
\subfigure{\includegraphics[width=0.45\textwidth]{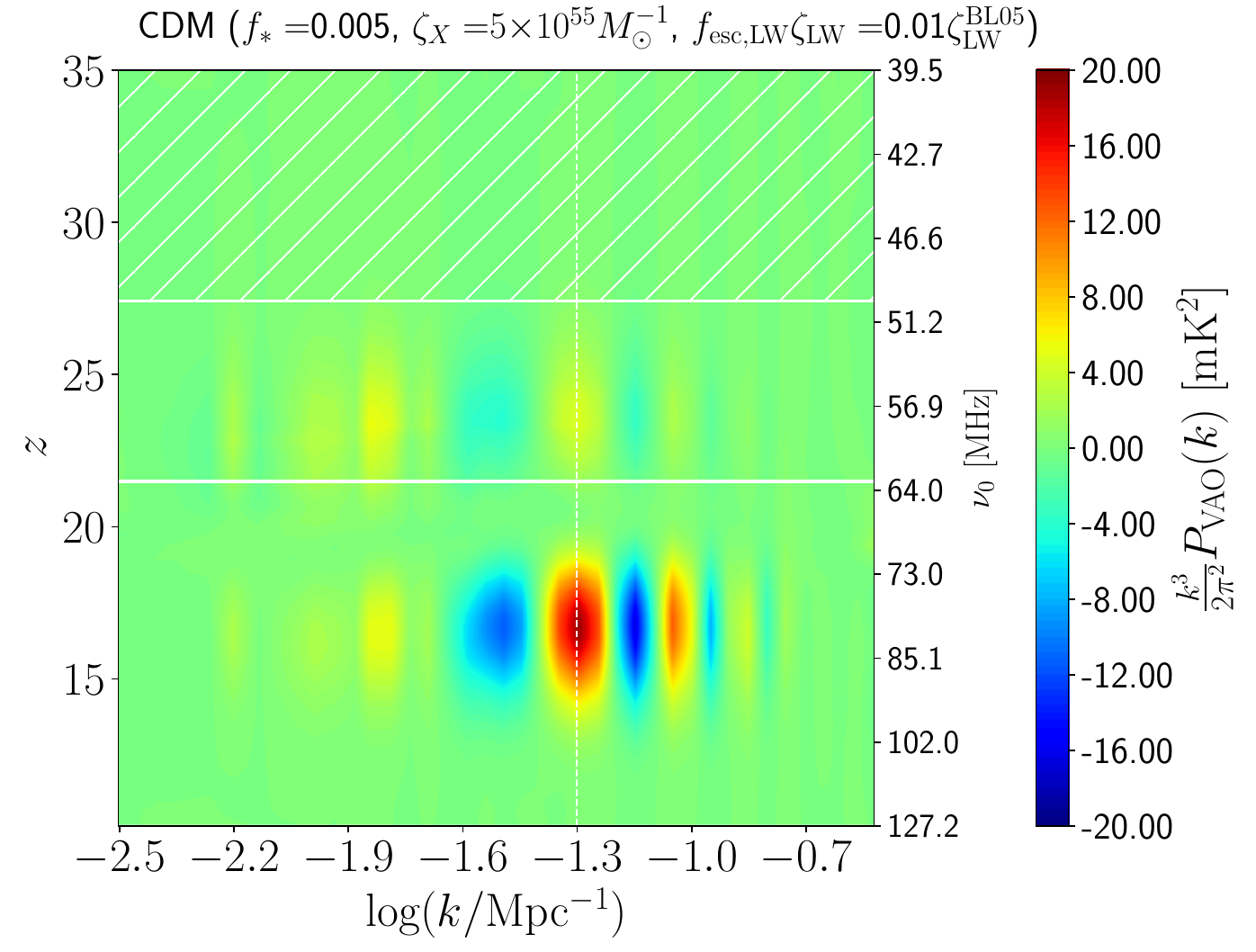 }}
\subfigure{\includegraphics[width=0.45\textwidth]{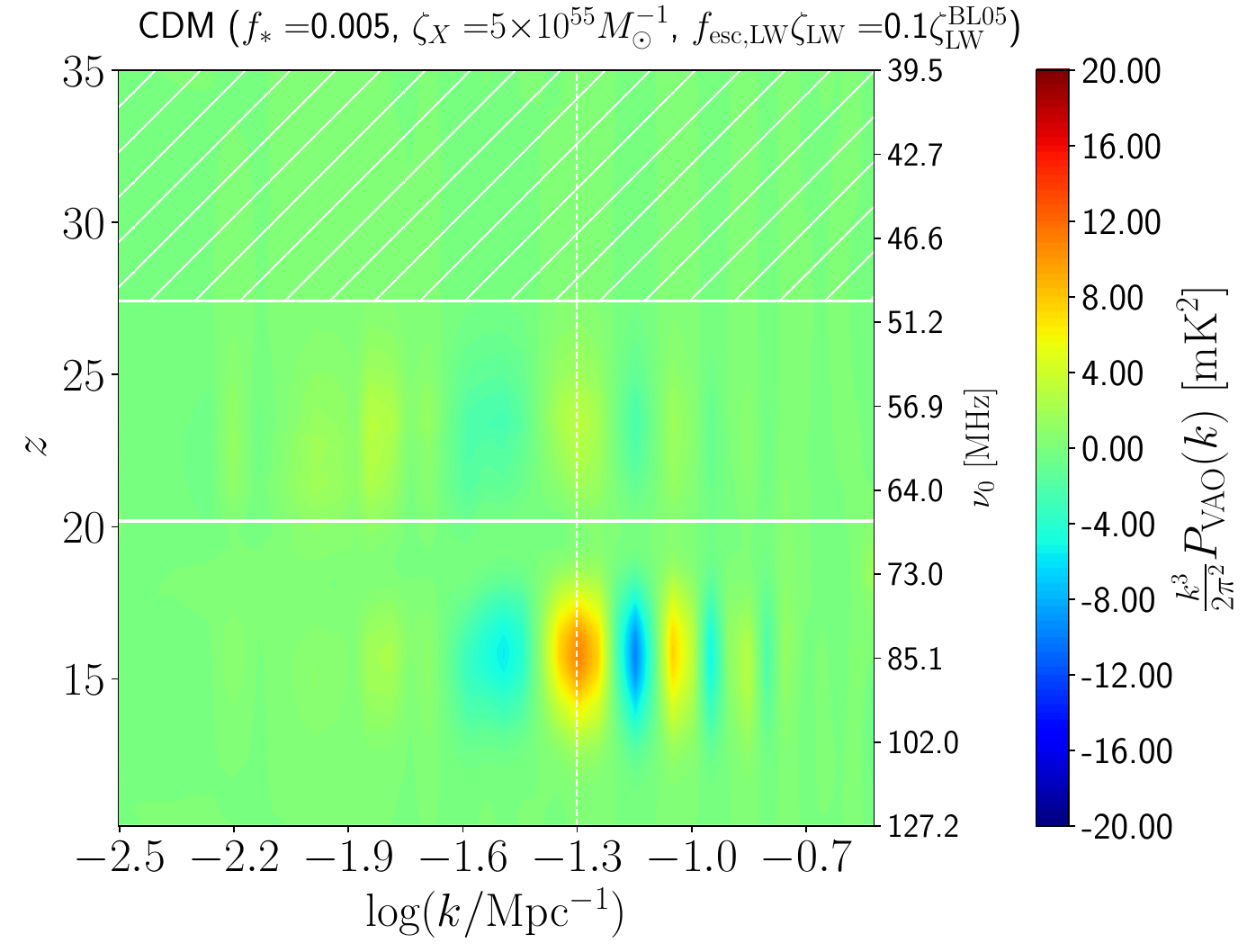 }}
\subfigure{\includegraphics[width=0.45\textwidth]{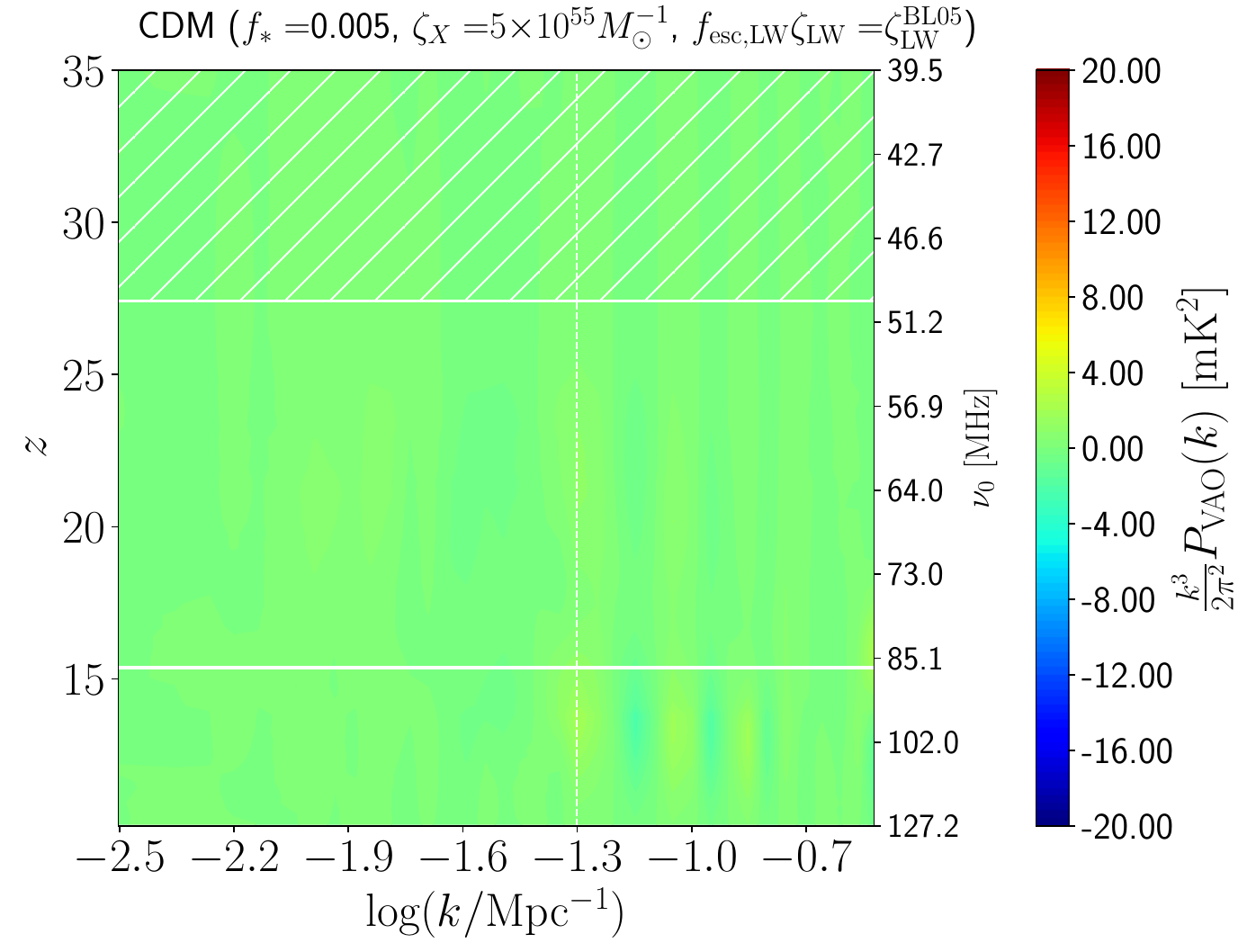 }}
\caption{
The 21 cm VAO wiggles at different redshifts for CDM models with different $f_{\rm esc,LW}\zeta_{\rm LW}$. Here we use Eq. (\ref{eq:M_min_S21}) as the critical mass for Pop III formation in minihalos.
Since we need to compare the VAO wiggles of different models, we set the same colorbar ranges for all panels. The three panels correspond to $f_{\rm esc,LW}\zeta_{\rm LW}=0.01 \zeta_{\rm LW}^{\rm BL05}$, $0.1 \zeta_{\rm LW}^{\rm BL05}$ and $\zeta_{\rm LW}^{\rm BL05}$ respectively.
}
\label{fig:ps_contour_S21}
}
\end{figure*}

In Fig. \ref{fig:ps_contour_S21} we show the extracted VAO wiggles for $f_{\rm esc,LW}\zeta_{\rm LW}=0.01 \zeta_{\rm LW}^{\rm BL05}$, $0.1\zeta_{\rm LW}^{\rm BL05}$  and  $\zeta_{\rm LW}^{\rm BL05}$ respectively, using Eq. (\ref{eq:M_min_S21}) as the critical minihalo mass for Pop III star formation. Obviously, the amplitude of the wiggles decreases with increasing LW feedback strength. For $f_{\rm esc,LW}\zeta_{\rm LW}=0.01 \zeta_{\rm LW}^{\rm BL05}$, the maximum peak is $\sim 10$ mK. However, when $f_{\rm esc,LW}\zeta_{\rm LW}=\zeta_{\rm LW}^{\rm BL05}$, the maximum peak is $\sim 2$ mK, comparable to the SKA2-low uncertainties level with $\Omega_{\rm survey}=200$ deg$^2$. So 
for the \citet{Schauer2021MNRAS} critical mass, when $f_{\rm esc,LW}\zeta_{\rm LW}\gtrsim  \zeta_{\rm LW}^{\rm BL05}$, the VAO features on the 21 cm power spectrum are not detectable, unless increasing the observation time.  

\citet{Fialkov2012MNRAS} proposed that in the presence of relative streaming velocities however the LW radiation is negligible, the critical mass $M_{\rm crit}(v_{\rm db},z| J_{\rm LW,21}=0)$ is the virial mass corresponding to the circular velocity
\begin{equation}
V_{\rm circ}(v_{\rm db}| J_{\rm LW,21}=0)=[V^2_{\rm circ}(v_{\rm db}=0)+(\alpha_{v} v_{\rm db})^2]^{1/2},
\label{eq:V_cool}
\end{equation}
where $V_{\rm circ}(v_{\rm db}=0)=3.714$ km s$^{-1}$, and $\alpha_v=4.015$ for the optimal fit. 

When the LW feedback works, analogous to \citet{Machacek2001ApJ}, the above critical mass is boosted by a factor $1+6.96(4\pi J_{\rm LW,21})^{0.47}$ and the final critical mass is \citep{Fialkov2013MNRAS}
\begin{align}
&M_{\rm crit}(J_{\rm LW,21},v_{\rm db},z)= \nonumber \\
&M_{\rm crit}(v_{\rm db},z|J_{\rm LW,21}=0) \times [1+6.96(4\pi J_{\rm LW,21})^{0.47}].
\label{eq:M_crit_F13}
\end{align}

\begin{figure*}
\centering{
\subfigure{\includegraphics[width=0.45\textwidth]{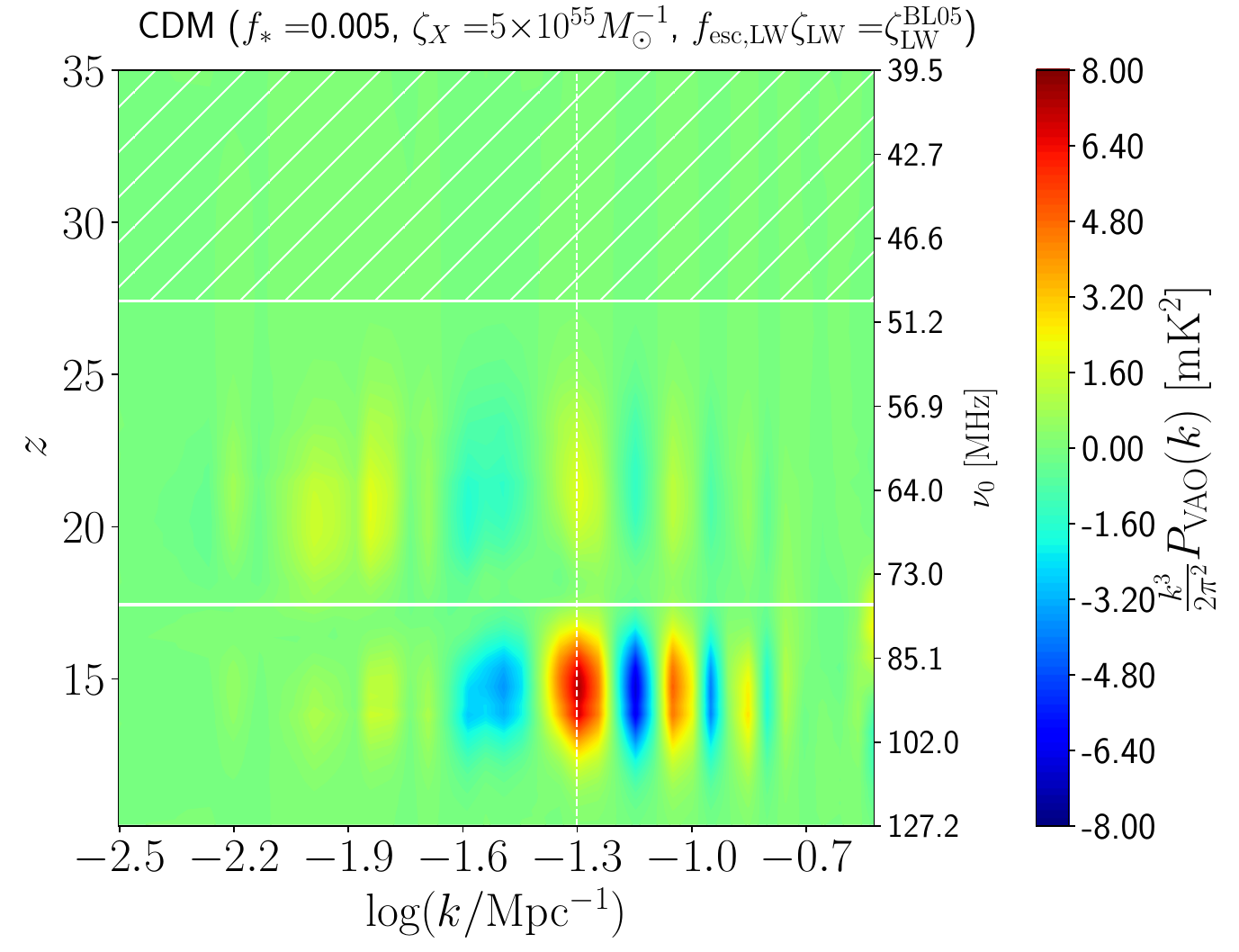 }}
\subfigure{\includegraphics[width=0.45\textwidth]{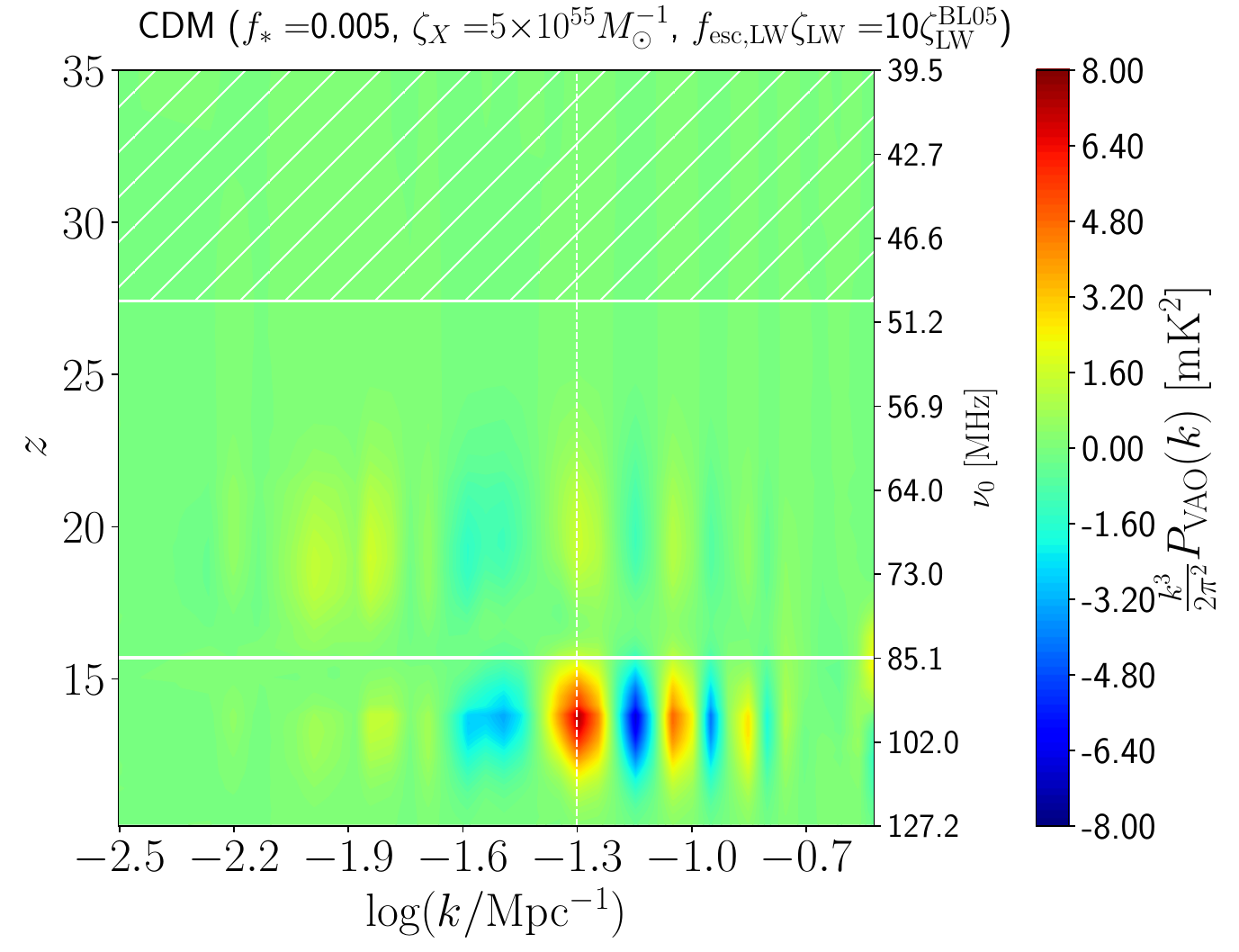 }}
\subfigure{\includegraphics[width=0.45\textwidth]{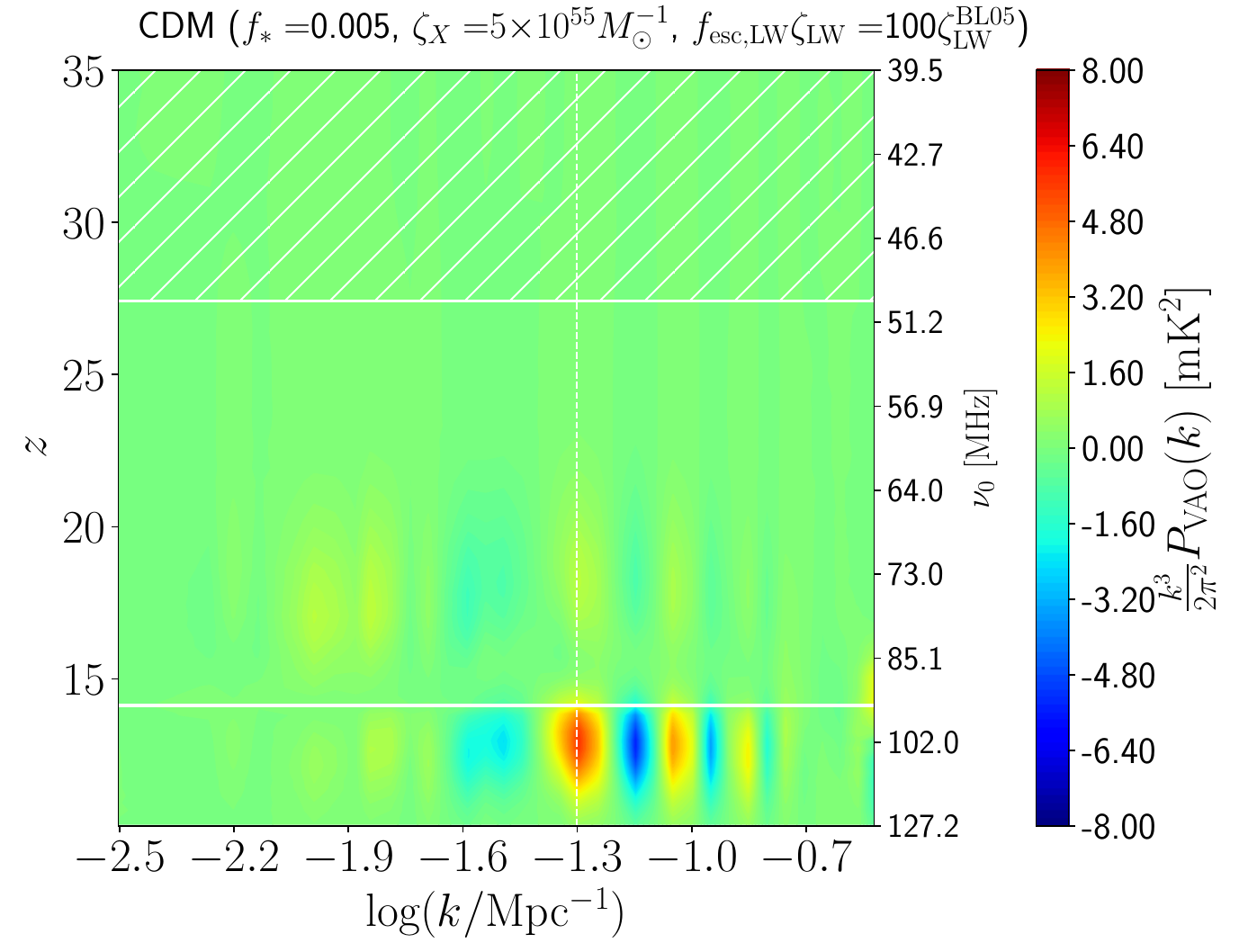 }}
\caption{
Similar to Fig. \ref{fig:ps_contour_S21}, however here we use the Eq. (\ref{eq:M_crit_F13}) as the critical mass for Pop III formation in minihalos.
The three panels correspond to $f_{\rm esc,LW}\zeta_{\rm LW}=\zeta_{\rm LW}^{\rm BL05}$, $10\zeta_{\rm LW}^{\rm BL05}$ and $100\zeta_{\rm LW}^{\rm BL05}$ respectively.
}
\label{fig:ps_contour_Fia}
}
\end{figure*}

Using the above critical mass, we show the extracted VAO wiggles for $f_{\rm esc,LW}\zeta_{\rm LW}=\zeta_{\rm LW}^{\rm BL05}$, 10$\zeta_{\rm LW}^{\rm BL05}$, and $100\zeta_{\rm LW}^{\rm BL05}$ respectively in Fig. \ref{fig:ps_contour_Fia}. The VAO features still exist even though the LW radiation field is strong, say $f_{\rm esc,LW}\zeta_{\rm LW}\sim 10 \zeta_{\rm LW}^{\rm BL05}$. This is partially because the Ly$\alpha$ radiation field is not just influenced by the instantaneous LW feedback. Actually, even for extremely strong LW radiation, say $J_{\rm LW,21}\sim \mathcal{O}(10)$, which suppresses the formation of Pop III stars in minihalos, there is still a large fraction of the Ly$\alpha$ photons from Pop III stars formed earlier when the LW feedback is still weak. However, when $f_{\rm esc,LW}\zeta_{\rm LW}\gtrsim 10 \zeta_{\rm LW}^{\rm BL05}$, the VAO signal is reduced significantly and not detectable for the SKA2-low.

Since Eq. (\ref{eq:M_crit_F13}) just simply re-scales the critical mass for zero-LW radiation, the relation  
\begin{equation}
\frac{M_{\rm crit}(v_{\rm db}\neq 0, J_{\rm LW,21}=0 )}{ M_{\rm crit}(v_{\rm db}= 0, J_{\rm LW,21}=0 )}=\frac{M_{\rm crit}(v_{\rm db}\neq 0, J_{\rm LW,21}\neq 0 )}{ M_{\rm crit}(v_{\rm db}= 0, J_{\rm LW,21}\neq 0 )}
\end{equation}
always holds, that the dependence of the critical mass on the relative streaming velocities is not influenced by the LW radiation. It may underestimate the role played by the LW feedback.

We can also define an alternative critical mass that is more sensitive to the LW feedback. From the simulations in \citet{Machacek2001ApJ}, in the presence of LW radiation however the relative motion is negligible, the critical mass for Pop III stars formation is
\begin{equation}
M_{\rm crit}(J_{\rm LW,21}|v_{\rm db}=0)=1.25\times10^5+8.7\times10^5(4\pi J_{\rm LW,21})^{0.47}~~~[M_\odot].
\label{eq:M_cool_2}
\end{equation}
The mass is then translated into circular velocity $V_{\rm circ,0}(J_{\rm LW,21}|v_{\rm db}=0)$. Then, analogous to Eq. (\ref{eq:V_cool}), the circular velocity for the critical mass in the presence of both LW radiation and relative streaming velocities is 
\begin{equation}
V_{\rm circ}(J_{\rm LW,21},v_{\rm db})=[V^2_{\rm circ}(J_{\rm LW,21}|v_{\rm db}=0)+(\alpha_v v_{\rm db})^2]^{1/2}.
\label{eq:V_cir2}
\end{equation} 
Then the viral mass corresponding to the above circular velocity is the new critical mass.

\begin{figure*}
\centering{

\subfigure{\includegraphics[width=0.45\textwidth]{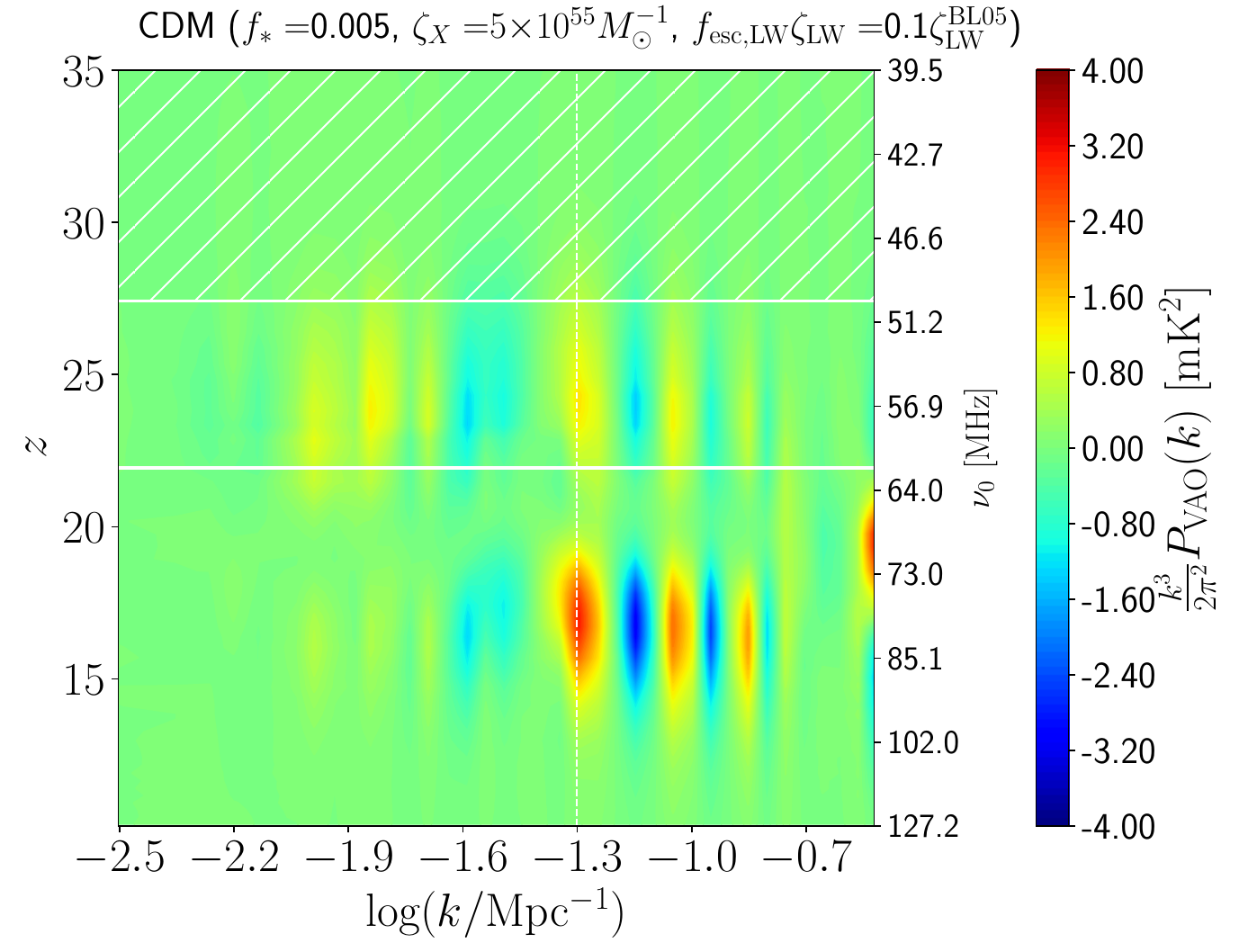 }}
\subfigure{\includegraphics[width=0.45\textwidth]{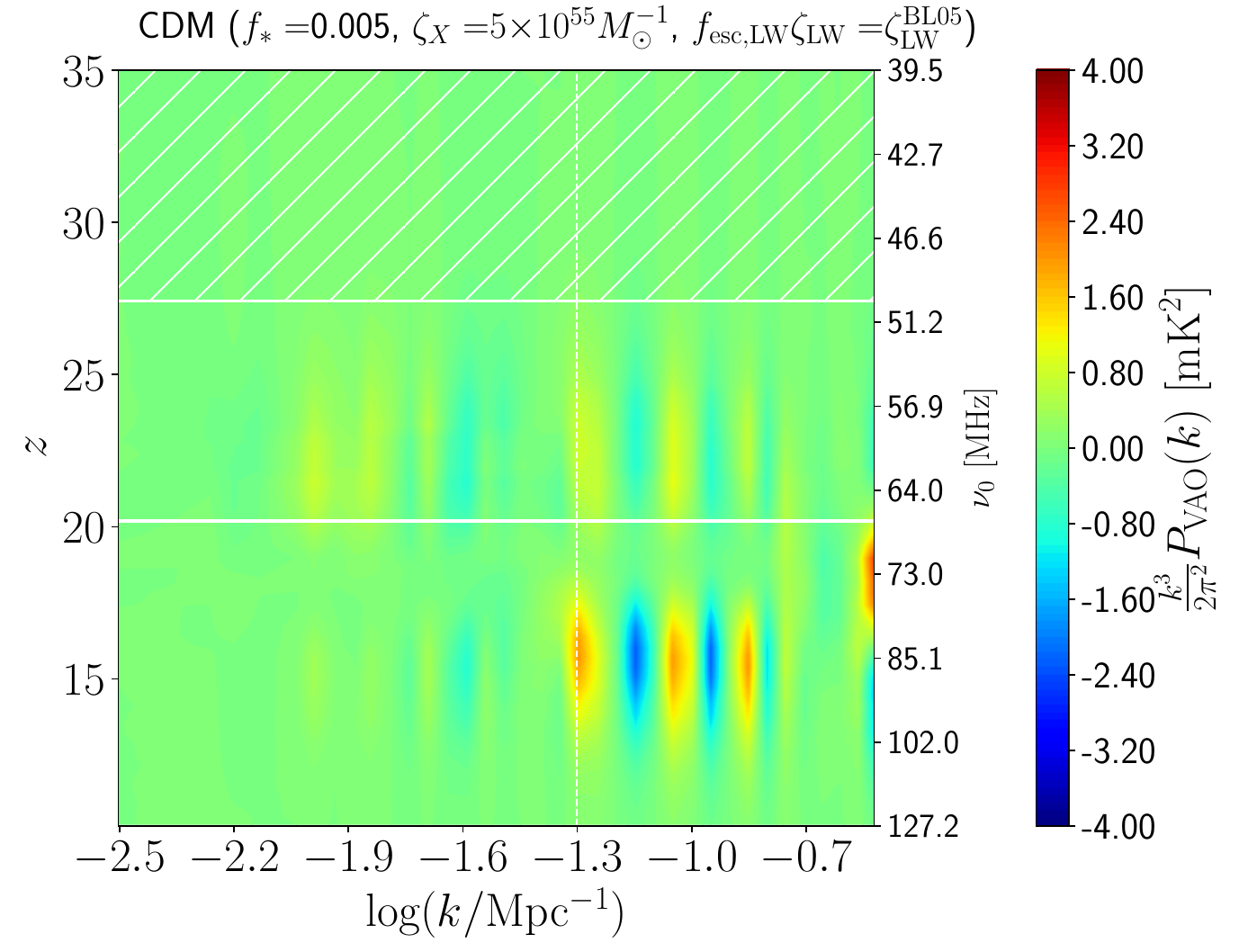 }}
\subfigure{\includegraphics[width=0.45\textwidth]{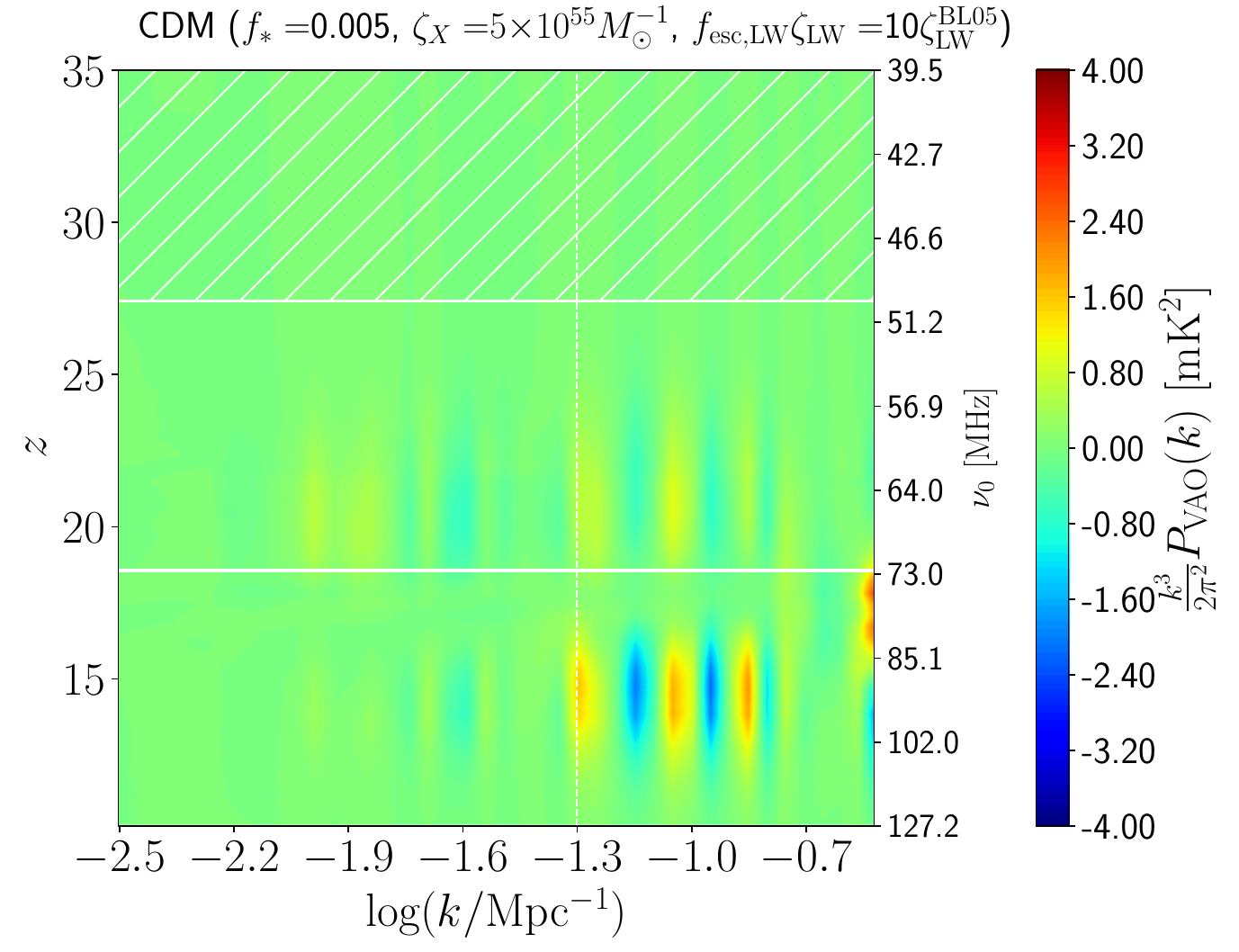 }}

\caption{
Similar to Fig. \ref{fig:ps_contour_S21}, however here we use the virial mass corresponding to circular velocity Eq. (\ref{eq:V_cir2}) as the critical mass for Pop III formation in minihalos.
The three panels correspond to $f_{\rm esc,LW}\zeta_{\rm LW}=0.1\zeta_{\rm LW}^{\rm BL05}$, $\zeta_{\rm LW}^{\rm BL05}$ and $10\zeta_{\rm LW}^{\rm BL05}$ respectively.
}
\label{fig:ps_contour_Mc2}
}
\end{figure*}

This critical mass is more sensitive to the LW radiation compared with Eq. (\ref{eq:M_crit_F13}).
The extracted VAO wiggles for $f_{\rm esc,LW}\zeta_{\rm LW}=0.1\zeta_{\rm LW}^{\rm BL05}$, $\zeta_{\rm LW}^{\rm BL05}$ and 10$\zeta_{\rm LW}^{\rm BL05}$ are shown in Fig. \ref{fig:ps_contour_Mc2}. We see that for this new critical mass, the VAO wiggles are reduced significantly when $f_{\rm esc,LW}\zeta_{\rm LW}\gtrsim \zeta_{\rm LW}^{\rm BL05}$.

\begin{figure}
\centering{
\subfigure{\includegraphics[width=0.45\textwidth]{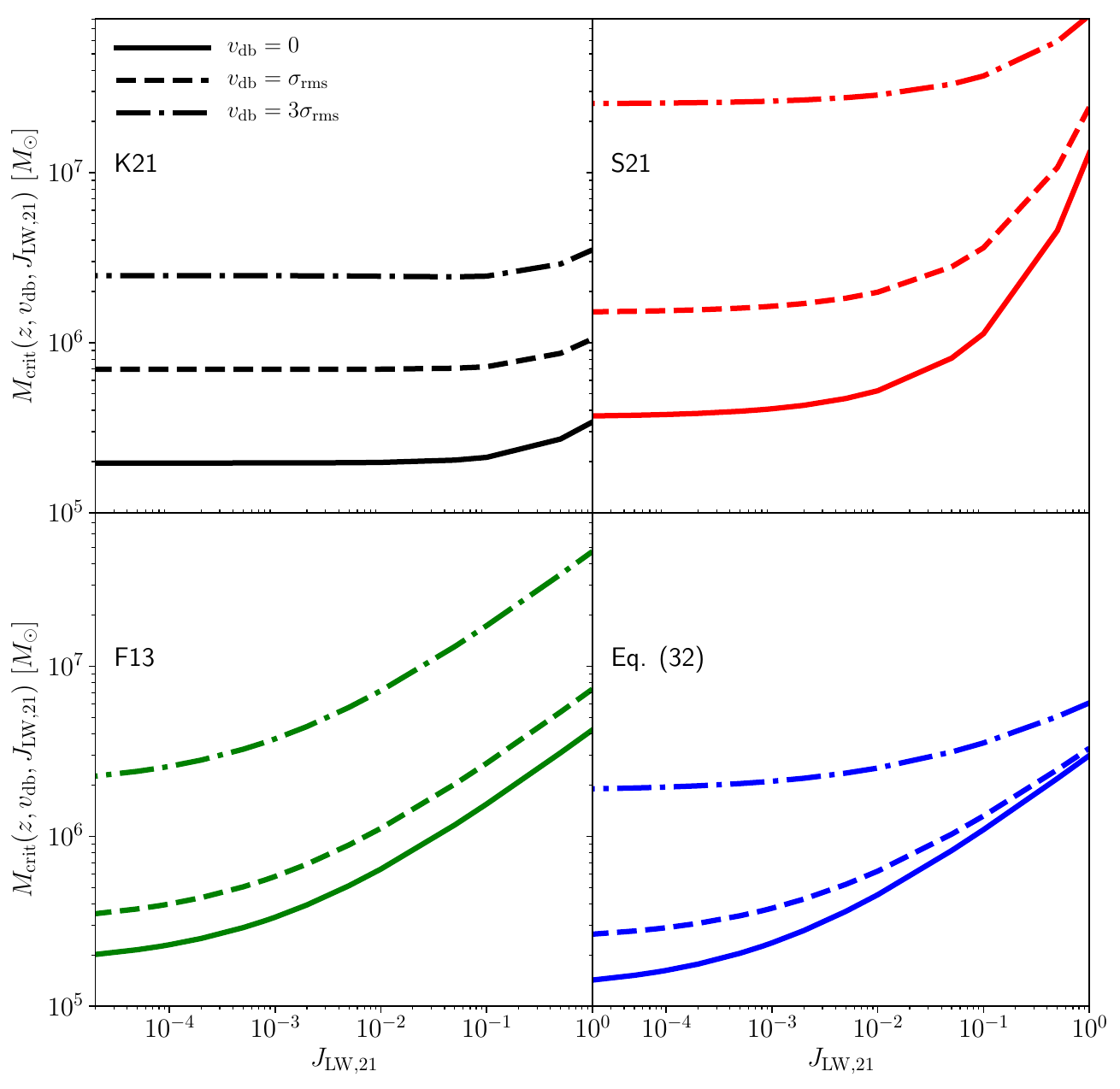}}
\caption{
The critical masses at $z=20$ as a function of $J_{\rm LW,21}$ for $v_{\rm db}=0$, $\sigma_{\rm rms}$ and $3\sigma_{\rm rms}$ respectively. The four panels refer to \citet{Kulkarni2021ApJ} (K21), \citet{Schauer2021MNRAS} (S21), \citet{Fialkov2013MNRAS} (F13) and the virial mass corresponding to Eq. (\ref{eq:V_cir2})  (Eq. (32)) respectively.
}
\label{fig:crit_masses}
}
\end{figure}

As a summary, in Fig. \ref{fig:crit_masses} we show the four critical masses as a function of $J_{\rm LW,21}$ for various relative streaming velocities. Their behaviors are quite different. For example, the \citet{Kulkarni2021ApJ} and \citet{Fialkov2013MNRAS} critical masses are sensitive to the relative streaming velocities even the $J_{\rm LW,21}$ is as high as $\sim 1$. So the models using those critical masses will predict stronger VAO wiggles. The \citet{Schauer2021MNRAS} critical mass is sensitive to the relative streaming velocities when the $J_{\rm LW,21} \lesssim 0.1$. So for this critical mass, the VAO wiggles are strong for weak LW radiation.  For the critical mass corresponding to Eq. (\ref{eq:V_cir2}), it is sensitive to the relative streaming velocities when $J_{\rm LW,21}\lesssim 0.01$. For higher LW specific intensity, the dependence on relative streaming velocities decreases gradually (curves for different velocities are approaching each other). As a result, the VAO wiggles are reduced by the  LW feedback.

To correctly calculate the VAO wiggles, it is crucial to model the critical mass appropriately.
Currently, the effects of LW feedback are still not yet fully understood, particularly the self-shielding effect. Nevertheless, the \citet{Kulkarni2021ApJ} and \citet{Schauer2021MNRAS} critical masses are the first results from simulations including both LW feedback and relative streaming velocities.  Based on our results using these critical masses, it is likely that for broad ranges of astrophysical parameters, there are VAO features on the 21 cm power spectrum.

\subsection{The VAO signal in various dark matter models}

\begin{figure}
\centering{
\subfigure{\includegraphics[width=0.45\textwidth]{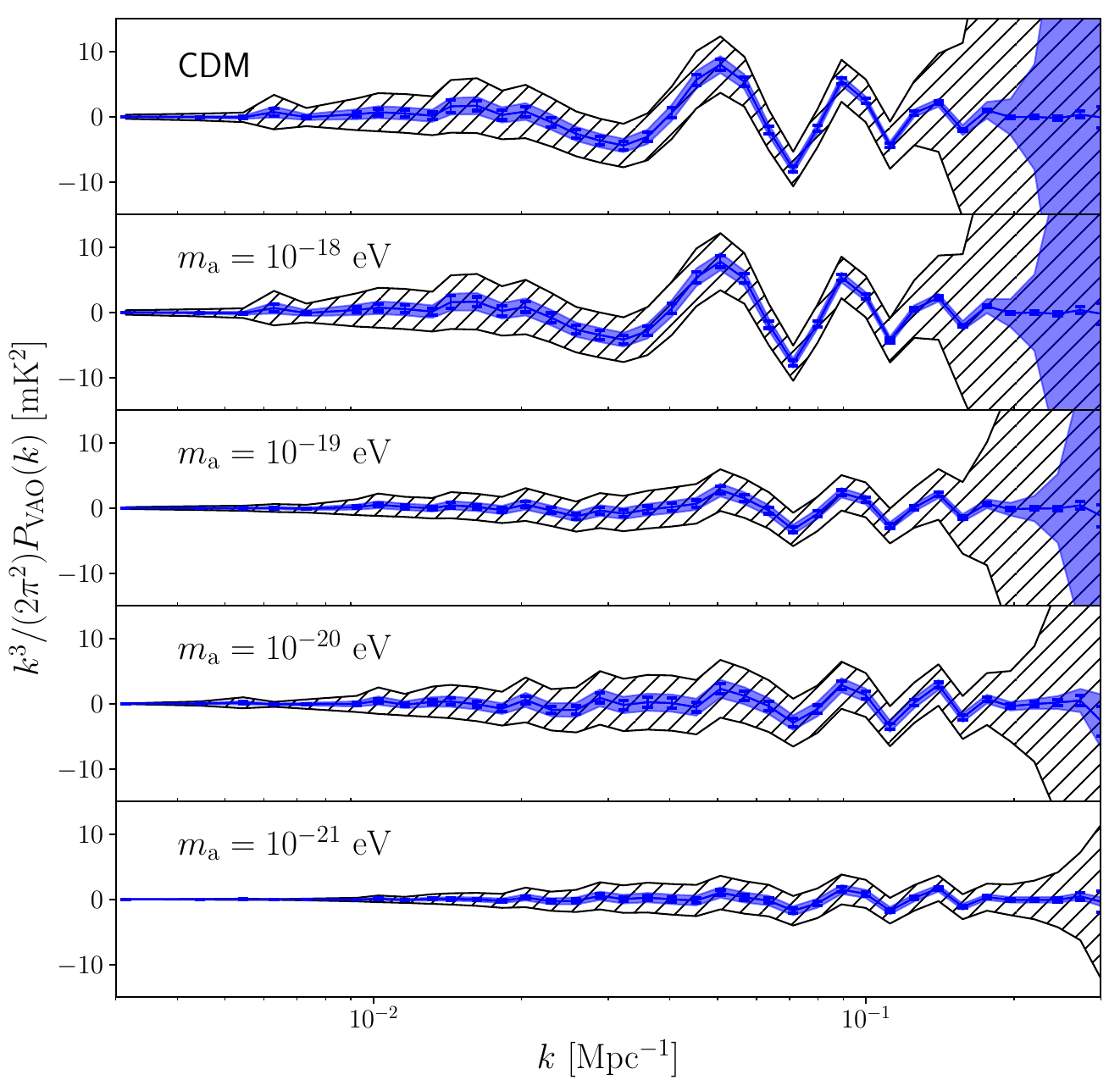}}
\caption{The VAO wiggles in different dark matter models. 
The hatched regions refer to the uncertainties for SKA1-low observation with survey area 20 deg$^2$, while the color-filled regions refer to SKA2-low observation with survey area 200 deg$^2$. For all of them we take $t_{\rm obs}=2000$ hour and $\Delta \nu=20$ MHz.
}
\label{fig:VAO_wiggles_vs_ma}
}
\end{figure}

The locations of the wiggles are determined by the large-scale distribution of the relative velocity field, so in the axion model their amplitudes are smaller but the locations are the same as that of the CDM model. In Fig. \ref{fig:VAO_wiggles_vs_ma} we plot the VAO wiggles in different dark matter models, with the SKA1-low uncertainties for survey area 20 deg$^2$ and 
SKA2-low uncertainties for survey area 200 deg$^2$ respectively.

\begin{figure}
\centering{
\subfigure{\includegraphics[width=0.45\textwidth]{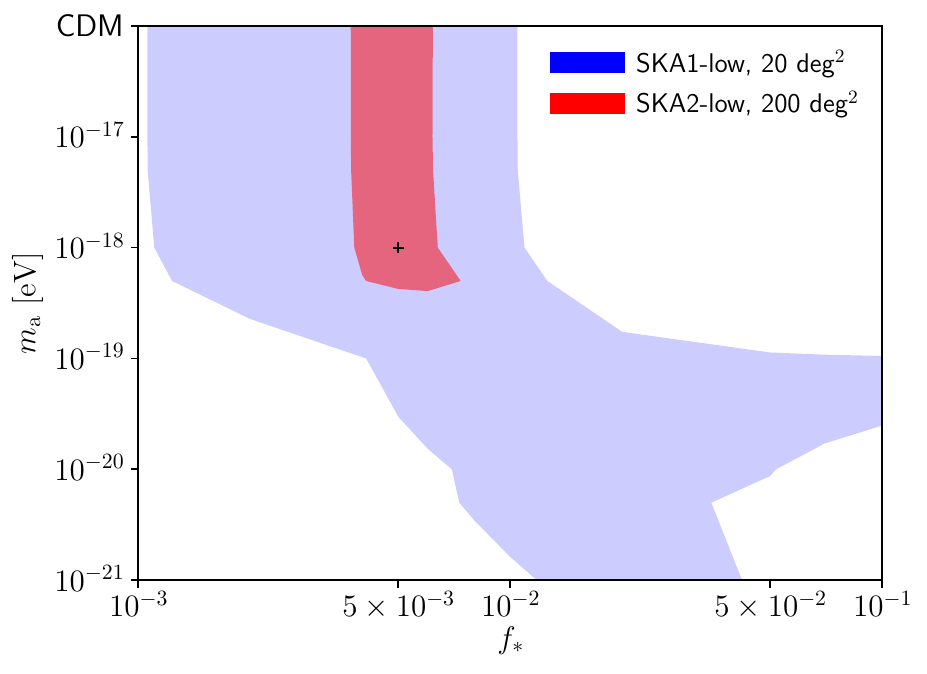}}
\caption{The predicted 3$\sigma$ confidence level (99.7\%) constraints on $m_{\rm a}-f_*$ by SKA1-low with 20 (light blue filled region) and SKA2-low with 200 (red filled region) deg$^2$ survey area respectively. 
We set the bandwidth $\Delta \nu =20$ MHz and observation time $t_{\rm obs}=2000$ hour.
The input parameters $m_{\rm a}=10^{-18}$ eV and $f_*=0.005$. We fix the astrophysical parameters $\zeta_X=5\times10^{55}~M_\odot^{-1}$ and $f_{\rm esc,LW}\zeta_{\rm LW}=\zeta_{\rm LW}^{\rm BL05}$ and the cosmology.
}
\label{fig:constraints_F_STAR_ma}
}
\end{figure}

In principle, to make the forecast for the parameter constraints, we should set all astrophysical and cosmological parameters as free parameters, and involve the {\tt 21cmFAST} in the MCMC procedure \citep{21cmmc}. It will require quite a lot of computational cost. This is not the focus of this paper. In this paper, we focus on inspecting the existence of VAO wiggles in the CDM model with different astrophysical parameters, and the axion models with different masses. 
However, if we only focus on two free parameters and freeze other astrophysical and cosmological parameters (suppose they can be derived from other observations or theories), we show simple demonstrations of the constraints on parameters $m_{\rm a}$ and $f_*$ in Fig. \ref{fig:constraints_F_STAR_ma}, 
from the suspected SKA1-low observations with survey area 20 deg$^2$ and SKA2-low observations with 200 deg$^2$ respectively.
$\Delta \nu =20 $MHz,
and $t_{\rm obs}=2000$ hour.

For SKA1-low, the main constraints are from $0.01~{\rm Mpc^{-1}}\lesssim k\lesssim 0.1$ Mpc$^{-1}$. At smaller scale the instrumental noise is large, at larger scale the cosmic variance is large. 
In the bottom panel of Fig. \ref{fig:21cm_ps} and in Fig. \ref{fig:VAO_wiggles_vs_ma}, seemingly at smaller $k$ the uncertainties are smaller, however, the relative uncertainties increase rapidly with decreasing $k$, due to the cosmic variance.
In such constraints, we do not involve the influence of the foreground, so this is an optimistic forecast. The foreground may bias the measured large-scale power spectrum, however, it can be somewhat corrected for, see the discussions in Sec. \ref{sec:foreground}. 
There is a degeneracy between the $m_{\rm a}$ and $f_*$ in the SKA1-low constraints, since both lower $f_*$ and lower $m_{\rm a}$ can result in weaker VAO wiggles.
Nevertheless, assuming input parameter $m_{\rm a}=10^{-18}$ eV and $f_*=0.005$, the SKA1-low  can help to rule out the axion models with $m_{\rm a} \lesssim 10^{-19}$ eV and $f_* \lesssim 5\times10^{-3}$, or with $m_{\rm a} \gtrsim 10^{-19}$ eV and $f_* \gtrsim 10^{-2}$. 
The SKA2-low can break the degeneracy and tightly constrain the $f_*$ and lower limit of $m_{\rm a}$.
We remind again that this is a demonstration because we freeze the parameters $\zeta_X$, $f_{\rm esc,LW}\zeta_{\rm LW}$ and the cosmology, however it is still useful.

\begin{figure*}
\centering{
\subfigure{\includegraphics[width=0.85\textwidth]{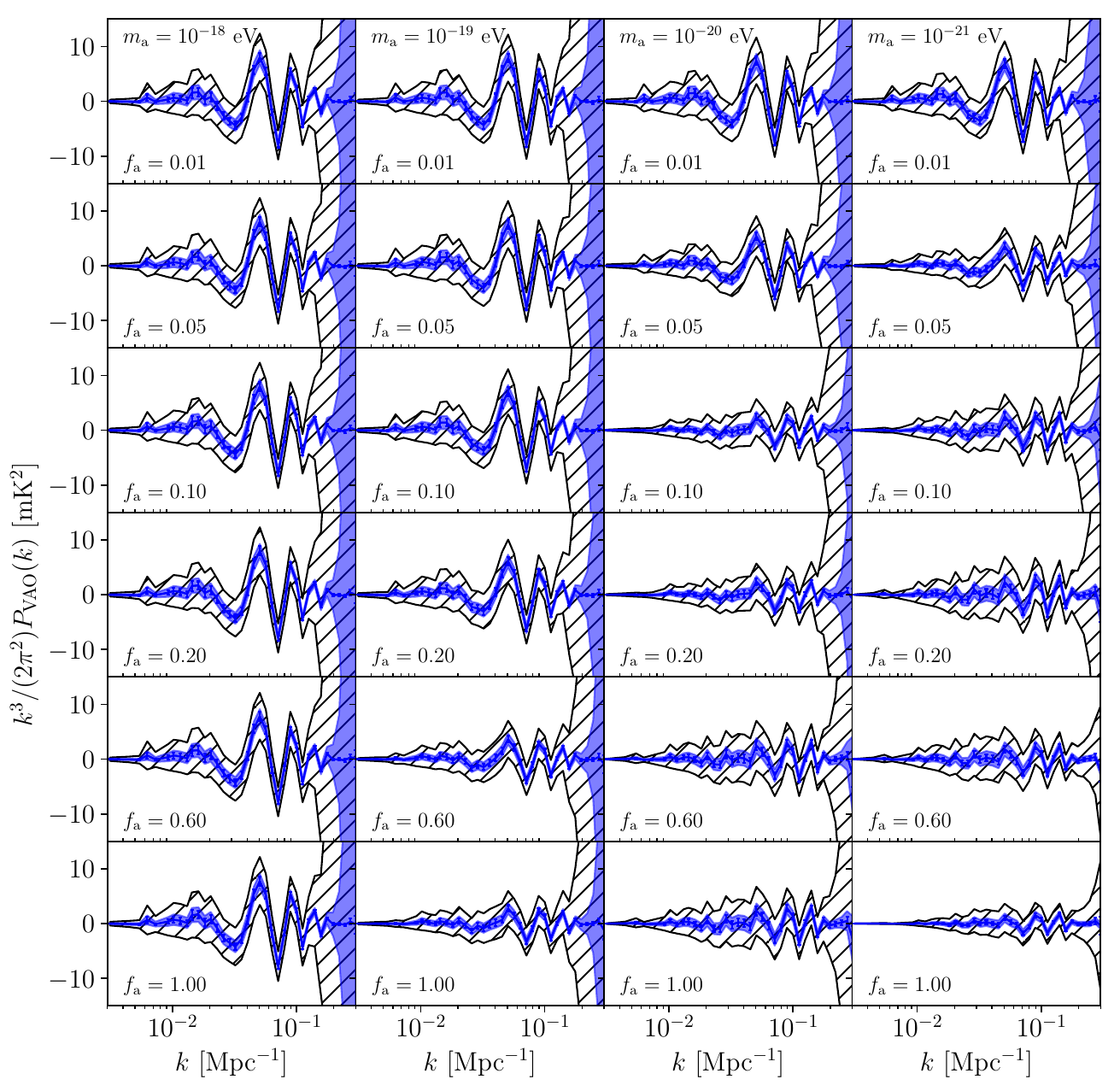}}
\caption{The VAO wiggles in dark matter models where CDM is mixed with different fractions of axions with various masses. 
The hatched regions refer to the uncertainties for SKA1-low observation with survey area 20 deg$^2$, while the color-filled regions refer to SKA2-low with 200 deg$^2$. 
}
\label{fig:VAO_wiggles_vs_fa}
}
\end{figure*}

\begin{figure}
\centering{
\subfigure{\includegraphics[width=0.45\textwidth]{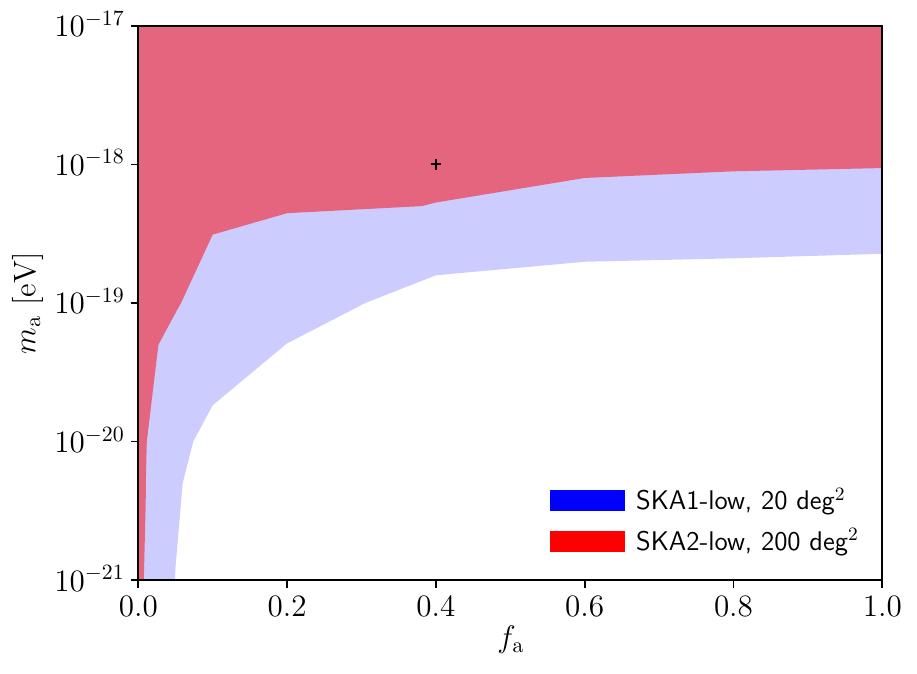}}
\caption{
Similar to Fig. \ref{fig:constraints_F_STAR_ma}, however here we assume the CDM is mixed with axions, so the constrained parameters are $m_{\rm a}$ and $f_{\rm a}$.
The input parameters are $m_{\rm a}=1\times10^{-18}$ eV and $f_{\rm a}=0.4$.
We fix $f_*=0.005$, $\zeta_X=5\times10^{55}~M_\odot^{-1}$ and $f_{\rm esc,LW}\zeta_{\rm LW}=\zeta_{\rm LW}^{\rm BL05}$.
}
\label{fig:constraints_fa_ma}
}
\end{figure}

For mixed dark matter models, in Fig. \ref{fig:VAO_wiggles_vs_fa} we show the VAO wiggles in dark matter models with CDM and different fractions of various axions. For axion models with $m_{\rm a}=10^{-21}, 10^{-20}$ and $10^{-19}$ eV, the critical $f_a$ above which VAO signal disappears  are $\sim10\%$, $\sim20\%$, $\sim60\%$ respectively.  For $m_{\rm a}=10^{-18}$ eV there is always VAO signal even when $f_{\rm a}$ reaches up $100\%$. Similar to Fig. \ref{fig:constraints_F_STAR_ma}, if we use fixed values for $f_*$, $\zeta_X$ and $f_{\rm esc,LW}\zeta_{\rm LW}$, the predicted constraints on $m_{\rm a}-f_{\rm a}$ is shown in Fig. \ref{fig:constraints_fa_ma}.   

\begin{figure}
\centering{
\subfigure{\includegraphics[width=0.45\textwidth]{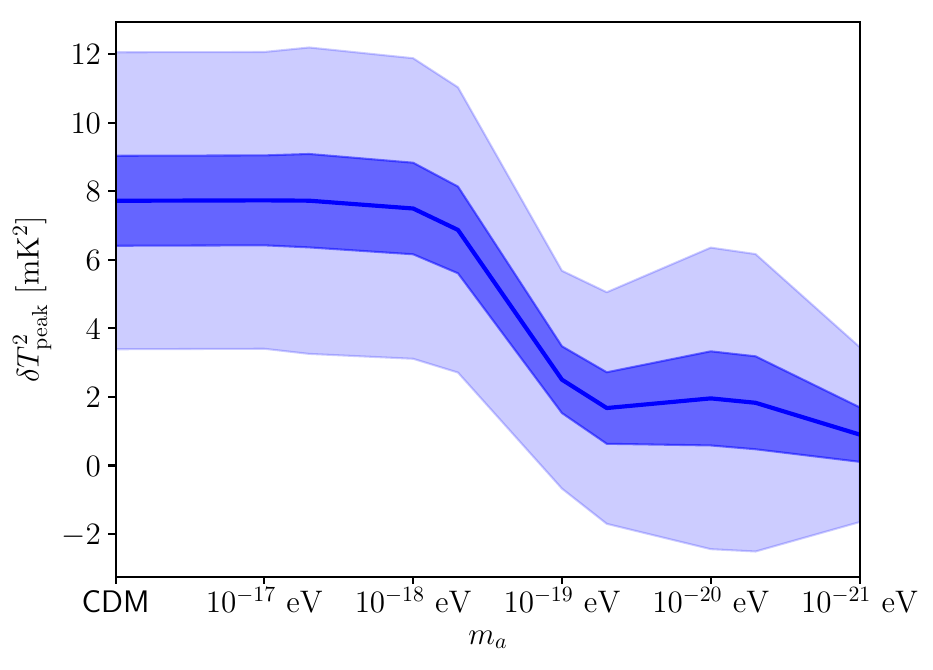}}
\caption{
The amplitude of the first peak at $k=0.05$ Mpc$^{-1}$ in different dark matter models. The region filled by light color refers to SKA1-low uncertainties for 20 deg$^2$ survey area, while the region filled by darker color refers to SKA2-low uncertainties for survey ara 200 deg$^2$.  
}
\label{fig:first_peak_vs_ma}
}
\end{figure}

\begin{figure}
\centering{
\subfigure{\includegraphics[width=0.45\textwidth]{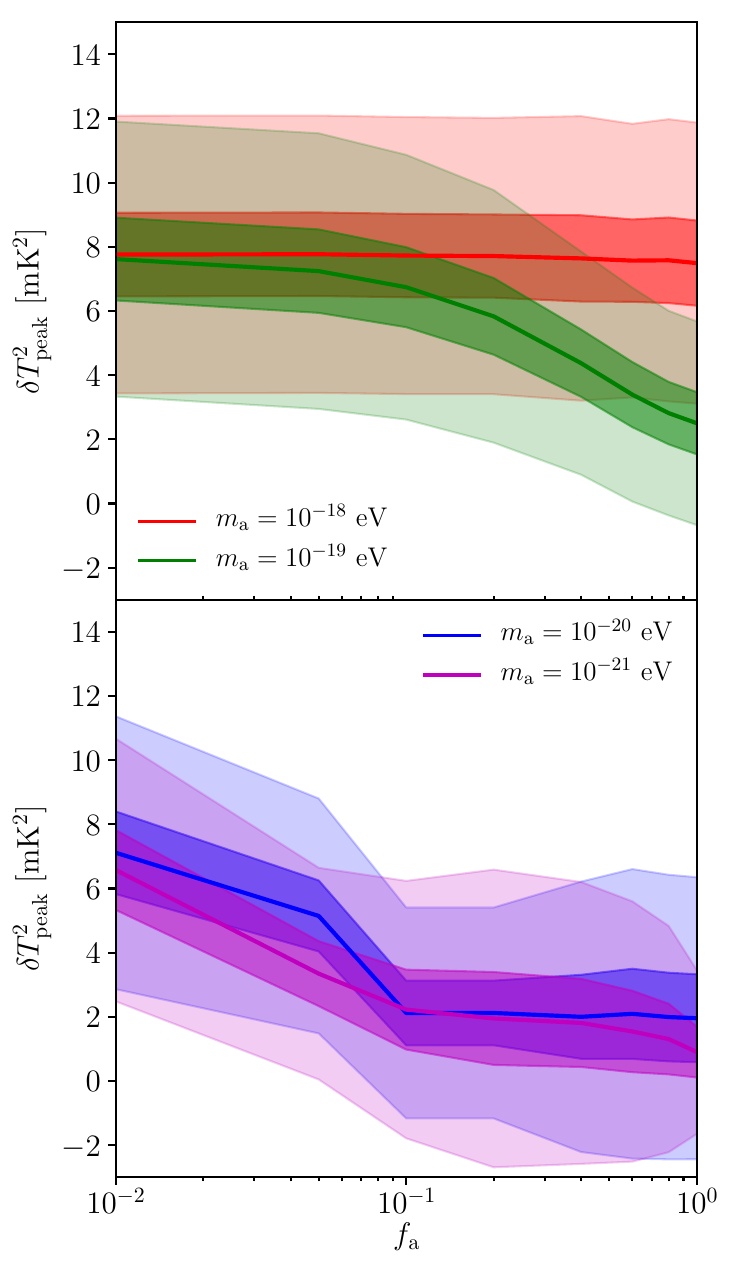}}
\caption{
The amplitude of the first peak at $k=0.05$ Mpc$^{-1}$ for different axion massess with various fractions. The meanings of the filled regions are same to Fig. \ref{fig:first_peak_vs_ma}. Since the panel will be too crowded if we plot all curves together, we use two panels.  
}
\label{fig:first_peak_vs_fa}
}
\end{figure}

If in different dark matter models the VAO wiggles have the similar shape, only the normalizations are different, then the amplitude of the strongest peak at $k=0.05$ Mpc$^{-1}$ is a very useful indicator for dark matter properties. In Fig. \ref{fig:first_peak_vs_ma} we show this amplitude vs. dark matter models. It shows more straightforwardly that the VAO signal almost disappears when $m_{\rm a}\lesssim 10^{-19}$ eV. In Fig. \ref{fig:first_peak_vs_fa} we show the amplitude of the strongest peak vs. $f_{\rm a}$, for different axion masses. To avoid a crowded panel, we use two panels.

For $\zeta_X=5\times10^{55}~M_\odot^{-1}$ and $f_{\rm esc,LW}\zeta_{\rm LW}=\zeta_{\rm LW}^{\rm BL05}$, we provide an approximation for the amplitude of the strongest VAO peak, $\delta T^2_{\rm peak}=\frac{k^3}{2\pi^2}P_{\rm VAO}(k=0.05~{\rm Mpc^{-1}})$, as a function of $f_*$, $m_{\rm a}$ and $f_{\rm a}$,
\begin{equation}
\frac{\delta T^2_{\rm peak}}{{\rm (mK)^2}}\approx g(f_*)[1-f(m_{\rm a})]\tilde{f}_{\rm voigt}(f_{\rm a},\sigma_{\rm a},\gamma_{\rm a})+ f(m_{\rm a})g(f_*),
\label{eq:first_peak_appro}
\end{equation}
where $\tilde{f}_{\rm voigt}$ is the normalized  Voigt function, $\tilde{f}_{\rm voigt}(0,\sigma_{\rm a},\gamma_{\rm a})=1$.  
\begin{equation}
f(m_{\rm a})= 0.5 \{1+{\rm erf}[ 1.5({\rm log}m_{\rm a}+18.8)  ]\},
\end{equation}
\begin{equation}
g(f_*)=79.6+49.4{\rm \log}f_*+7.9({\rm log}f_*)^2,
\end{equation}
\begin{equation}
\sigma_{\rm a}={\rm max}[1.0+3({\rm log}m_{\rm a}+18),0.0],
\end{equation}
and
\begin{equation}
\gamma_{\rm a}={\rm max}[0.05+0.23({\rm log}m_{\rm a}+20),0.035].
\end{equation}
Eq. (\ref{eq:first_peak_appro}) can be used to quickly estimate the VAO signal.

\section{Conclusion and discussions}\label{sec:conclusion}

\subsection{The conclusion}

The relative streaming velocities between dark matter particles and baryon particles have large-scale coherent structures. Such velocity can suppress the formation of Pop III stars in minihalos, reducing their production of Ly$\alpha$, X-ray, and ionizing photons, therefore modifying the 21 cm brightness temperature field. Therefore the structures of the streaming velocities are encoded in the 21 cm signal. Intuitively, the streaming velocities generate the VAO wiggles on the 21 cm power spectrum of the Cosmic Dawn. We modified the {\tt 21cmFAST} code to add the streaming velocities and the LW feedback, then used it to investigate the VAO signal in CDM and axion dark matter models, and the axion-CDM mixed model. We found:

\begin{itemize}
\item In the CDM model, there are always VAO wiggles on the 21 cm power spectrum for wide ranges of Pop III star parameters. The wiggles are detectable for SKA2-low with $\sim$ 2000 hour integration time;

\item In the axion model, when $m_{\rm a} \lesssim10^{-19}$ eV, the VAO wiggles become negligible;

\item In the mixed models with $m_{\rm a}=10^{-21}, 10^{-20}$ and $10^{-19}$ eV, when $f_{\rm a }\gtrsim 10\%$, 20\% and 60\%, the VAO wiggles are neglibilbe. For $m_{\rm a}=10^{-18}$ eV, there are always obvious VAO wiggles even when $f_{\rm a}$ is 100\%; 

\item Using the SKA2-low, the VAO signal can be used as a good tool to distinguish dark matter models with different particle properties. 
\end{itemize}

Considering the uncertainties of astrophysical parameters, to detect the small-scale density fluctuations and distinguish the CDM and the axion models reliably, for the SKA-low it is necessary to survey the full frequency range $50~{\rm MHz} \lesssim \nu_0 \lesssim 130$ MHz.

\subsection{The discussions}

\subsubsection{The influence of X-ray}

We did not model the X-ray feedback in our paper, as this effect cannot be simply described by the change of critical minihalo mass. We checked that, however, in all our cases, when then VAO wiggles reach the maximum amplitude, the kinetic temperature of the IGM is still smaller than or comparable to the CMB. Furthermore, at this time the Jeans mass is boosted by $\lesssim 10$, and the filtering mass \citep{Gnedin2000ApJ} is boosted by $\lesssim1.3$ by X-ray heating. Since both the Jeans mass and the filtering mass are just the minimal mass for minihalos that can hold gas, they are much smaller than the critical mass for Pop III star formation (e.g. \citealt{Barkana2001PhR,Glover2013ASSL}). If they are just boosted by modest factors, star formation in minihalos above the critical mass is unlikely influenced. We believe that the X-ray feedback will not change our conclusion. However, to fully model such an effect, future hydrodynamic simulations including both the relative streaming velocities, the LW feedback, and the X-ray feedback simultaneously will be required.

\subsubsection{The convergence of simulations with different resolutions}\label{sec:convergence}

In this paper, all the fiducial simulations have the cell size 16.7 Mpc. This is much larger than the coherence scale of the relative streaming velocity field, which is $\sim$3 Mpc (e.g. \citealt{Tseliakhovich2010PhRvD}). Since the initial conditions of the density field and velocity field are generated in Fourier space, we can reproduce their structures above the cell size, but lose the structures within the cell. When calculating the collapse fraction, it implies that for each cell we replace $\mean{f_{\rm coll}(v_{\rm db})}_{\rm cell}$ with the approximation $f_{\rm coll}( \mean{v_{\rm db}}_{\rm cell})$, where $\mean{}_{\rm cell}$ denotes the average inside the cell. Although such a kind of approximation is quite popular in practice, it may result in some errors.

\begin{figure}
\centering{
\subfigure{\includegraphics[width=0.45\textwidth]{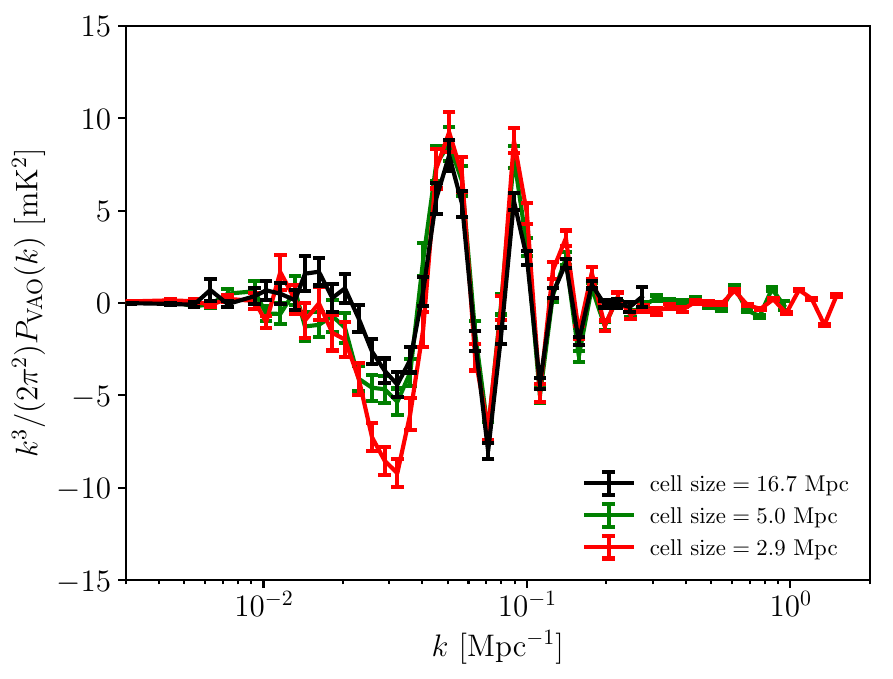}}
\caption{
The VAO wiggles in simulations with different resolutions. The redshift is $\approx$17.
}
\label{fig:resolution}
}
\end{figure}

In this section, we check for the convergence of our simulations for various resolutions. In Fig. \ref{fig:resolution}, we show the VAO wiggles at the same redshift $\approx 17$ in simulations with cell size $16.7$ Mpc (our fiducial simulation), 5 Mpc, and 2.9 Mpc, respectively. All simulations have the same parameters except the resolutions. From this figure, we see that in the overlapped $k$ range and $k\gtrsim 0.03~{\rm Mpc}^{-1}$, all simulations, including our fiducial simulation, have results consistent with the simulation with cell size down to the coherence scale of the relative streaming velocity field, $\approx 3$ Mpc. At smaller $k$, however, the discrepances become more obvious.

\subsubsection{The impact of the foreground}\label{sec:foreground}

To evaluate the impact of the foreground on the VAO signal, we generate mock 21 cm data cubes with survey area $\Omega_{\rm survey}=100$ deg$^2$,  beam size $\theta_{\rm beam}=390''$ and frequency channel width $\delta \nu =0.2$ MHz. The 21 cm maps are centered at $z=17$, when the VAO wiggles are strongest in our fiducial model. We generate two samples with bandwidth $\Delta \nu =20$ MHz and 40 MHz respectively. To focus on the performance of foreground removal, we ignore the signal evolution within the bandwidth. We also ignore the beam effects. We take the foreground from the {\tt GSM} sky model\footnote{\url{https://github.com/telegraphic/PyGSM}} \citep{GSM2008,GSM2016} near the Locakman hole where the Galactic foreground is lowest, and add it to the 21 cm cubes. We also add the thermal noise of SKA2-low for $t_{\rm obs}=2000$ hour.

\begin{figure}
\centering{
\subfigure{\includegraphics[width=0.45\textwidth]{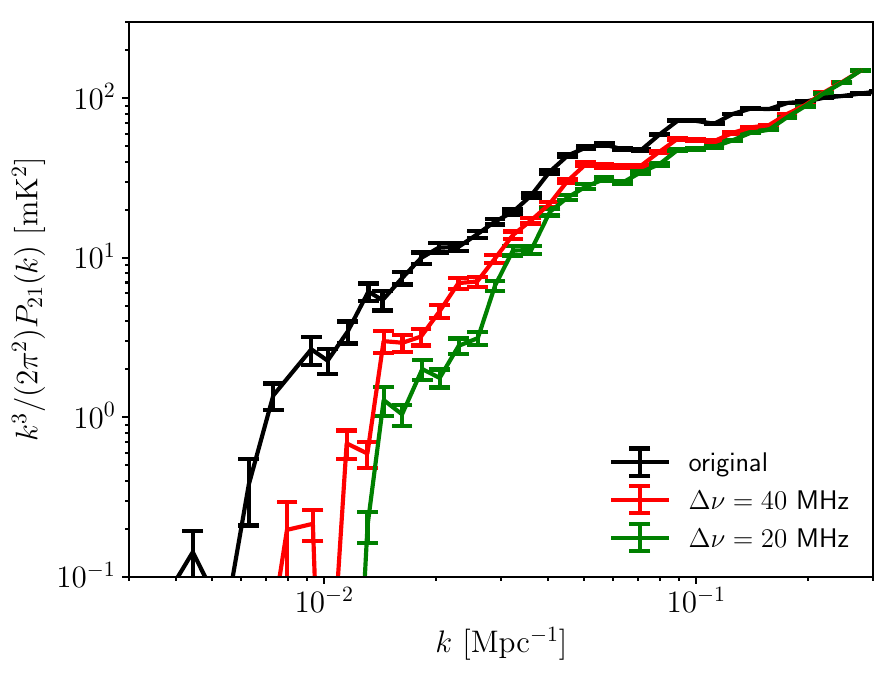}}

\subfigure{\includegraphics[width=0.45\textwidth]{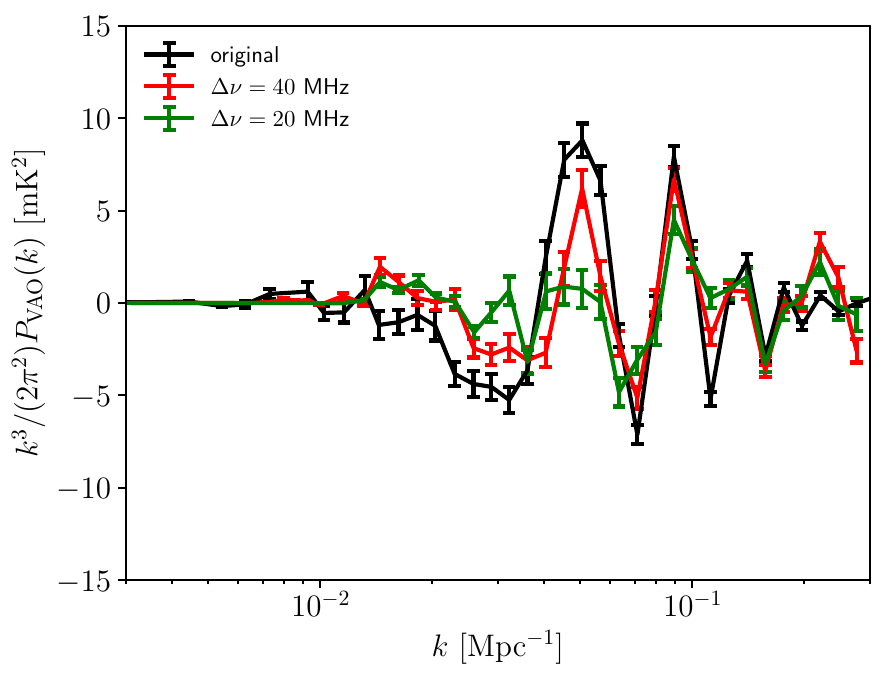}}

\caption{
{\it  Top:} The original 21 cm power spectrum and the power spectra of the foreground-removed mock data cubes for bandwidth 20 MHz and 40 MHz respectively. 
{\it Bottom:} The VAO wiggles extracted from the power spectra in the top panel.
}
\label{fig:foreground}
}
\end{figure}

Since the foreground overwhelmingly dominates over the 21 cm signal and is spectrally smooth, 
we use the SVD algorithm (e.g. \citealt{Yue2015MNRAS}) to find the principal components of the spectrum and remove the first 3 strongest principal components from each line of sight. The original 21 cm power spectrum and the power spectrum of the foreground-removed data cubes, and the corresponding VAO wiggles, are plotted in Fig. \ref{fig:foreground}. We find that, the SVD foreground removal algorithm generates some bias (the first peak is systematically underestimated) on the recovered VAO wiggles, and the performance depends on bandwidth. For larger bandwidth, the bias is smaller. However, we realize that the foreground removal algorithms are complicated and still in development. For example, although the PCA method may underestimate the power spectrum, the power loss can be corrected by using a transfer function (e.g.  \citealt{Cunnington2023MNRAS}). Machine learning may also help to compensate for the power loss (e.g. \citealt{deep21,Zhou2023}). On the other hand, however, the foreground may be not that smooth in frequency space, because of the polarization leakage (e.g. \citealt{Jelic2010,Gao2023}). For this reason, in this paper, we only involve the instrumental noise and cosmic variance when making the forecast for constraints, as optimistic estimates. We leave the impact of the foreground to further explicit investigations.

\begin{figure*}
\centering{
\subfigure{\includegraphics[width=0.45\textwidth]{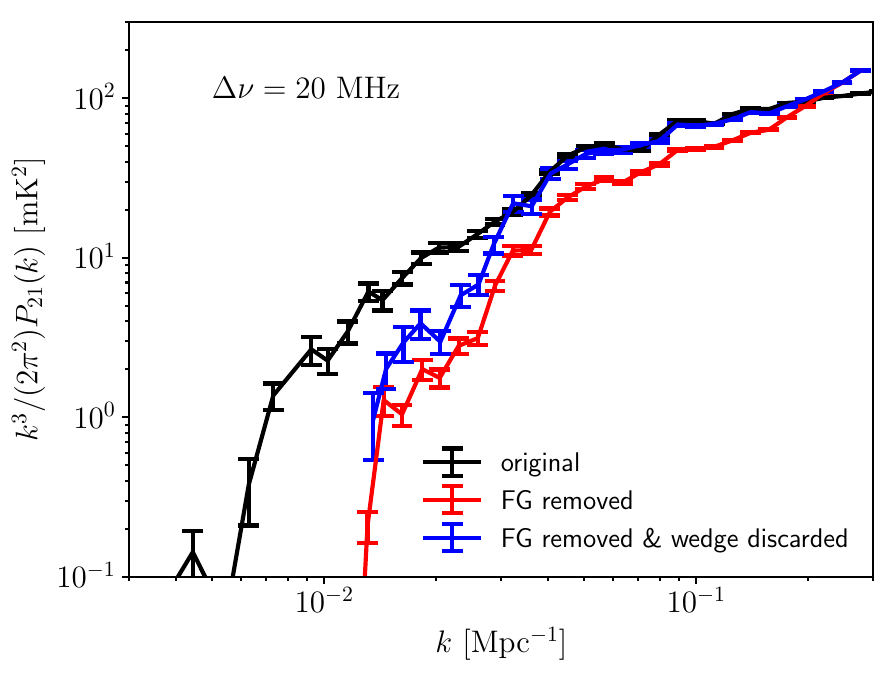}}
\subfigure{\includegraphics[width=0.45\textwidth]{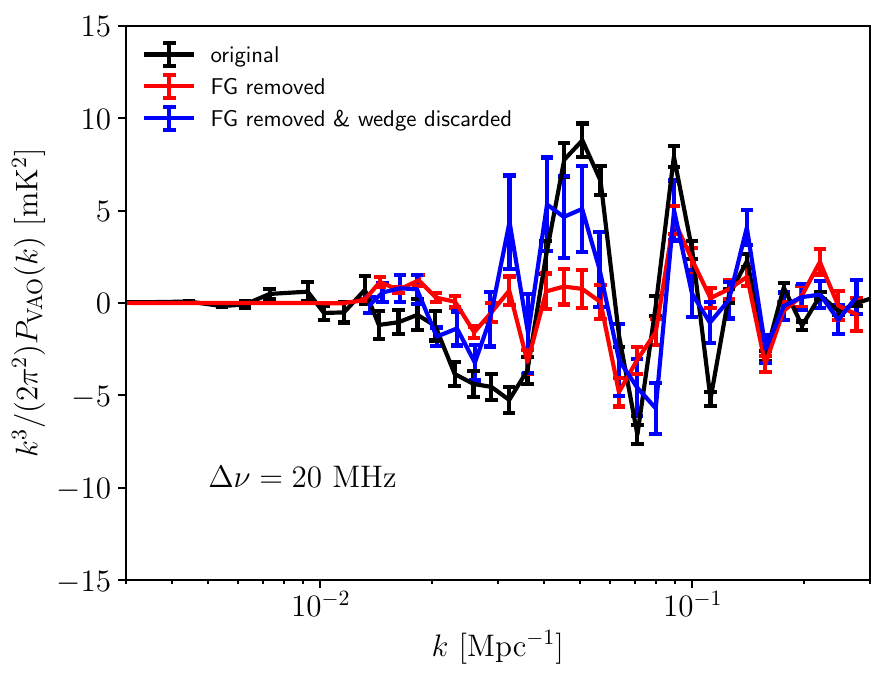}}
\subfigure{\includegraphics[width=0.45\textwidth]{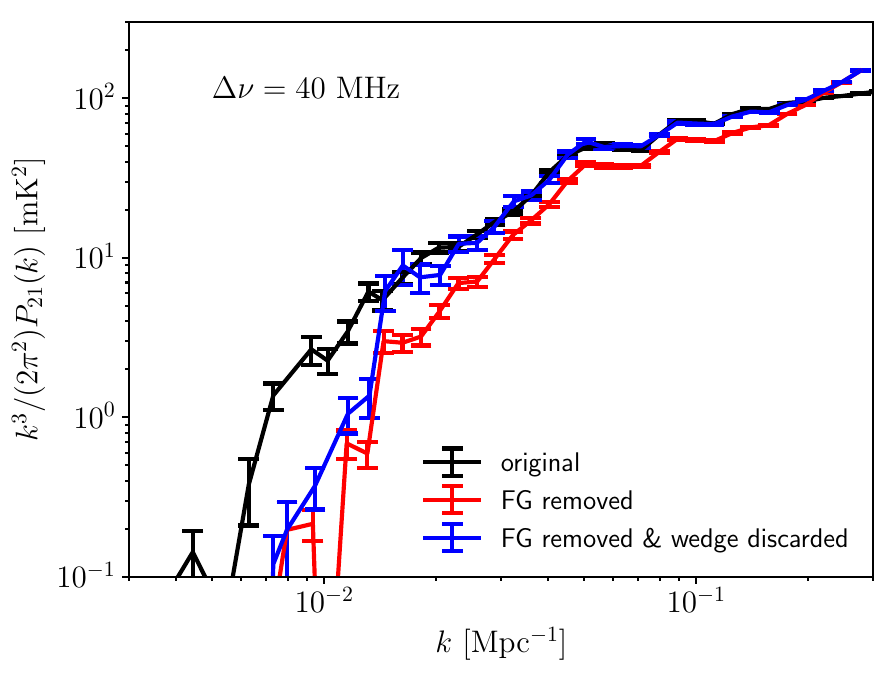}}
\subfigure{\includegraphics[width=0.45\textwidth]{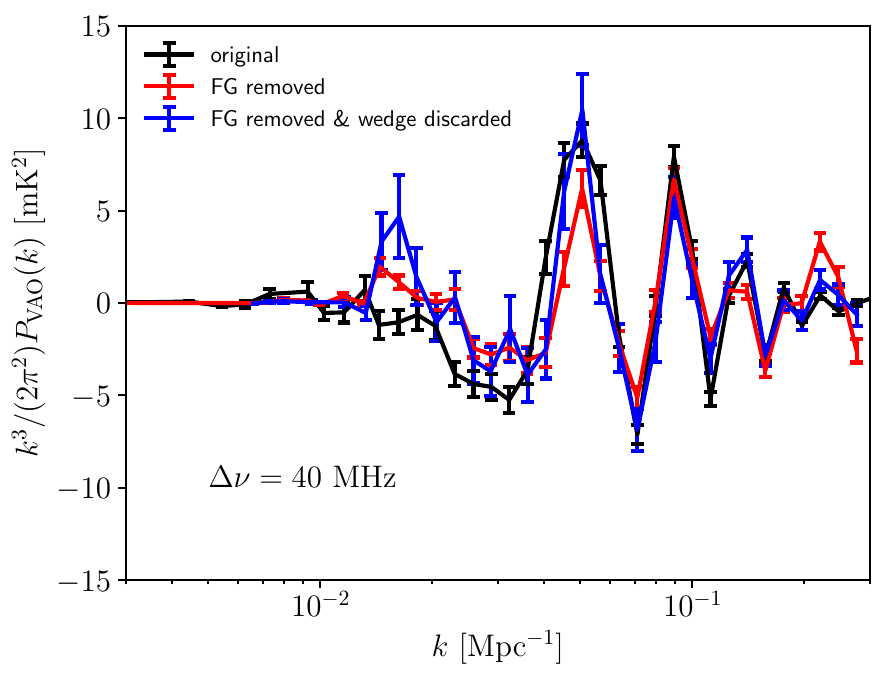}}
\caption{
{\it Left:} The 21 cm power spectra of the mock data cubes if we remove the foreground, and if we remove the foreground and discard the modes in the foreground contaminated wedge. 
{\it Right:} The extracted VAO wiggles from the left panels.
{\it Top row:} bandwidth 20 MHz. {\it Bottom row:}  bandwidth 40 MHz. 
}
\label{fig:wedge}
}
\end{figure*}

Another effect is the foreground wedge. Because of the chromatic instrument response, the foreground leakage contaminates the cosmological signal with Fourier modes in a wedge in the $(k_\perp,k_\parallel)$ space defined as (\citealt{Datta2010ApJ,Morales2012ApJ,Liu2014aPhRvD,Liu2014bPhRvD,Jensen2016MNRAS} and references therein)
\begin{equation}
k_\parallel \lesssim \sin(\theta_{\rm FoV}) k_\perp \left( \frac{D_M(z)E(z)}{D_H(1+z)}  \right), 
\end{equation}
where $k_\perp$ and $k_\parallel$ are Fourier wavenumbers perpendicular and parallel to the line of sight respectively.
$\theta_{\rm FoV}$ is the field of view angular size. $D_H=c/H_0$ and $D_M$ is the transverse comoving distance, where $c$ is the light speed and $H_0$ is the Hubble constant. $E(z)=\sqrt{\Omega_m(1+z)^3+\Omega_\Lambda}$.
Although in principle one can carefully perform foreground removal to reduce the foreground contamination in the wedge, the simplest way is just to discard all modes in the wedge.  
Foreground removal is necessary because the first peak of the VAO wiggles appears at $k\sim 0.05~{\rm Mpc}^{-1}$, however, the foreground dominates over the cosmological signal at least for $k_\parallel\lesssim 0.1~{\rm Mpc}^{-1}$, see e.g. \citealt{Chapman2016MNRAS}.
In Fig. \ref{fig:wedge} we plot the power spectra and the extracted VAO wiggles of the foreground removed and wedge discarded mock data cubes. 
Surprisingly and interestingly, 
we find that if we perform foreground removal
first and then discard the modes suspected to be contaminated in the wedge, the power loss discussed in the last paragraph is somewhat corrected. At $k\gtrsim 0.01~{\rm Mpc}^{-1}$ the recovered 21 cm power spectra and the extracted VAO wiggles are quite close to the original ones. Probably because the power loss in foreground removal is mainly in the modes with small $k_\parallel$. When we discard the modes in the wedge, we also discard many modes with small $k_\parallel$. As a result, they will not bias the recovered power spectra. For this reason, we believe that in our previous predictions, if we evaluate the influence of the foreground by such a method, the results do not change much.
However, we note that this is because we assume the cosmological signal is isotropic. The real 21 cm power spectrum could be anisotropic, discarding the wedge modes may bias the isotropically averaged power spectrum \citep{Jensen2016MNRAS}.

\section*{Acknowledgements}

This work is supported by National SKA Program of China Nos. 2020SKA0110402 and 2020SKA0110401, and the National Natural Science Foundation of China Grant No. 11973047. Y.G. acknowledges the support of National Key R\&D Program of China grant Nos. 2022YFF0503404, the CAS Project for Young Scientists in Basic Research (No. YSBR-092). This work is also supported by science research grants from the China Manned Space Project with grant Nos. CMS-CSST-2021-B01 and CMS- CSST-2021-A01.

%\appendix

% 
%\renewcommand\thefigure{A\arabic{figure}} 
%\setcounter{figure}{0}  
%

\bibliography{refe}{}
\bibliographystyle{aasjournal}



\end{document}